\documentclass[aps,prc,twocolumn,superscriptaddress,showpacs,preprintnumbers,nofootinbib]{revtex4-1}
\usepackage{epsfig,bm,epsf,float,footnote}
\usepackage{amsmath}
\usepackage{amsthm}
\usepackage[dvips]{color}
\usepackage[breaklinks=true]{hyperref}

\begin{document}

\preprint{YITP-15-112}

\title{Structure of the $\Lambda(1405)$ and the $K^- d\rightarrow
\pi\Sigma n$ reaction}


\author{Shota Ohnishi}
\email[]{s\_ohnishi@nucl.sci.hokudai.ac.jp}
\affiliation{Department of Physics, Hokkaido University, Sapporo 060-0810, Japan}
\author{Yoichi Ikeda}
\affiliation{RIKEN Nishina Center, Wako, Saitama 351-0198, Japan}

\author{Tetsuo Hyodo}
\affiliation{Yukawa Institute for Theoretical Physics, Kyoto University, Kyoto 606-8502, Japan}

\author{Wolfram Weise}
\affiliation{ECT*, Villa Tambosi, I-38123 Villazzano (Trento), Italy}
\affiliation{Physik Department, Technische Universit\"{a}t M\"{u}nchen, D-85747 Garching, Germany}


\date{\today}

\begin{abstract}
The $\Lambda(1405)$ resonance production reaction
 is investigated within the framework of the coupled-channels
 Alt-Grassberger-Sandhas (AGS) equations.
We perform full three-body calculations for the  $\bar{K}NN$-$\pi YN$
 amplitudes 
 on the physical real energy axis and investigate how the signature of
 the $\Lambda(1405)$ appears in the cross sections of the $K^-
 d\rightarrow \pi\Sigma n$ reactions, also in view of the planned 
 E31 experiment at J-PARC.
 Two types of meson-baryon interaction models are considered: an
 energy-dependent interaction based on chiral $SU(3)$ effective
 field theory, and an energy-independent version that has been used 
 repeatedly  in phenomenological approaches.
 These two models have different off-shell properties that imply
 correspondingly different behavior in the three-body system.
 We investigate how these features show up in differential cross sections of
$K^-  d\rightarrow \pi\Sigma n$ reactions. Characteristic patterns distinguishing
between the two models are found in the invariant mass spectrum of the
 final $\pi\Sigma$ state.
The  $K^-d\rightarrow \pi\Sigma n$ reaction, with different ($\pi^{\pm}\Sigma^{\mp}$ 
and $\pi^{0}\Sigma^{0}$) charge combinations in the final state, is thus demonstrated to be 
a useful tool for investigating the subthreshold behavior of the $\bar{K}N$ interaction.
\end{abstract}


\maketitle

\section{Introduction}
Understanding the structure of the $\Lambda(1405)$ 
is a long-standing issue in hadron physics. The nominal location of the 
$\Lambda(1405)$ mass, 27 MeV below the $K^- p$ threshold, deviates
prominently from the expected naive quark model pattern and indicates a more complex 
structure. Following early work by Dalitz {\it et al.} more than half a century
ago~\cite{Dalitz:1959dn,Dalitz:1960du}, the $\Lambda(1405)$ began to be considered as a
quasi-bound $\bar{K}N$ state embedded in the $\pi\Sigma$ continuum.
Motivated by such a picture, phenomenological $\bar{K}N$ potential models were designed to 
reproduce the $\Lambda(1405)$ mass together with two-body scattering data ~\cite{Akaishi:2002bg,Shevchenko:2011ce}. 

A more systematic framework emerged with developments
of meson-baryon effective field theory based on the spontaneous breaking of
chiral $SU(3)_L\times SU(3)_R$ symmetry in low-energy QCD. In this theory the kaon is
part of the pseudoscalar octet of Nambu-Goldstone bosons, but with an important explicit chiral symmetry breaking term introduced by its mass, $m_K \sim 0.5$ GeV, that reflects the relatively large mass of the strange quark, $m_s \sim 0.1$ GeV. Over the years, chiral $SU(3)$ dynamics, as the synthesis of chiral effective field theory and coupled channels methods~\cite{Kaiser:1995eg,Oset:1997it,Oller:2000fj,Hyodo:2011ur}, has turned out to be a highly successful approach to deal with $\bar{K}N$ interactions and the $\Lambda(1405)$. 

Even though the phenomenological and the chiral SU(3) $\bar{K}N$ interactions produce comparable
results at and above $\bar{K}N$ threshold, they differ significantly in their extrapolations to subthreshold energies~\cite{Hyodo:2007jq}. 
The phenomenological $\bar{K}N$ interactions are constructed to describe the $\Lambda(1405)$ as a single
pole of the scattering amplitude around 1405~MeV, corresponding to
a quasi-bound state of the $\bar{K}N$ system with a binding energy of about 30~MeV.
On the other hand, the $\bar{K}N$-$\pi\Sigma$ coupled-channels amplitude resulting from chiral SU(3)
dynamics has two poles, one of which is located around
1420 MeV~\cite{Oller:2000fj,Jido:2003cb} while the other pole represents a broad structure above
the $\pi\Sigma$ threshold. 
The pole at 1420 (rather than 1405) MeV corresponds to a $\bar{K}N$ quasi-bound system with 
a binding energy of 15~MeV, about half the binding produced with the purely phenomenological $\bar{K}N$ potentials. These differences in the pole structures come from different off-shell properties. The $\bar{K}N$ interaction based on chiral SU(3) dynamics
is necessarily energy-dependent: the Nambu-Goldstone boson nature of the $\bar{K}$ dictates that the leading-order $\bar{K}N$ $s$-wave interaction is proportional to the time derivative of the antikaon field and thus varies linearly with the $\bar{K}$ energy. Consequently, as one extrapolates deeper into the subthreshold region, the attraction generated by this interaction becomes progressively weaker than the one proposed by the energy-independent phenomenological potentials. At the same time, corresponding differences occur in the strong $\bar{K}N \leftrightarrow \pi\Sigma$ channel couplings. 

Hence the $\bar{K}N$ binding energies
predicted by interactions based on chiral $SU(3)$ dynamics are systematically
smaller than those suggested by the phenomenological models.
These differences are further enhanced in the so-called few-body kaonic nuclei, such as the strange
dibaryon resonance under discussion in the $\bar{K}NN$-$\pi YN$ coupled
system~\cite{Yamazaki:2002uh, Shevchenko:2006xy, Ikeda:2007nz,
Shevchenko:2007ke, Yamazaki:2007cs, Dote:2008in, Dote:2008hw,
Wycech:2008wf, Ikeda:2008ub, Ikeda:2010tk, Barnea:2012qa}.
How a possible signature of this strange dibaryon resonance shows up
in a suitable production reaction is of great interest as it reflects the two-body
dynamics in the $\Lambda(1405)$ channel~\cite{Ohnishi:2013rix}.

Exploring the structure of the $\Lambda(1405)$ requires a 
precise determination of the $\bar{K}N$-$\pi\Sigma$ interaction.
The data base available to constrain these interactions includes the old $K^- p$ scattering cross
sections~\cite{Humphrey:1962zz,Sakitt:1965kh,Kim:1965zz,Kittel:1966zz,Evans:1983hz}, 
the $\bar{K}N$ threshold branching ratios~\cite{Tovee:1971ga,Nowak:1978au}, 
and the kaonic hydrogen measurements~\cite{Iwasaki:1997wf,Ito:1998yi,Beer:2005qi}
with special emphasis on more recent accurate SIDDHARTA data~\cite{Bazzi:2011zj,Bazzi:2012eq}.
These latter data strongly constrain the $\bar{K}N$ input, as shown by
the systematic study of chiral SU(3) dynamics using next-to-leading
order driving interactions~\cite{Ikeda:2011pi,Ikeda:2012au}. 
The experimental data just mentioned are collected at and above the $\bar{K}N$ threshold. Since $\pi\Sigma$ elastic scattering cannot be performed, the subthreshold energy region is only accessible  by measuring mass spectra of decay products in reactions producing the 
$\Lambda(1405)$. The relevant $\pi\Sigma$ spectra have recently been measured in photoproduction
reactions by the LEPS Collaboration at
SPring-8~\cite{Ahn:2003mv,Niiyama:2008rt} and by the CLAS Collaboration at JLab~\cite{Moriya:2013eb,Moriya:2013hwg},
and in $pp$ collisions by the HADES Collaboration at GSI~\cite{Agakishiev:2012xk}.
The importance of accurately determined $\pi\Sigma$ spectra as constraints for the subthreshold 
$\bar{K}N$ interaction has also been emphasized in Refs.~\cite{Roca:2013av,Roca:2013cca,Guo:2012vv,Mai:2014xna}.

Yet another process of prime interest is the
$K^-d\rightarrow\pi\Sigma n$ reaction. It was studied long
ago by Braun $et~al.$~\cite{Braun:1977wd} in a bubble-chamber
experiment at $K^{-}$ momenta between 686 and 844 MeV. A new experiment
is ongoing at J-PARC (E31~\cite{Noumi}) with a 1 GeV $K^-$
beam~\footnote{In this paper, we focus on the in-flight reactions with
relatively energetic incident kaons. The same $K^-d\rightarrow\pi\Sigma
n$ process at lower energy has been studied in
Ref.~\cite{Tan:1973at}. For theoretical studies with this kinematics,
see Refs.~\cite{Jido:2010rx,Revai:2012fx}.}. In the E31 experiment, the
$\pi\Sigma$ production cross
sections will be measured separately for all combinations of charges, i.e.,
$\pi^+ \Sigma^-$, $\pi^- \Sigma^+$, and $\pi^0\Sigma^0$.
It is therefore important to establish a theoretical framework for a detailed analysis of
this reaction.
Theoretical investigations of $K^- d\rightarrow \pi \Sigma n$ with comparable kinematics have previously been performed in simplified models assuming a two-step process\,\cite{Jido:2009jf,Miyagawa:2012xz,Jido:2012cy,YamagataSekihara:2012yv}.
To extract the information of the subthreshold $\bar{K}N$ interaction from the experimental spectrum, an improved framework for the reaction mechanism is called for.

In this work a full three-body calculation of the
$\bar{K}NN$-$\pi YN$ amplitude is performed employing the coupled-channels
Alt-Grassberger-Sandhas (AGS) equations. We investigate how the  
$\Lambda(1405)$ resonance manifests itself in the differential cross section of the
$K^- d\rightarrow \pi\Sigma n$ reaction. At J-PARC it is planned to observe the $\Lambda(1405)$ in the $\pi\Sigma$ mass spectrum measured by detecting the forward kicked-out
neutron~\cite{Noumi}. Our calculation focuses on this observable.
One of the aims is to study the role of different off-shell properties of the underlying interactions
as they are realized in chiral $SU(3)$ dynamics versus phenomenological potential models. We
thus employ two different types of $\bar{K}N$-$\pi\Sigma$ interactions,
i.e., energy-dependent (E-dep.) and energy-independent (E-indep.), and
examine how the different off-shell properties of these interactions show up in the three-body dynamics.

In Sec.~\ref{sec:three-body}, we introduce the AGS equations for the three-body
$\bar{K}NN$-$\pi YN$ system and derive the cross section for the $K^-d\rightarrow \pi\Sigma n$ reaction. The two-body interactions used in this work are summarized in Sec.~\ref{sec:two-body}.
The numerically computed differential cross sections are presented and discussed in Sec.~\ref{sec:result}. A summary follows in Sec.~\ref{sec:summary}.

\section{Three-Body Equations}
\label{sec:three-body}
\subsection{Alt-Grassberger-Sandhas equations for the $K^- d\rightarrow \pi\Sigma n$ 
  reaction}

We begin by constructing the three-body amplitudes relevant to the $K^-d\rightarrow \pi\Sigma n$ reaction.
Throughout this paper it is assumed 
that the three-body processes take place via
separable two-body interactions given by
the following forms 
in the two-body center-of-mass (c.m.) frame:
\begin{align}
V_{\alpha\beta}^{(I)}(\bm{q}_i',\bm{q}_i ; E) = 
g_{\alpha }^{\ast(I)}(\bm{q}_i ') \,\,\lambda_{\alpha\beta}^{(I)}(E)
 \,\,g_{\beta }^{(I)}(\bm{q}_i) ~,
\label{eq:v_sepa}
\end{align}
where $g_{\alpha}^{(I)}(\bm{q}_i)$ is a vertex (cutoff) factor
of the two-body channel $\alpha$ with relative momentum $\bm{q}_i$ and
isospin $I$. The interaction matrix $\lambda_{\alpha\beta}^{(I)}(E)$ is a function of
the total energy $E$ in the two-body system.
In the three-body system, we define the two-body energy as
$E=\sqrt{(W-E_i(\bm{p}_i))^2-p_i^2}$ with the three-body energy $W$ and the spectator
particle energy $E_i(\bm{p}_i)$, where
$\bm{p}_i$ is the relative momentum of the spectator particle $i$.
The explicit forms of the relevant two-body interactions are presented in detail in Sec.~\ref{sec:two-body}.

\begin{table*}[thb]
\caption{
 Indices specifying the two-body subsystems (``isobars'').
 Symbols $Y$ denote hyperons $\Lambda$ and $\Sigma$.
 The isospins in parentheses are allowed for $Y=\Sigma$.
 Mass splittings in the isospin multiplets are neglected.
}
\label{isobar}      
\begin{ruledtabular}
\begin{tabular}{lcccccc}
Isobar& Allowed isospin(s) & Spectator particle & Three-body Fock space\\
\hline
$Y_K={\bar K_3 N_2},\bar{K}_3N_1$ & 0, 1  & $N_1,N_2$ &$\left|
		 N_1N_2\bar{K}_3\right\rangle$\\
$Y_\pi=\pi_3Y_2,\pi_3Y_1$ & (0), 1 & $N_1,N_2$
	     &$\left|N_1Y_2\pi_3\right\rangle,\left|Y_1 N_2\pi_3\right\rangle$\\
$d=N_1N_2$ & 0 & $\bar K_3$ &$\left| N_1N_2\bar{K}_3\right\rangle$\\
$N^*=\pi_3 N_2,\pi_3N_1$  & 1/2, (3/2) & $Y_1,Y_2$ &$\left| Y_1 N_2\pi_3\right\rangle,\left|N_1Y_2\pi_3\right\rangle$\\
$d_y=Y_1 N_2,Y_2 N_1$ & 1/2, (3/2) & $\pi_3$ &$\left| Y_1 N_2\pi_3\right\rangle,\left|N_1Y_2\pi_3\right\rangle$\\
\end{tabular}
\end{ruledtabular}
\end{table*}

The ansatz (\ref{eq:v_sepa})
 specifies strongly interacting two-body subsystems in the three-body processes. We refer to these 
meson-baryon or dibaryon subsystems conveniently as ``isobars''. The
 three-body dynamics can then
be described as quasi-two-body scattering of 
an isobar and a spectator particle in all possible coupled
isobar-spectator channels.
The quasi-two-body amplitudes, $X_{\alpha\beta}^{(I)(I')}(\bm{p}_i, \bm{p}_j; W)$, 
are determined by solving
the AGS equations \cite{PhysRev.132.485,Alt:1967fx},
\begin{align}
	X&_{\alpha\beta}^{(I)(I')}({\bm{p}}_i,{\bm{p}}_j,W)\nonumber\\
 &=(1-\delta_{ij})\,
	Z_{\alpha\beta}^{(I)(I')}({\bm{p}}_i,{\bm{p}}_j,W)
\nonumber\\
&~~~+\sum_{\gamma,\delta}\sum_{I''}\sum_{n\ne i}\int d^3{\bm{p}}_n\,\, Z_{\alpha\gamma}^{(I)(I'')}({\bm{p}}_i,{\bm{p}}_n,W)
\nonumber\\
&~~~\times
	\tau^{(I'')}_{\gamma\delta}\left(W-E_n(\bm{p}_n),\bm{p}_n\right) X_{\delta\beta}^{(I'')(I')}
({\bm{p}}_n,{\bm{p}}_j,W)~.
	\label{AGS}
\end{align}
Here, $\alpha$ and $\beta$ denote two-particle subsystems forming ``isobars'' with
isospins $I$ and $I'$, respectively;
the subscripts $i,j,n$ represent the spectator particles which include,
respectively, $N$, $\Sigma$, $\Lambda$, $\bar{K}$, or $\pi$.
The notations for the isobars are summarized in Table~\ref{isobar}.
As pointed out in Sec.~\ref{sec:two-body}, in this work 
we include only the $^3S_1$ partial wave for the $NN$ interaction,
thus only the isospin $I=0$ state appears for the $NN$ subsystem (the isobar denoted $d$).
For later purposes the partial wave projections of the amplitudes
$X$ of Eq.~(\ref{AGS}) are needed.
They are given as:
\begin{align}
X_{\alpha\beta,L}^{(I)(I')}&(p_i, p_j; W) \nonumber\\
 &= \frac{1}{2}\int_{-1}^1 d\cos\theta~ X_{\alpha\beta}^{(I)(I')}(\bm{p}_i, \bm{p}_j; W)~P_L(\cos\theta)
\label{eq:z-diagram-s}
\end{align}
with $\cos\theta = \hat {\bm{p}}_i \cdot \hat {\bm{p}}_j$
and the notation 
$\hat {\bm{p}} \equiv \bm{p}/|\bm{p}|$.
Here, $P_L$ is the Legendre polynomial with orbital angular
momentum $L$ between isobar and spectator particle.
After the partial wave projections
the AGS equations~(\ref{AGS})
are written as:
\begin{align}
	X&_{\alpha\beta,L}^{(I)(I')}(p_i,p_j,W)\nonumber\\
 &=(1-\delta_{ij})\,
	Z_{\alpha\beta,L}^{(I)(I')}(p_i,p_j,W)
\nonumber\\
&~~~+\sum_{\gamma,\delta}\sum_{I''}\sum_{n\ne i}\int dp_np_n^2\,\, Z_{\alpha\gamma,L}^{(I)(I'')}(p_i,p_n,W)
\nonumber\\
&~~~\times
	\tau^{(I'')}_{\gamma\delta}\left(W-E_n(\bm{p}_n),\bm{p}_n\right) X_{\delta\beta,L}^{(I'')(I')}
(p_n,p_j,W)~.
	\label{AGS_partial}
\end{align}

The driving term $Z_{\alpha\beta}^{(I)(I')}({\bm{p}}_i,{\bm{p}}_j;W)$ describes a particle-exchange interaction process connecting two-body channels, $\beta \rightarrow \alpha$, and corresponding spectators as illustrated in Fig.~\ref{z-diagram}(a). It is given by:
\begin{align}
Z_{\alpha\beta}^{(I)(I')}&({\bm{p}}_i,{\bm{p}}_j;W)\nonumber\\
& =
\frac{g_{\alpha }^{(I)}(\bm{q}_i)g_{\beta}^{\ast(I')}(\bm{q}_j)}{W - E_i(\bm{p}_i) - E_j(\bm{p}_j) - E_k (\bm{p}_k) + i\epsilon},
\label{eq:z-diagram}
\end{align}
where $E_i(\bm{p}_i)$ and $E_j(\bm{p}_j)$ are the energies of the
spectator particles $i$ and $j$, respectively;
$E_k(\bm{p}_k)$ with $\bm{p}_k = - \bm{p}_i - \bm{p}_j$ is the energy of the exchanged particle $k$;
$\bm{q}_i$ ($\bm{q}_j$) is the relative momentum between the exchange-particle 
and the spectator-particle $j$ ($i$).
Using relativistic kinematics we have 
$E_{n}(\bm{p}_n) = \sqrt{m_n^2 + \bm{p}_{n}^2}$ ($n=i,j,k$)
and $q_i=|\bm{q}_i|$ is defined as
\begin{align}
 q_i &= \sqrt{\left(\frac{M_{jk}^2+m_j^2-m_k^2}{2M_{jk}}\right)^2-m_j^2}~~,\\
 M_{jk}(\bm{q}_i) &= \sqrt{(E_j(\bm{p}_j) +E_k
 (\bm{p}_k))^2-\bm{p}_i^2}~~\nonumber\\
 &= E_j(\bm{q}_i) + E_k(\bm{q}_i)~~.
\end{align}
%
%

\begin{widetext}
The isobar amplitudes, $\tau _{\alpha\beta}^{(I)}\left(W-E_i(\bm{p}_i),\bm{p}_i\right)$ as illustrated in 
Fig.~\ref{z-diagram} (b), are determined by solving
the Lippmann-Schwinger equations with the two-body interaction~(\ref{eq:v_sepa}),
\begin{align}
 \tau_{\alpha\beta}^{(I)}(W-E_i(\bm{p}_i),\bm{p}_i)= 
  \lambda_{\alpha\beta}^{(I)}
  +\sum_{\gamma}\int d{ q}_i q_i^2
  \frac{\lambda_{\alpha\gamma}^{(I)}|g_{\gamma}^{(I)}({ q}_i)|^2}
  {W-E_i({\bm{p}}_i)-E_{jk}({\bm{p}}_i,{\bm{q}}_i)}~\tau_{\gamma\beta}^{(I)}(W-E_i(\bm{p}_i),\bm{p}_i)~~.
\end{align}
Here, $E_{jk}({\bm{p}}_i,{\bm{q}}_i)$ is the energy of the interacting pair ($jk$),
$
E_{jk}({\bm{p}}_i,{\bm{q}}_i) = \sqrt{M_{jk}^2(\bm{q}_i)+\bm{p}_i^2}
$.

\begin{figure*}[tbh]
 \includegraphics[width=0.30\textwidth,clip]{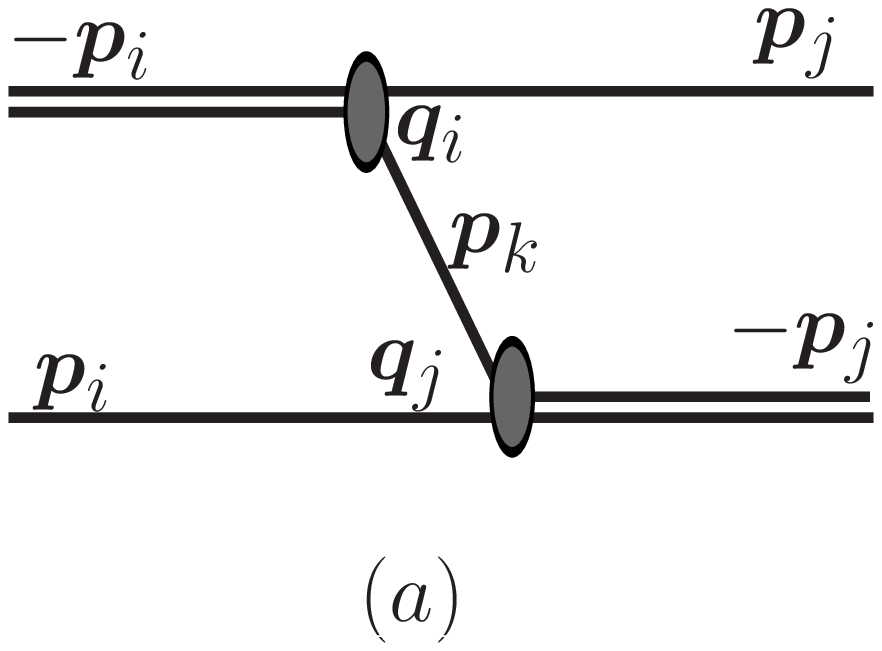}
 \includegraphics[width=0.30\textwidth,clip]{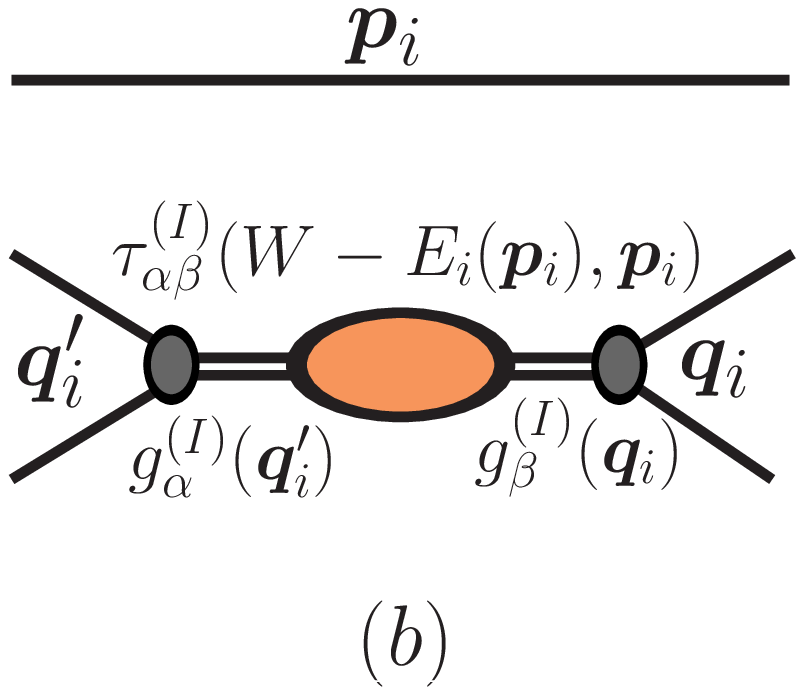}
 \caption{(a) One particle exchange interaction $Z_{\alpha\beta,L}^{(I)(I')}({ p}_i,{
 p}_j,W)$. (b) Isobar propagator $\tau_{\alpha\beta}^{(I)}(W-E_i(\bm{p}_i),\bm{p}_i)$.}
 \label{z-diagram}
\end{figure*}

After antisymmetrization of the two-nucleon states in the three-body system, 
the coupled-channels AGS matrix integral equations (\ref{AGS_partial}) are
 written formally and symbolically 
(suppressing sums, integrals and all indices other than the isobar assignments) as:
\begin{align}
 \begin{pmatrix}
   X_{Y_K d}\\
   X_{Y_\pi  d}\\
   X_{d d}\\
   X_{N^*  d}\\
   X_{d_y  d}
  \end{pmatrix}
 =
 \begin{pmatrix}
  2Z_{Y_K  d}\\
  0\\
  0\\
  0\\
  0
 \end{pmatrix}
-
  \begin{pmatrix}
  Z_{Y_K Y_K}\tau_{Y_K Y_K}&Z_{Y_K Y_K}\tau_{Y_K Y_\pi}&2Z_{Y_K d}\tau_{d d}&0&0\\
  0&0&0&Z_{Y_\pi  N^*}\tau_{N^* N^*}&Z_{Y_\pi  d_y}\tau_{d_y d_y}\\
  Z_{d Y_K}\tau_{Y_K Y_K}&Z_{d Y_K}\tau_{Y_K Y_\pi}&0&0&0\\
  Z_{N^*  Y_\pi}\tau_{Y_\pi  Y_K}&Z_{N^*  Y_\pi}\tau_{Y_\pi  Y_\pi}&0&0&Z_{N^*  d_y}\tau_{d_y d_y}\\
  Z_{d_y  Y_\pi}\tau_{Y_\pi  Y_K}&Z_{d_y  Y_\pi}\tau_{Y_\pi  Y_\pi}&0&Z_{d_y  N^*}\tau_{N^*  N^*}&0
\end{pmatrix}
 \begin{pmatrix}
   X_{Y_K d}\\
   X_{Y_\pi  d}\\
   X_{d  d}\\
   X_{N^*  d}\\
   X_{d_y  d}
 \end{pmatrix}.\label{coupled-AGS}
\end{align}

\subsection{Cross sections for $K^- d \rightarrow \pi\Sigma
  n$}
In this subsection, following Ref.~\cite{Ohnishi:2013rix},
we present formulas for computing cross sections of
the two-body-to-three-body reaction, $K^- d\rightarrow \pi\Sigma n$.
By using the anti-symmetrized AGS amplitudes~(\ref{coupled-AGS}), the breakup amplitudes
for $K^- d\rightarrow \pi\Sigma n$ are
given as 
\begin{align}
  &T_{K^- d\rightarrow \pi\Sigma n}(\bm{q}_N,\bm{p}_N,\bm{p}_{\bar{K}},W) \nonumber\\
 &=
 \frac{1}{\sqrt{2}}\sum_{I,L}\nonumber\\
 &\times\Big[\langle \pi\Sigma n|[[\pi\otimes \Sigma]_{Y_\pi}\otimes N]_\Gamma;I,L\rangle~
 g_{Y_\pi}^{(I)}(q_N)~\tau_{Y_\pi Y_K}^{(I)}(W-E_N(\bm{p}_N),\bm{p}_N)~X_{Y_K d,L}^{(I)(I=0)}(p_N,p_{\bar{K}},W)\nonumber\\
 & +\langle \pi\Sigma n|[[\pi\otimes \Sigma]_{Y_\pi}\otimes N]_\Gamma;I,L\rangle~
 g_{Y_\pi}^{(I)}(q_N)~\tau_{Y_\pi Y_\pi}^{(I)}(W-E_N(\bm{p}_N),\bm{p}_N)~X_{Y_\pi d,L}^{(I)(I=0)}(p_N,p_{\bar{K}},W) \nonumber\\
 &+\langle \pi\Sigma n|[[\pi\otimes N]_{N^*}\otimes \Sigma]_\Gamma;I,L\rangle~
 g_{N^*}^{(I)}(q_\Sigma)~\tau_{N^*N^*}^{(I)}(W-E_\Sigma(\bm{p}_\Sigma),\bm{p}_\Sigma)~X_{N^*d,L}^{(I)(I=0)}(p_\Sigma,p_{\bar{K}},W)\nonumber\\
 & +\langle \pi\Sigma n|[[\Sigma\otimes N]_{d_y}\otimes \pi]_\Gamma;I,L\rangle~
 g_{d_y}^{(I)}(q_\pi)~\tau_{d_yd_y}^{(I)}(W-E_\pi(\bm{p}_\pi),\bm{p}_\pi)~X_{d_yd,L}^{(I)(I=0)}(p_\pi,p_{\bar{K}},W)
 \Big]\nonumber\\
 &\times\langle [d\otimes \bar{K}]_{\Gamma'};I=0,L|d K^- \rangle\sqrt{R_d}~~.
\label{eq:T_Kd_piSn}
\end{align}
Here
$R_d$ is the residue of the two-body $NN$ propagator,
$\tau_{dd}^{(I=0)}$ at the deuteron pole, with its proper binding
energy, i.e., $\sqrt{R_d}$
normalizes the initial state deuteron wave function.
Note again that all two-body subsystems listed in Table~\ref{isobar},
including both hyperons $Y=\Sigma$ and $\Lambda$, contribute to  $T_{K^-
d\rightarrow \pi\Sigma n}$ when permitted by selection rules.
The following notations
are used for the expressions appearing in Eq.~(\ref{eq:T_Kd_piSn}):\\

$|ABC\rangle$: plane wave state of the three-body system;\\

$|[[A\otimes B]_\alpha \otimes C]_\Gamma ;I,L\rangle$:
three-body system in the $LS$ coupling scheme, with $\alpha$, $\Gamma$, $I$, and $L$ being the isobar
quantum number,
the total quantum number,
the isospin of the isobar and its angular momentum relative to the spectator,
respectively.\\

The projection $\langle ABC|[[A\otimes B]_\alpha \otimes C]_\Gamma ;I,L\rangle$ involves the
product of spherical harmonics and spin-isospin Clebsch-Gordan
coefficients.
The $T$ matrix calculated in the isospin basis is then
decomposed into the $\pi^+\Sigma^- n$, $\pi^0\Sigma^0 n$, and $\pi^-\Sigma^+ n$ final states
using isospin CG coefficients.
The momenta $\bm{p}_\pi$ and $\bm{p}_\Sigma$ are related to the momenta
$\bm{p}_N$ and $\bm{q}_N$ by a Lorentz boost
\begin{align}
 \bm{p}_\pi &= \bm{q}_N
 -\frac{\bm{p}_N}{M_{\pi\Sigma}(\bm{q}_N)}\left[E_\pi(\bm{q}_N)
 -\frac{\bm{p}_N\cdot
 \bm{q}_N}{E_{\pi\Sigma}(\bm{p}_N,\bm{q}_N)+M_{\pi\Sigma}(\bm{q}_N)}\right]~~,\\
 \bm{p}_\Sigma &= -\bm{q}_N
 -\frac{\bm{p}_N}{M_{\pi\Sigma}(\bm{q}_N)}\left[E_\Sigma(\bm{q}_N)
 +\frac{\bm{p}_N\cdot
 \bm{q}_N}{E_{\pi\Sigma}(\bm{p}_N,\bm{q}_N)+M_{\pi\Sigma}(\bm{q}_N)}\right]~~.
\end{align}
With the $T$ matrix Eq.~(\ref{eq:T_Kd_piSn}) the cross sections
of interest are derived as
\begin{align}
 \sigma (W)
 &= \frac{(2\pi)^4}{v}\int d^3\bm{p}_Nd^3\bm{q}_N\sum_{\bar{i}f}
  \delta(W-E_N(\bm{p}_N)-E_{\pi\Sigma}(\bm{p}_N,\bm{q}_N))
 |T_{K^- d\rightarrow\pi\Sigma n} (\bm{q}_N,\bm{p}_N,\bm{p}_{\bar{K}},W)|^2 \nonumber\\
&= (2\pi)^4\frac{E_dE_{\bar{K}}}{Wp_{\bar{K}}}
\int dM_{\pi\Sigma} d\hat{\bm p}_N d\hat{\bm q}_N\frac{E_N(\bm{p}_N) E_\Sigma(\bm{p}_\Sigma)
 E_\pi(\bm{p}_\pi)}{W}
 p_Nq_N \sum_{\bar{i}f}|T_{K^-d\rightarrow\pi\Sigma n} (\bm{q}_N,\bm{p}_N,\bm{p}_{\bar{K}},W)|^2~~\label{eq:cross}
\end{align}
with the initial relative velocity
$v=\frac{W}{E_dE_{\bar{K}}}p_{\bar{K}}$,
the $\pi\Sigma$ invariant/missing mass
$M_{\pi\Sigma}=E_\pi(\bm{q}_N)+E_\Sigma(\bm{q}_N) =
\sqrt{(W-E_N(\bm{p}_N))^2-\bm{p}_N^2}$,
and the $K^-d$ total energy
$W=E_{\bar{K}}(\bm{p}_{\bar{K}})+E_d(\bm{p}_{\bar{K}})
=\sqrt{m_{\bar{K}}^2+\bm{p}_{\bar{K}}^2}+\sqrt{m_d^2+\bm{p}_{\bar{K}}^2}$.
In the second line of Eq.~(\ref{eq:cross}), the momenta $p_N$ and $q_N$
are the on-shell momenta for given energies $W$ and $M_{\pi\Sigma}$. Angular integrations are denoted by
$\int d\hat{\bm p} \equiv \int d\cos\theta_p\,d\phi_p$\,.
The differential cross sections are
\begin{align}
 \frac{d^2\sigma}{dM_{\pi\Sigma} d \cos\theta_{p_N}} =&
 (2\pi)^4\frac{E_dE_{\bar{K}}}{Wp_{\bar{K}}}
 \int d\phi_{p_N}d\hat{\bm q}_N\frac{E_N(\bm{p}_N) E_\Sigma(\bm{p}_\Sigma)
 E_\pi(\bm{p}_\pi)}{W}
 p_Nq_N \sum_{\bar{i}f}|T_{K^- d\rightarrow\pi\Sigma n}
 (\bm{q}_N,\bm{p}_N,\bm{p}_{\bar{K}},W)|^2~~,\label{eq:differential}\\
 \frac{d\sigma}{dM_{\pi\Sigma}} =& (2\pi)^4\frac{E_dm_{\bar{K}}}{Wp_{\bar{K}}}
 \int d\hat{\bm p}_N d \hat{\bm q}_N\frac{E_N(\bm{p}_N) E_\Sigma(\bm{p}_\Sigma)
 E_\pi(\bm{p}_\pi)}{W}
 p_Nq_N \sum_{\bar{i}f}|T_{K^- d\rightarrow\pi\Sigma n}
(\bm{q}_N,\bm{p}_N,\bm{p}_{\bar{K}},W)|^2~~\label{eq:differential2},
\end{align}
where
\begin{align}
 \cos \theta_{p_N}=\hat{\bm p}_N \cdot \hat{\bm q}_N ~~.
\end{align} 
The symbol $\sum_{\bar{i}f}$ stands as usual for averaging of
initial states and sum of final states subject
to conservation laws.
\end{widetext}

\section{Two-Body Interactions}
\label{sec:two-body}

\begin{figure*}[tbh]
 \includegraphics[width=0.329\textwidth,clip]{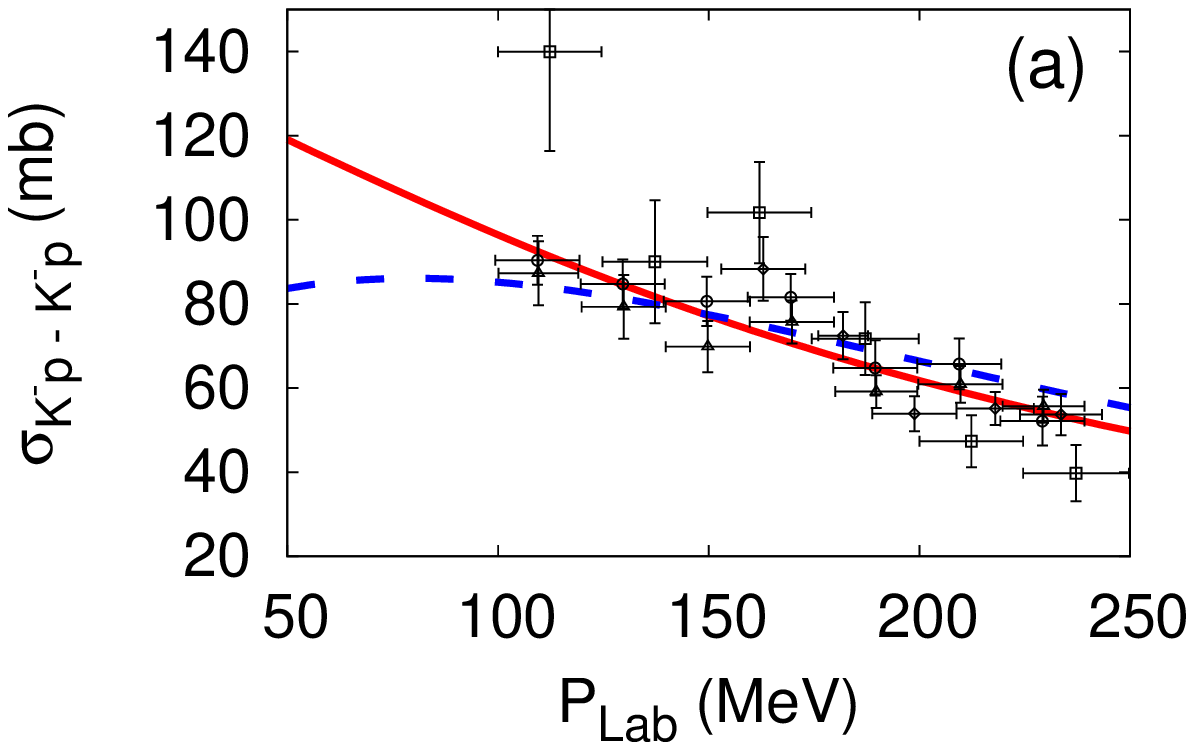}
 \includegraphics[width=0.329\textwidth,clip]{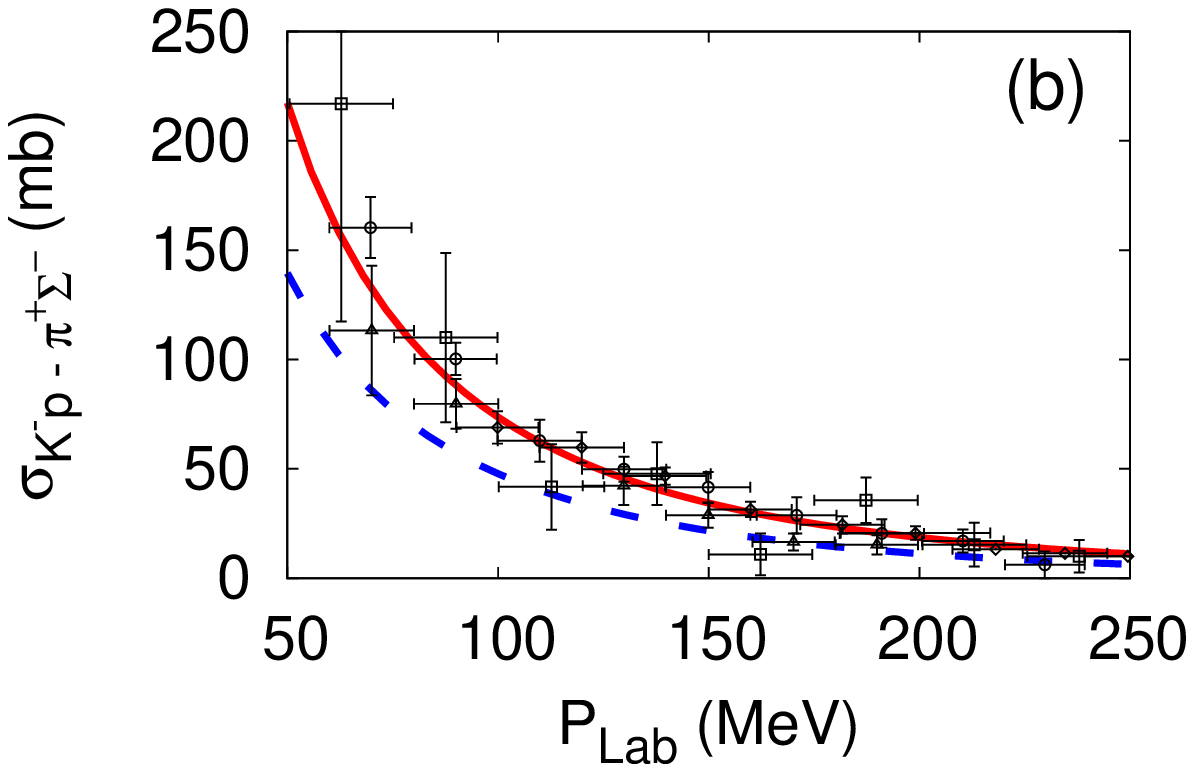}
 \includegraphics[width=0.329\textwidth,clip]{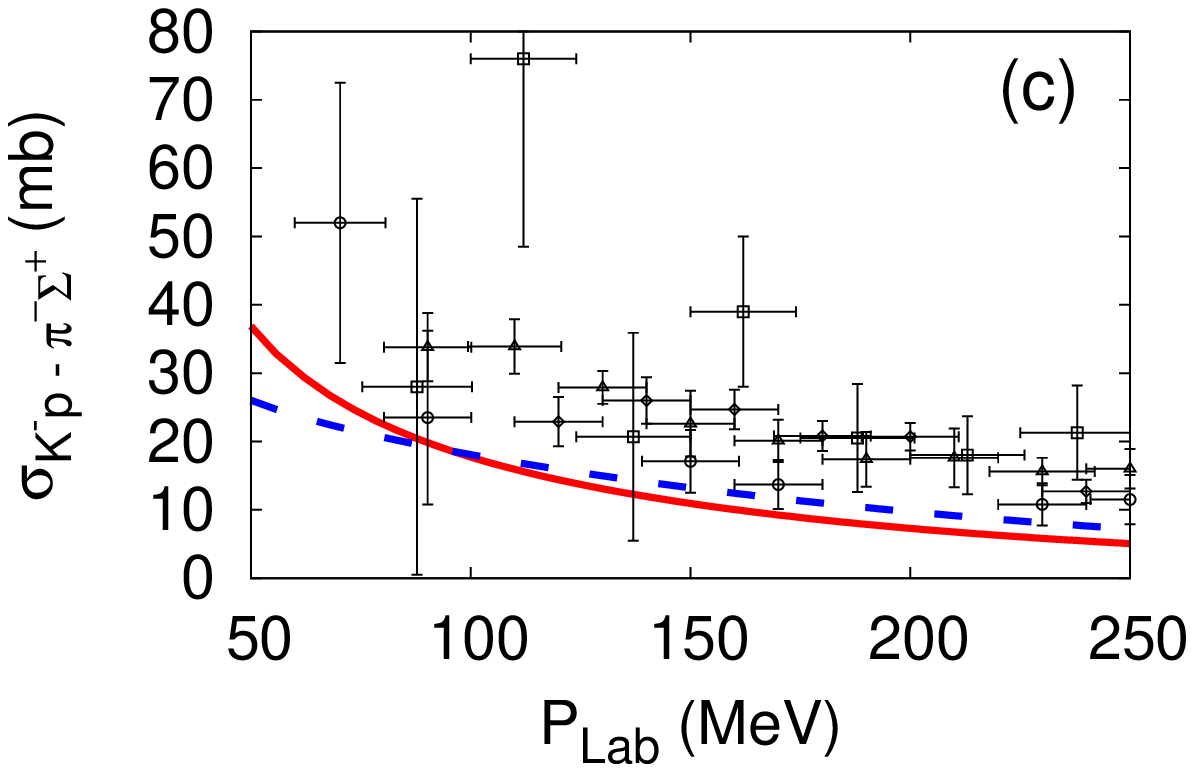}\\
 \includegraphics[width=0.329\textwidth,clip]{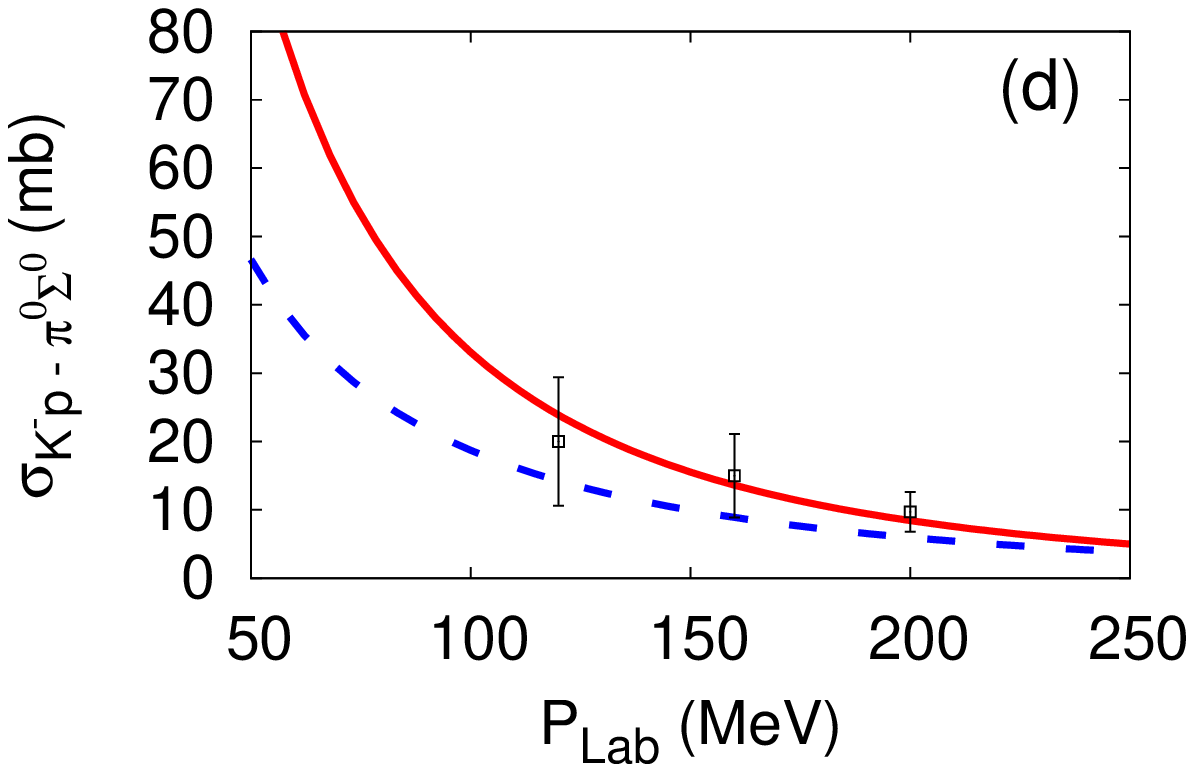}
 \includegraphics[width=0.329\textwidth,clip]{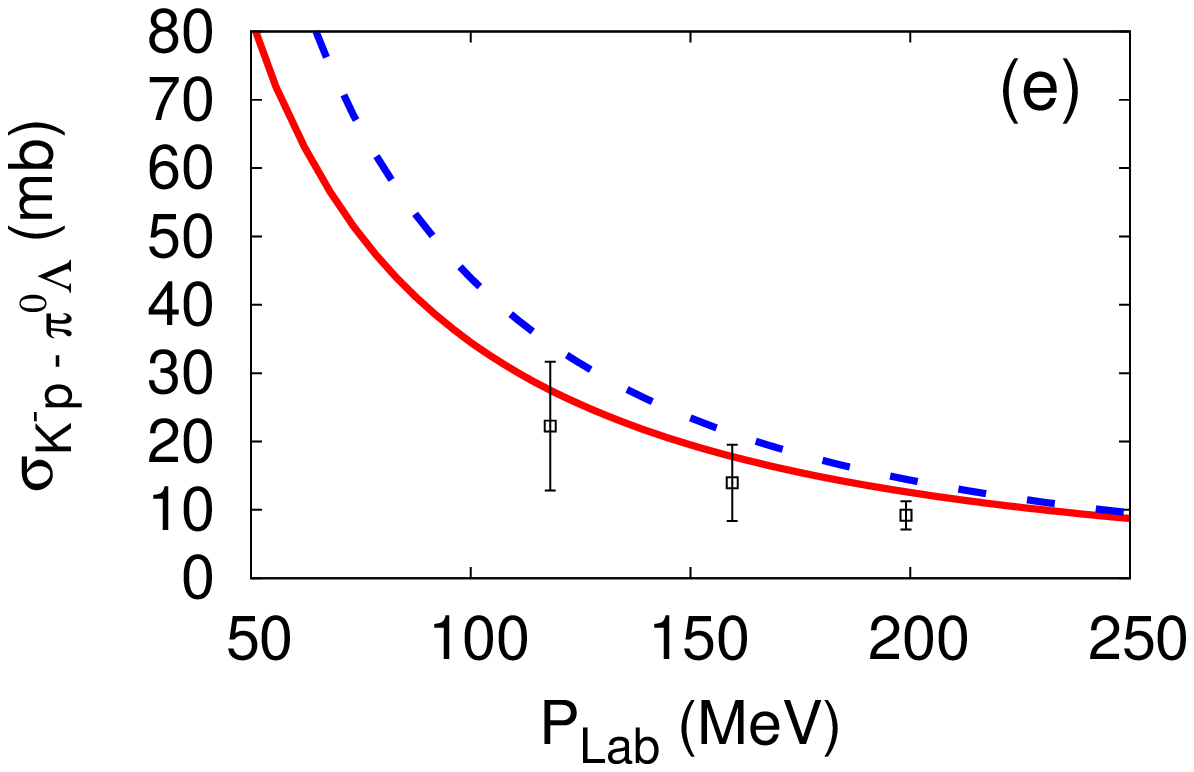}
 \caption{(Color online)
Results of the fit with the E-dep. model (solid lines) and
 E-indep. model (dashed lines).
Total cross sections of 
(a) $K^-p\rightarrow K^-p$, 
(b) $K^-p\rightarrow \pi^+\Sigma^-$, 
(c) $K^-p\rightarrow \pi^-\Sigma^+$,
(d) $K^-p\rightarrow \pi^0\Sigma^0$, and 
(e) $K^-p\rightarrow \pi^0\Lambda$.
 Data are from
 Refs.~\cite{Humphrey:1962zz,Sakitt:1965kh,Kim:1965zz,Kittel:1966zz,Evans:1983hz}.}
 \label{cross_fit}
\end{figure*}

\begin{table*}[thb]
\begin{center}
\caption{Cutoff parameters of the $\bar{K}N$-$\pi Y$ interaction.}
\label{cutoff}       
\begin{ruledtabular}
\begin{tabular}{lccccc}
 \vspace{-4mm}\\
 &$\Lambda^{(I=0)} _{Y_{K}}$ (MeV) &$\Lambda^{(I=0)}
 _{Y_{\pi}=\pi\Sigma}$ (MeV) &$\Lambda^{(I=1)} _{Y_{K}}$ (MeV)
 &$\Lambda^{(I=1)} _{Y_\pi=\pi\Sigma}$ (MeV)&$\Lambda^{(I=1)}
 _{Y_\pi=\pi\Lambda}$ (MeV) \\
 \vspace{-4mm}\\
\hline
E-dep.&1100 & 1100 & 800&800&800\\
E-indep.&1160 & 1100 & 1100&850&1250\\
\end{tabular} 
\end{ruledtabular}
\end{center}
\end{table*}

We refer back to Eq.\,(\ref{eq:v_sepa}) and explain the explicit forms of the 
meson-baryon interactions (Sec.~\ref{sec:mb-int})
and baryon-baryon interactions (Sec.~\ref{sec:bb-int}) used in this work.
In this section spectator indices are suppressed for simplicity.

\subsection{Meson-baryon interaction}
\label{sec:mb-int}

Two models for the $s$-wave meson-baryon
interactions used in
Refs.~\cite{Ikeda:2007nz,Ikeda:2008ub,Ikeda:2010tk,Ohnishi:2013rix}
are employed in this work.
Both are derived from the leading order chiral
Lagrangian (the Weinberg-Tomozawa term~\cite{Weinberg:1966kf,Tomozawa:1966jm}) but have different off-shell behavior.
One of them is referred to as the energy dependent (E-dep.) model~\cite{Ikeda:2010tk},
\begin{align}
V&^{(I)\text{E-dep.}}_{\alpha \beta}(q_{\alpha},q_\beta;E)\nonumber\\
&=-{C^I_{\alpha\beta}\over 32\pi^2 f_\pi^2}
\frac{2E-M_\alpha -M_\beta}{\sqrt{\mathstrut \omega_\alpha \omega_\beta}}
g^{(I)}_{\alpha}(q_{\alpha})
g^{(I)}_{\beta}(q_\beta).
\label{eq:e-dep}
\end{align}
Here,
$\omega_\alpha=\sqrt{q_\alpha^2 +m_\alpha^2}$ is the meson energy of the channel
$\alpha=\bar{K}N$, $\pi\Sigma$, $\pi\Lambda$;
$m_\alpha$ $(M_\alpha)$ is the meson (baryon) mass;
$f_\pi = 92.4$ MeV is the pion decay constant; the coupling coefficients
$C^I_{\alpha\beta}$ are determined by the flavor SU(3) structure
constant (see Ref.~\cite{Ohnishi:2013rix}).
The vertex
factors $g^{(I)}_{\alpha}(q_{\alpha})$ are chosen as dipole form
factors with cutoff scales $\Lambda^{(I)}_{\alpha}$,
\begin{align}
g^{(I)}_{{\alpha}}(q_{\alpha})=\left({\Lambda^{(I)2}_\alpha \over \Lambda^{(I)2}_{\alpha}+q_{\alpha}^2}\right)^2~.
\nonumber
\end{align}

The characteristic energy dependence of the meson-baryon interaction (\ref{eq:e-dep}) is dictated by spontaneously broken chiral $SU(3)_L\times SU(3)_R$ symmetry. The corresponding Nambu-Goldstone bosons are identified with the pseudoscalar meson octet, and their leading $s$-wave couplings involve the time derivatives of the meson fields.

The other model, referred to here as the energy independent
(E-indep.) model~\cite{Ikeda:2007nz,Ikeda:2008ub}, is obtained by
fixing the two-body energy at each threshold energy, $2E=m_\alpha+M_\alpha+m_\beta+M_\beta$:
\begin{align}
V&^{(I)\text{E-indep.}}_{\alpha \beta}(q_{\alpha},q_\beta)\nonumber\\
&=-{C^I_{\alpha\beta}\over 32\pi^2 f_\pi^2}
\frac{m_\alpha +m_\beta}{\sqrt{\mathstrut \omega_\alpha \omega_\beta}}
g^{(I)}_{\alpha}(q_{\alpha})
g^{(I)}_{\beta}(q_\beta) .
\label{eq:e-indep}
\end{align}
While this restricted model with constant couplings is not consistent with Goldstone's theorem for low-energy pseudoscalar meson interactions, it is nonetheless a prototype of phenomenological potentials that have been used in the literature, and so we discuss it here for comparison with the energy-dependent approach
based on chiral $SU(3)_L\times SU(3)_R$ meson-baryon effective field theory.

The cutoff parameters for the
$\bar{K}N$-$\pi\Sigma$-$\pi\Lambda$ systems are determined
by fitting 
the $K^-p$ scattering cross sections~\cite{Humphrey:1962zz,Sakitt:1965kh,Kim:1965zz,Kittel:1966zz,Evans:1983hz}.
Results of the fit for the E-dep. and E-indep. models are presented in 
Fig.~\ref{cross_fit}.
The fitted cutoff values are listed in Table~\ref{cutoff}.

\begin{figure*}[tbh]
\begin{tabular}{c}
   \includegraphics[width=0.5\textwidth]{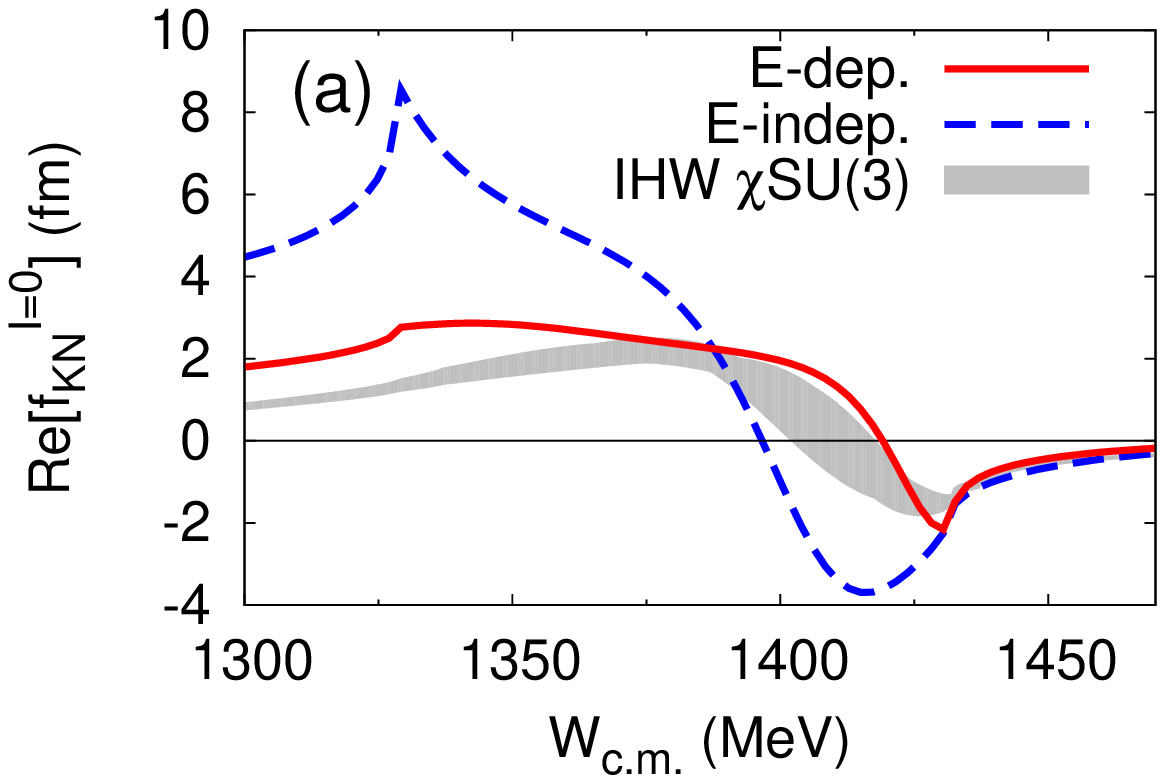}
   \includegraphics[width=0.5\textwidth]{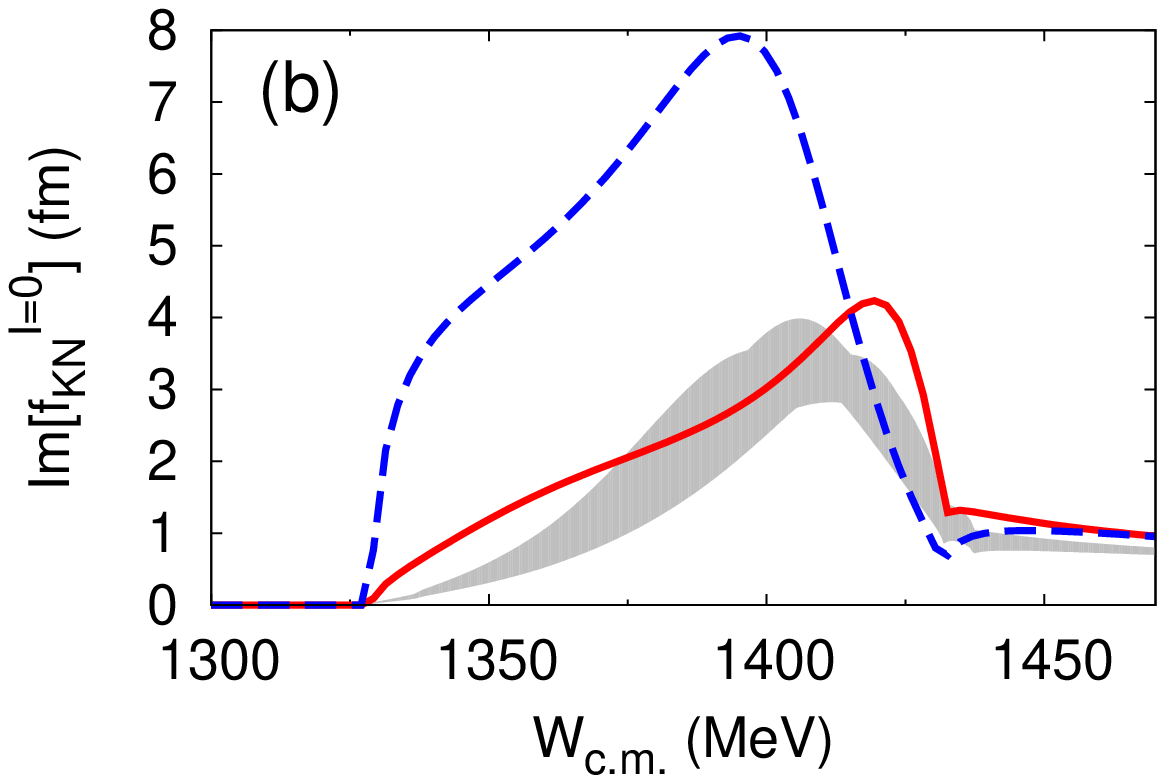}
\end{tabular}
 \caption{(Color online) 
 (a) Real and (b) imaginary parts of the $\bar{K}N$ amplitude in the
 isospin $I=0$ channel as functions of the total $\bar{K}N$
 center-of-mass energy.
 Solid curve: E-dep. model based on the chiral SU(3) potential~(\ref{eq:e-dep});
 dashed curve: E-indep. model using the potential~(\ref{eq:e-indep}).
 The shaded area show for comparison the $I=0$ $\bar{K}N$ amplitude from
 NLO chiral SU(3) dynamics including uncertainties as described in Ref.~\cite{Ikeda:2012au}.
 }
 \label{fig:amp}
\end{figure*}

The different off-shell behaviors of the two types of 
models leads to different analytic structure of the $\bar{K}N$
amplitudes.
We find that the E-dep. model has two poles on the $\bar{K}N$ physical and 
$\pi\Sigma$ unphysical sheet. 
The behavior of the subthreshold amplitudes is similar to that obtained with
the chiral SU(3) dynamics~\cite{Ikeda:2011pi,Ikeda:2012au} (see Fig.~\ref{fig:amp}),
and the scattering length is consistent with SIDDHARTA measurement in the E-dep. model. 
On the other hand, the E-indep. model has a single pole
corresponding to $\Lambda(1405)$. 
It shares this property with other phenomenological
potential models.
The behavior below $\bar{K}N$ threshold of the
amplitudes in the E-indep. model is very different compared with that
obtained from
chiral SU(3) dynamics (Fig.~\ref{fig:amp}).
In the E-indep. model, it is difficult to reproduce the $K^-p$ scattering
length in comparison with SIDDHARTA measurements although the 
cross sections are reproduced within experimental errors.

Table~\ref{pole_energy} lists the pole energies of the 
$\bar K N$ $s$-wave scattering
amplitudes in the complex energy plane between the $\bar{K}N$ and
$\pi\Sigma$ threshold energies and
the $K^-p$ scattering length. 
The primary purpose of this study is to clarify the influence of the subthreshold behavior of the $\bar{K}N$ interaction in the $\pi\Sigma$ spectrum.
 
\begin{table}[tb]
\caption{
Resonance energies $E_R$ of the $I=0$ $\bar{K}N$-$\pi\Sigma$ interaction
and the $K^-p$ scattering lengths $a_{K^-p}$
for the E-dep. and E-indep. models.
The scattering length $a_{K^-p}$ from the SIDDHARTA measurements are
 extracted using the improved Deser-Trueman formula~\cite{Meissner:2004jr}.
}
\label{pole_energy} 
\centering   
\begin{ruledtabular}  
\begin{tabular}{lccc}
 & $E_R$ (MeV) & $a_{K^-p}$ (fm) \\
\hline
E-dep. model & $1428.8-i~15.3$ & $-0.72+i~0.77$\\
 & $1344.0-i~49.0$&\\
E-indep. model & $1405.8-i~25.2$&$-0.54+i~0.46$\\
SIDDHARTA & & $-0.65(0.10)+i~0.81(0.15)$
\end{tabular}
\end{ruledtabular}
\end{table}

As for the cutoff parameters of $\pi N$ interactions, we have determined
them by fitting the $S_{11}$ and $S_{31}$ $\pi N$ scattering
lengths~\cite{Schroder:1999uq}.
The resulting values are $\Lambda^{(I=1/2)} _{N^*}=\Lambda^{(I=3/2)}
_{N^*}=500$ MeV for both the E-dep. and E-indep. models.

\subsection{Cutoff parameter dependence}

\begin{figure*}[tbh]
 \includegraphics[width=0.329\textwidth,clip]{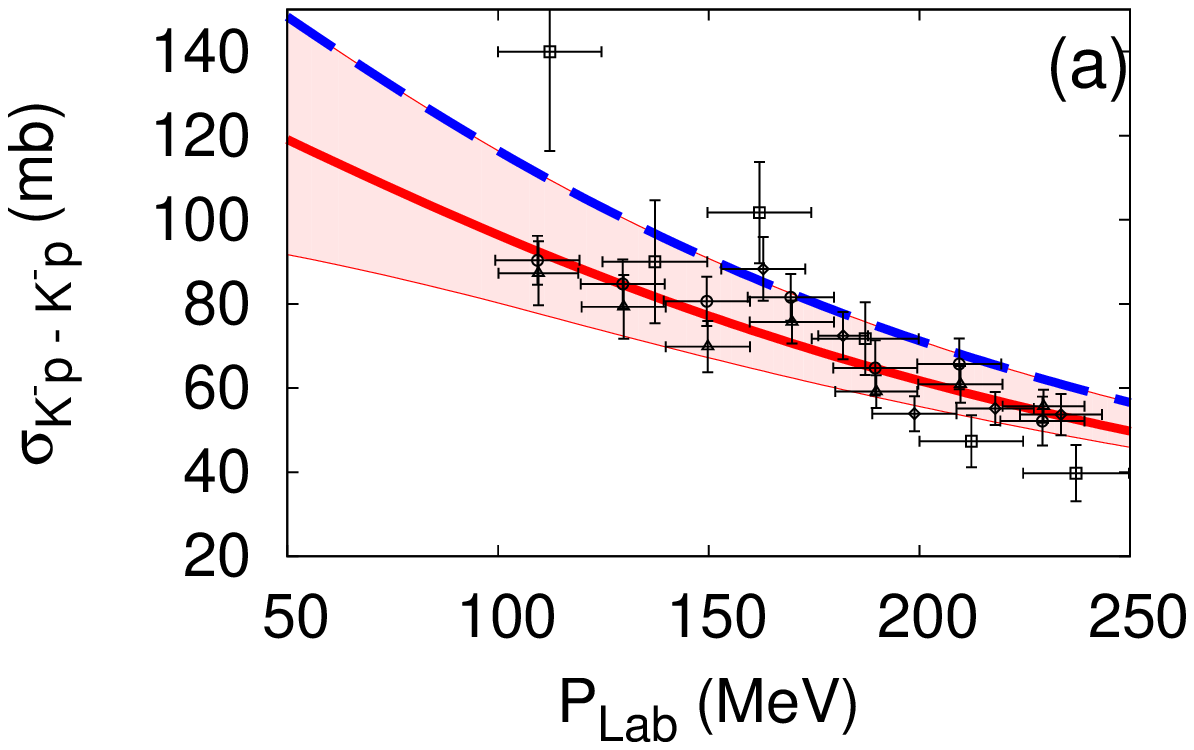}
 \includegraphics[width=0.329\textwidth,clip]{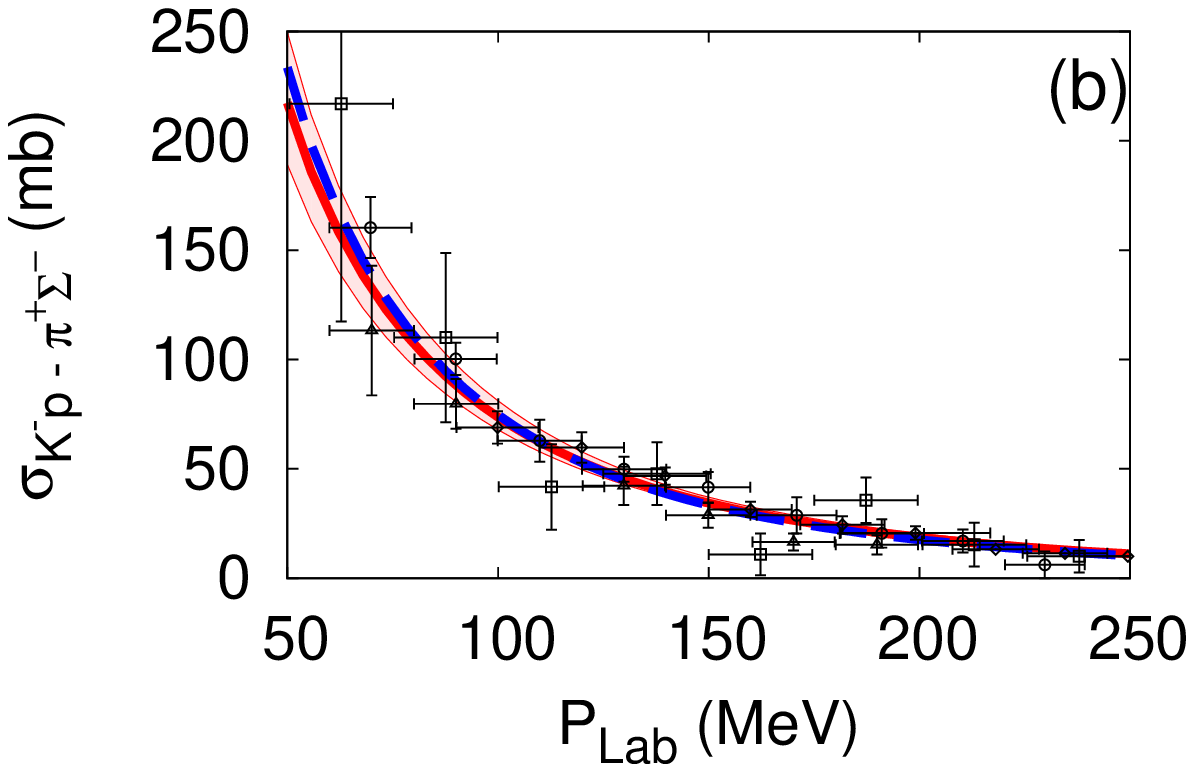}
 \includegraphics[width=0.329\textwidth,clip]{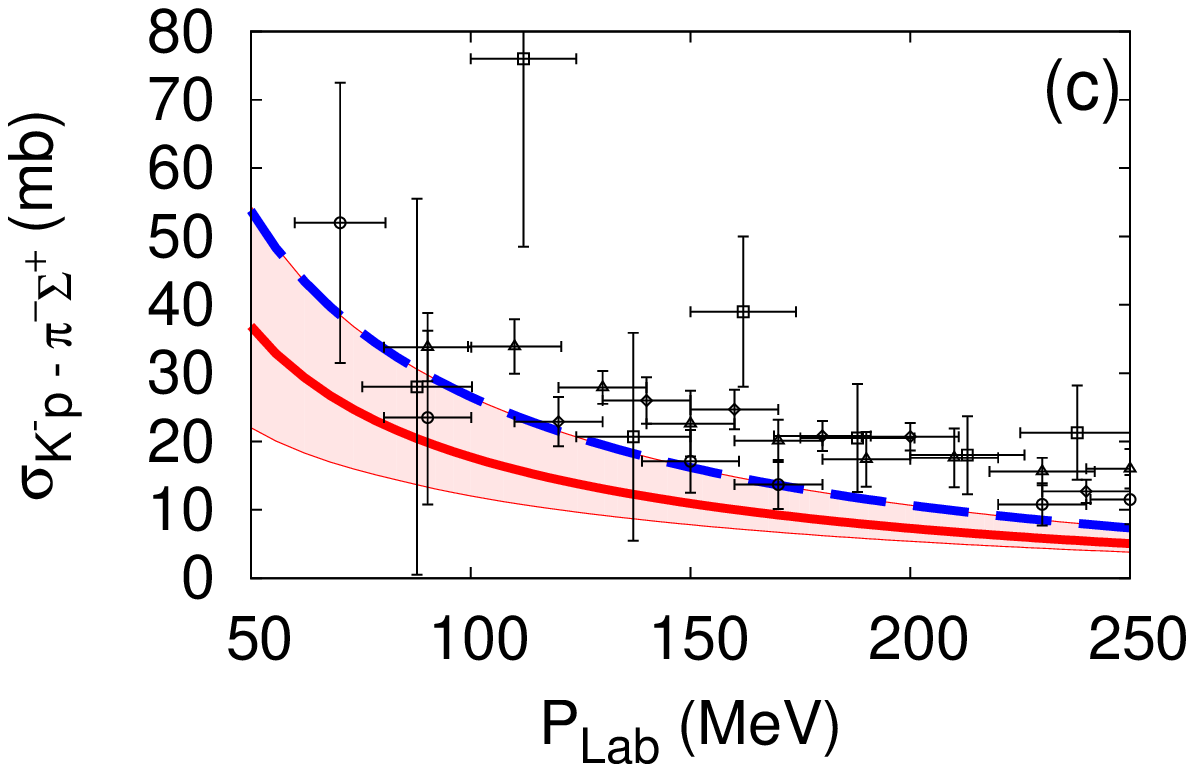}\\
 \includegraphics[width=0.329\textwidth,clip]{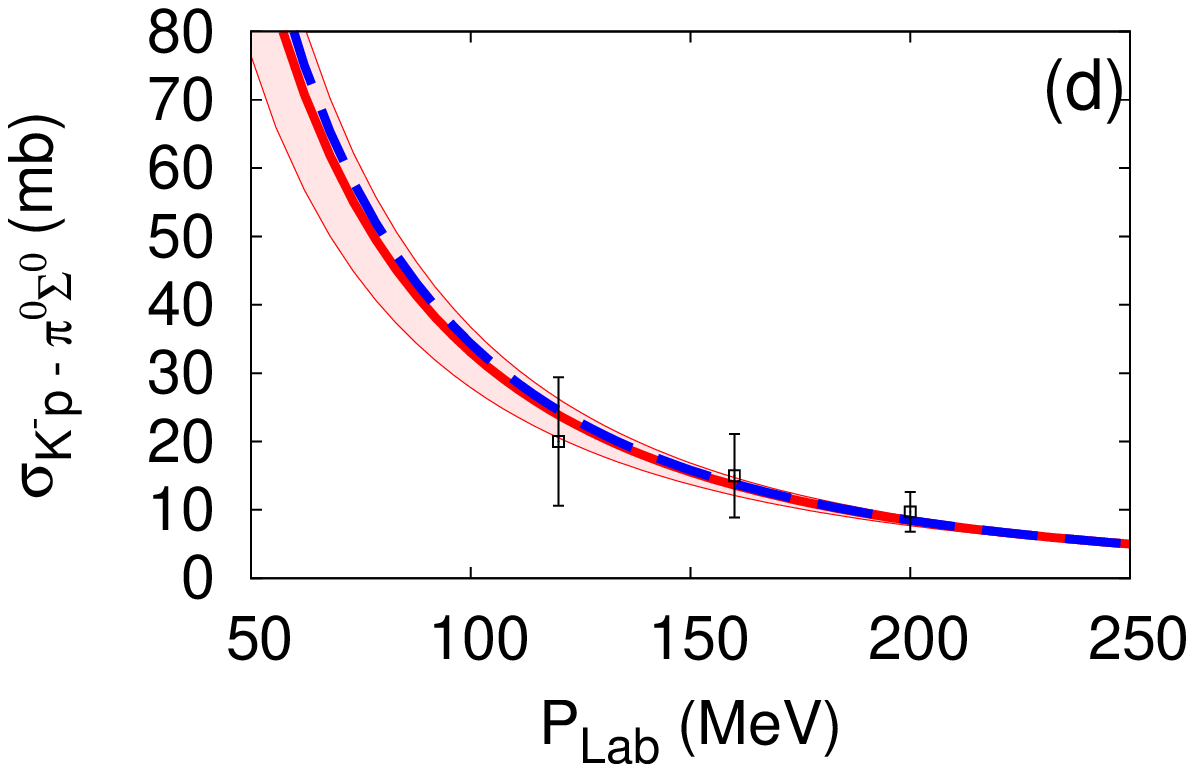}
 \includegraphics[width=0.329\textwidth,clip]{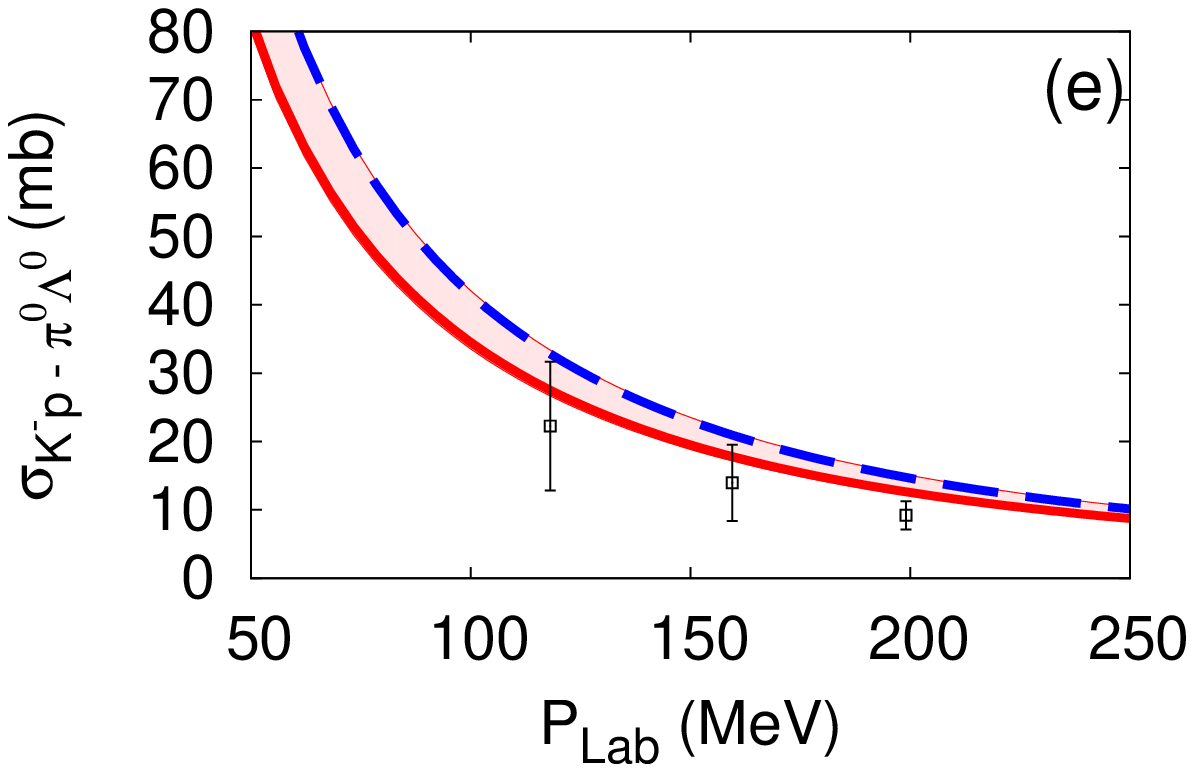}
 \caption{(Color online) 
Results of the fit with the E-dep. model.
Total cross sections of 
(a) $K^-p\rightarrow K^-p$, 
(b) $K^-p\rightarrow \pi^+\Sigma^-$, 
(c) $K^-p\rightarrow \pi^-\Sigma^+$,
(d) $K^-p\rightarrow \pi^0\Sigma^0$, and 
(e) $K^-p\rightarrow \pi^0\Lambda$.
 Data are from
 Refs.~\cite{Humphrey:1962zz,Sakitt:1965kh,Kim:1965zz,Kittel:1966zz,Evans:1983hz}.
 The shaded areas reflect variations of the cutoff
 $\Lambda^{(I)}_{\alpha}$ as listed in Table~\ref{cutoff_eb}.
 In (c) the dashed curve indicates the best consistent fit to the
 $K^-p \rightarrow \pi^-\Sigma^+$ cross section.
 Its implication for the other cross sections is shown by the dashed
 curves in subfigures (a), (b), (d), and (e).
 }
 \label{cross_eb}
\end{figure*}
\begin{table*}[thb]
\begin{center}
\caption{Ranges of cutoff parameters of the $\bar{K}N$-$\pi Y$
 interaction
 compatible with experimental errors.}
\label{cutoff_eb}       
\begin{ruledtabular}
\begin{tabular}{lccccc}
 \vspace{-4mm}\\
 &$\Lambda^{(I=0)} _{Y_{K}}$ (MeV) &$\Lambda^{(I=0)}
 _{Y_{\pi}=\pi\Sigma}$ (MeV) &$\Lambda^{(I=1)} _{Y_{K}}$ (MeV)
 &$\Lambda^{(I=1)} _{Y_\pi=\pi\Sigma}$ (MeV)&$\Lambda^{(I=1)}
 _{Y_\pi=\pi\Lambda}$ (MeV) \\
 \vspace{-4mm}\\
\hline
E-dep.&1070-1170 & 1070-1170 & 790-900&790-900&790-900\\
\end{tabular} 
\end{ruledtabular}
\end{center}
\end{table*}

Parameters of the two-body potential are the cutoffs
$\Lambda^{(I)}_{\alpha}$, determined
by fitting the $\bar{K}N$ reaction cross
sections within experimental errors.
Acceptable variations of these cutoffs are examined for the
E-dep. model.
The ranges of cutoff scales compatible with experimental errors are
listed in Table~\ref{cutoff_eb} and the corresponding fits to data are
presented in Fig.~\ref{cross_eb}.
The resulting $K^-p$ scattering length including uncertainties is
$a_{K^-p}=-(0.72^{+0.06}_{-0.12})+i~(0.77^{+0.19}_{-0.15})$~fm,
consistent with the scattering length deduced from the SIDDHARTA kaonic
hydrogen measurements.

As seen in Fig.~\ref{cross_eb} one might have the impression that the
$K^-p\rightarrow \pi^-\Sigma^+$ cross section is not optimally
reproduced. On the other hand, this is a relatively small cross section
with limited weight in the overall fitting procedure.
By examining the dashed curves in Fig.~\ref{cross_eb},
we have checked that optimizing the fit to this selected cross section
does not have a significant influence  on  the other cross sections within
uncertainties.

\subsection{Baryon-baryon interactions}
\label{sec:bb-int}

\begin{figure}[tbh]
\begin{center}
 \includegraphics[width=0.5\textwidth,clip]{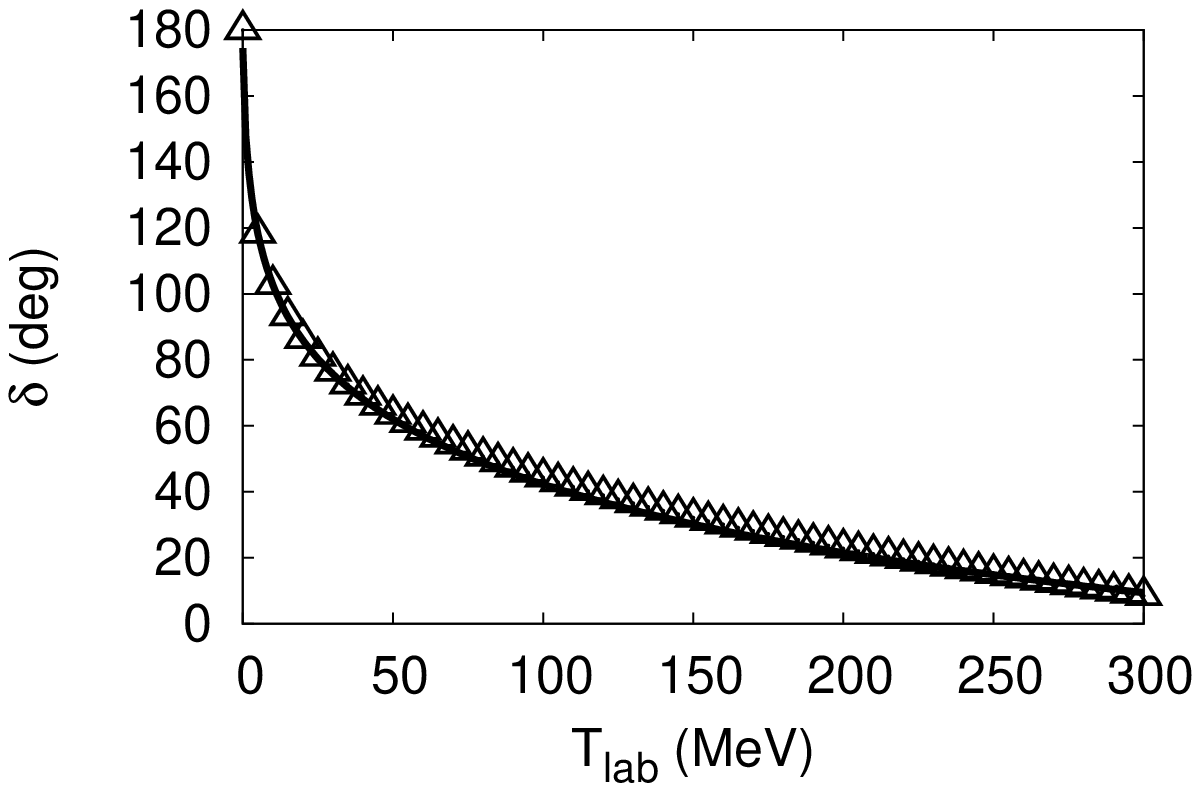}
\caption{Phase shifts of the $NN$ scattering in the $^3S_1$
 channel. The solid line shows the phase shift
 with our model, and the triangles show the phase shifts
 with the model of Ref.~\cite{Stoks:1994wp}.}
\label{phase_pn} 
\end{center}
\end{figure}

\begin{table}[tb]
\begin{center}
\caption{Parameters of the $NN$ interaction in $^3S_1$.}
\label{cut_pn}       
\begin{ruledtabular}
\begin{tabular}{cccc}
\vspace{-4mm}\\
$\Lambda_R$(MeV) &$\Lambda_A$(MeV) &$C_R$(MeV fm$^3$)&$C_A$(MeV fm$^3$) \\[3pt]
\hline
1350&321&1.41&$-5.59$ \\
\end{tabular}
\end{ruledtabular}
\end{center}
\end{table}

\begin{table}[tb]
\begin{center}
\caption{Coupling constants of the $Y N$ interactions.}
\label{coupl_yn}       
\begin{ruledtabular}
\begin{tabular}{cccc}
 \vspace{-4mm}\\
$C^{(I=1/2)}_{\Sigma N\Sigma N}$ &$C^{(I=1/2)}_{\Sigma N\Lambda N}$ &$C^{(I=1/2)}_{\Lambda N\Lambda N}$ &$C^{(I=3/2)}_{\Sigma N\Sigma N}$ \\[3pt]
\hline
1.51&0.40&1.08&$-1.11$\\
\end{tabular} 
\end{ruledtabular}
\end{center}
\end{table}

The following baryon-baryon interactions are commonly used for the
E-dep. and E-indep. meson-baryon models.
As for the $NN$ interaction in $^3 S_1$, we take the following
Yamaguchi-type two-term separable form:
\begin{equation}
V^{(I=0)}_{d,d}({ q}',{ q})=4\pi C_Rg_R({ q}')g_R({ q})+4\pi C_Ag_A({ q}')g_A({ q}).
\label{eq:NN_int_d}
\end{equation}
Here, $C_R$ ($C_A$) is the coupling strength of the repulsive (attractive) potential.
The form factors $g_{R,A}({ q})$ are defined by 
$g_{R,A}({ q})={\Lambda_{R,A}}^2/({ q}^2+{\Lambda_{R,A}}^2)$, with
$\Lambda_{R,A}$ being the cutoff parameters of the $NN$ interactions.
The coupling strengths $C_{R,A}$ and the cutoff parameters $\Lambda_{R,A}$ are determined by 
fitting the $^3S_1$ phase shifts of the Nijmegen 93
model~\cite{Stoks:1994wp}
(see Fig.~\ref{phase_pn} for the result of the fit).
The resulting values of the parameters are summarized in Table~\ref{cut_pn}.
The obtained deuteron binding energy is $2.23$ MeV.

As for the $s$-wave $YN$ interactions, we follow the form given in 
Ref.~\cite{Torres:1986mr}:
\begin{align}
	V^{(I)}_{\alpha\beta}(q_\alpha,q_\beta)=-4\pi \frac{C^{(I)}_{\alpha
	 \beta}}{2\pi^2}&(\mu_\alpha\mu_\beta \Lambda^{(I)}_{\alpha} \Lambda^{(I)}_{\beta})^{-1/2}\nonumber\\
	&\times g^{(I)}_{\alpha}(q_\alpha)g^{(I)}_{\beta}(q_\beta).
	\label{yn-pot}
\end{align}
Here, 
$\mu_{\alpha}$ is the reduced mass of the $YN$ system; 
the form factor $g^{(I)}_{\alpha}(q_\alpha)$ is defined as
$g^{(I)}_{\alpha}(q_\alpha)=\Lambda^{(I)2}_{\alpha}/(q_\alpha ^2+\Lambda^{(I)2}_{\alpha})$.
The coupling constants $C^{(I)}_{\alpha\beta}$ and the cutoff parameters
$\Lambda^{(I)}_{\alpha}$ are determined by fitting the $^3S_1$ phase
shifts of the J\"ulich'04 model~\cite{Haidenbauer:2005zh}.
The resulting values of the coupling constants $C^{(I)}_{\alpha\beta}$
are summarized in Table~\ref{coupl_yn}. The cutoff parameters 
$\Lambda^{(I)}_{\alpha}$ are
$\Lambda^{(I=1/2)}_{\Sigma N}=261$ MeV, 
$\Lambda^{(I=3/2)}_{\Sigma N}=540$ MeV, 
 and $\Lambda^{(I=1/2)}_{\Lambda N}=285$ MeV.

\section{Results and Discussion}
\label{sec:result}
\begin{figure*}[tbh]
\begin{tabular}{c}
   \includegraphics[width=0.5\textwidth]{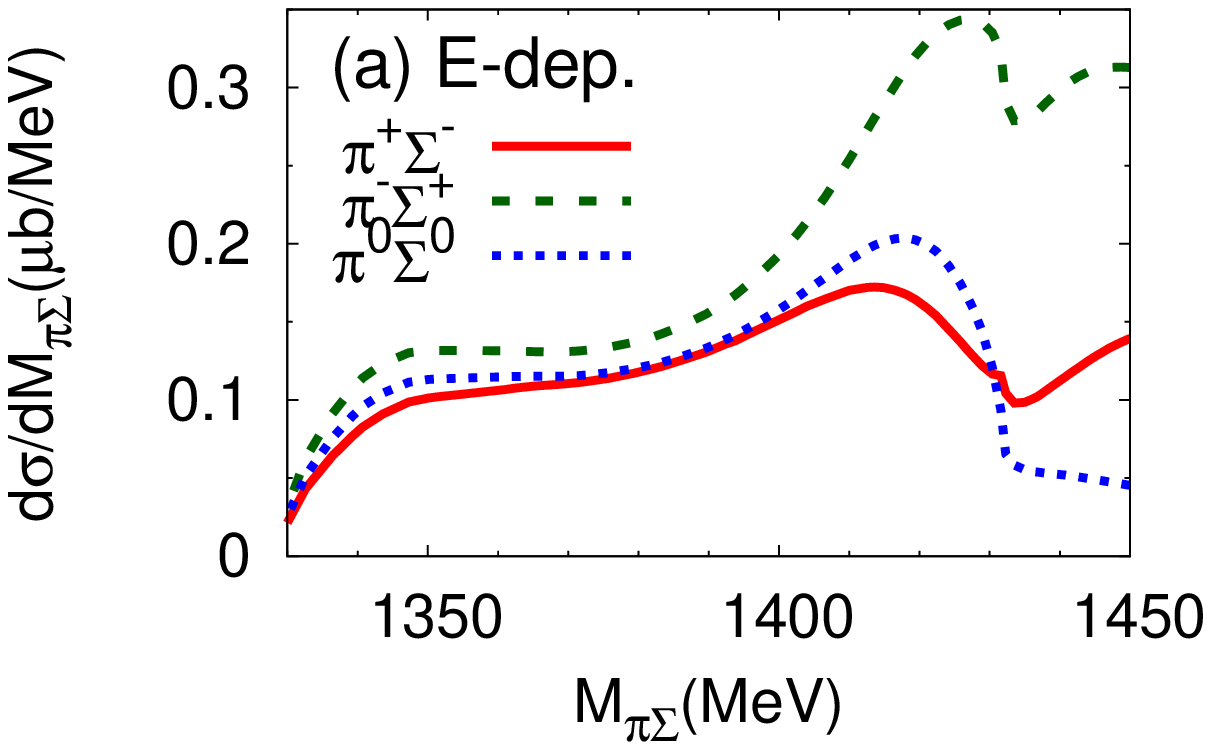}
   \includegraphics[width=0.5\textwidth]{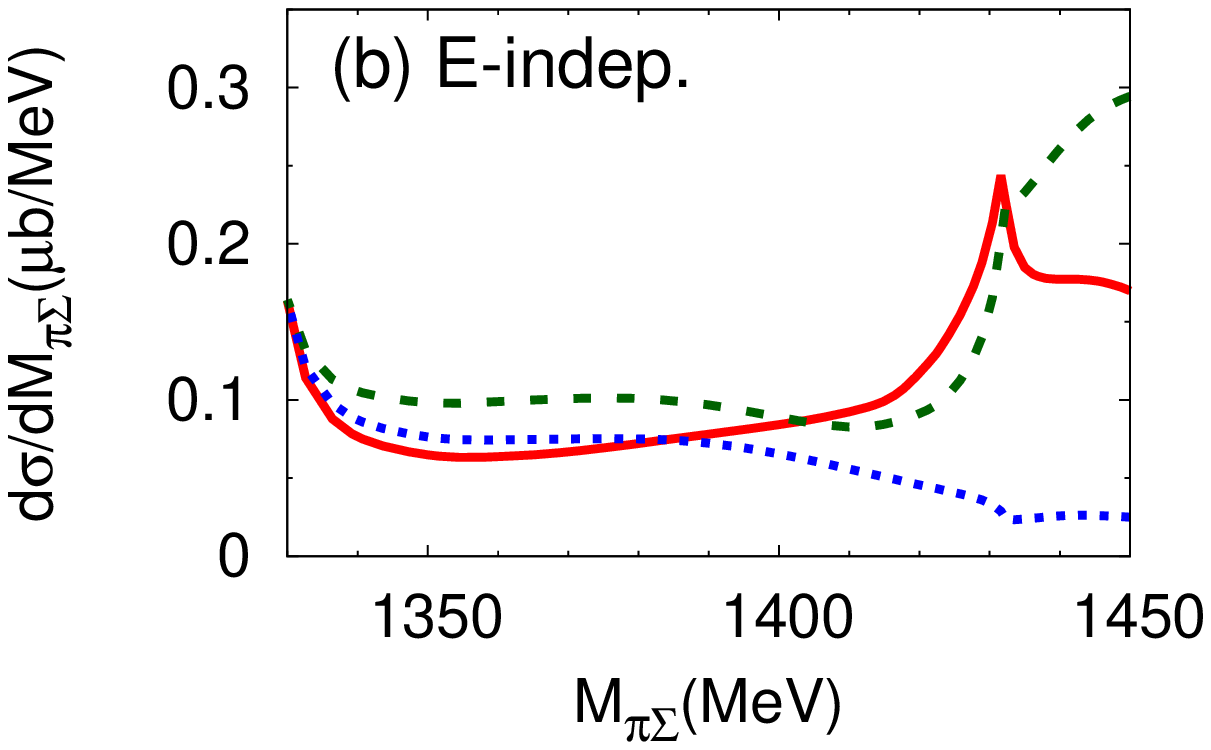}
\end{tabular}
 \caption{(Color online) Differential cross sections 
 {$d\sigma/dM_{\pi\Sigma}$} 
 for $K^-  d\rightarrow \pi\Sigma n$.
 The initial kaon momentum is set to $p_{\rm lab}= 1$~GeV.
 (a) E-dep. model; (b) E-indep. model.
 Solid curves: $\pi^+\Sigma^-n$;
dashed curves: $\pi^-\Sigma^+n$;
dotted curves: $\pi^0\Sigma^0n$ in the final
state, respectively.
 }
 \label{fig:inv_mas}
\end{figure*}

\subsection{Differential cross section of the $K^-  d \rightarrow \pi
   \Sigma  n$ reaction}

\begin{figure*}[tbh]
 \begin{center}
 \begin{tabular}{c}
 \includegraphics[width=0.33\textwidth]{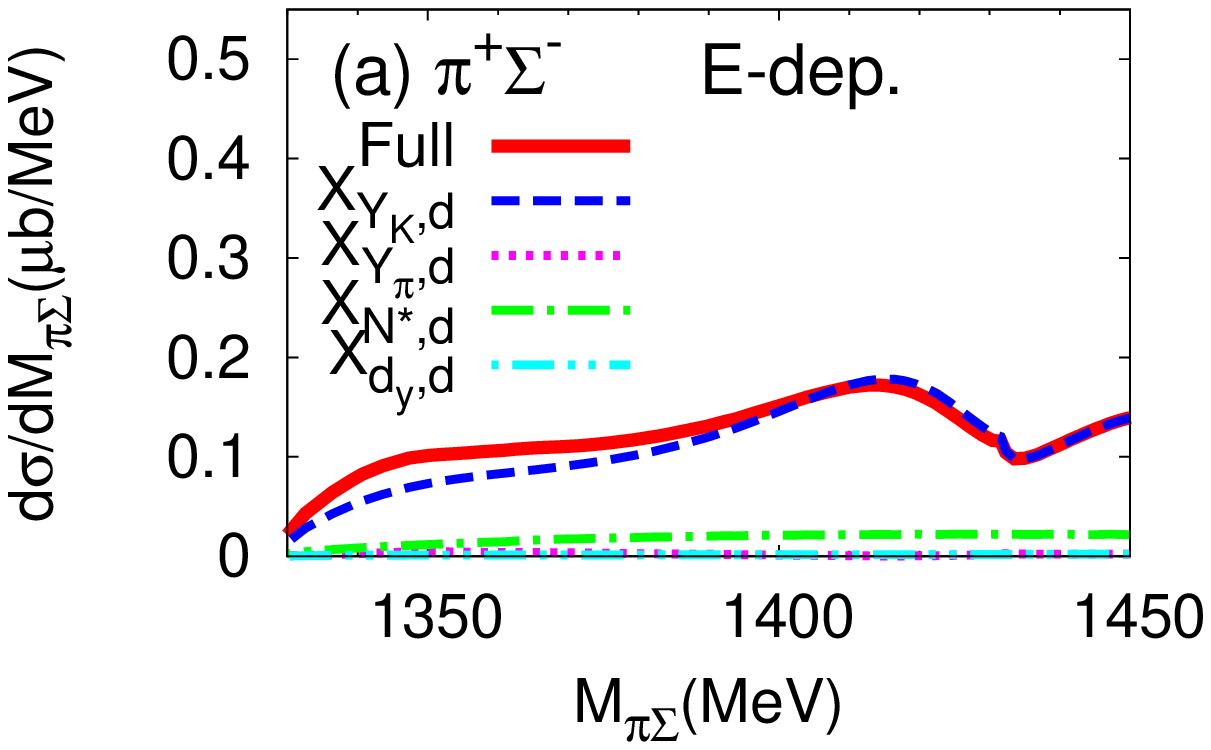}
 \includegraphics[width=0.33\textwidth]{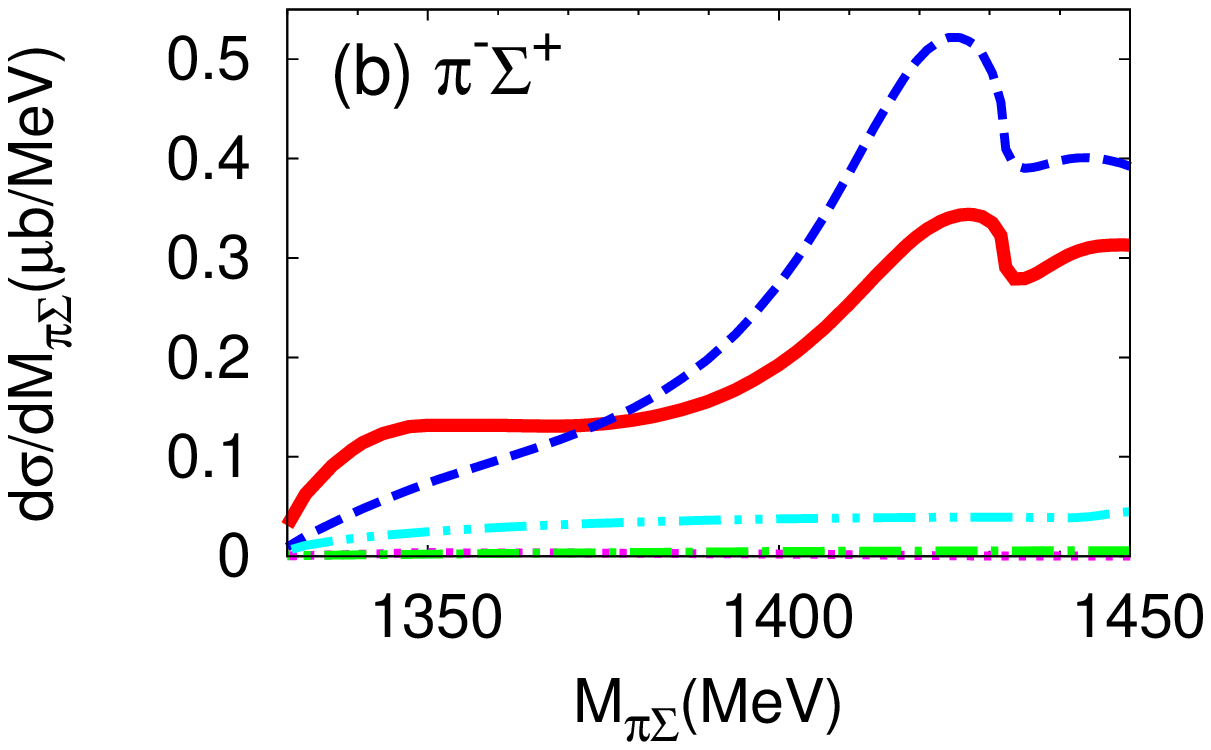}
 \includegraphics[width=0.33\textwidth]{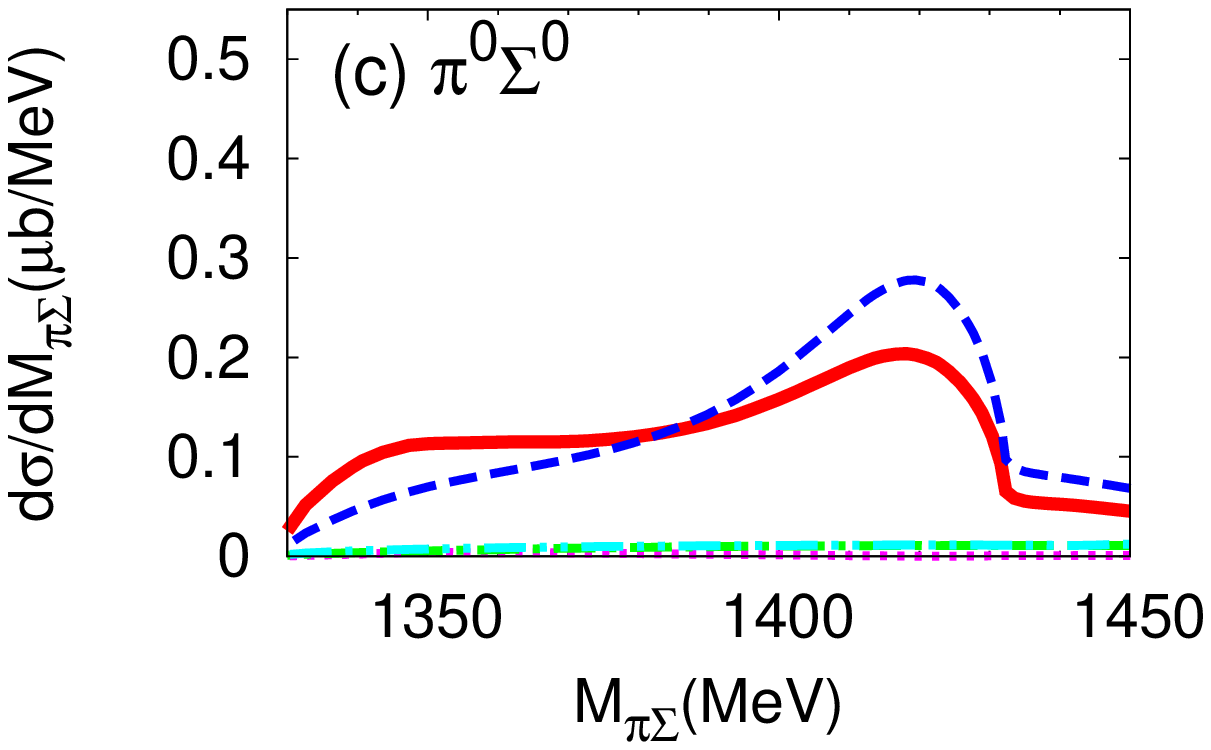}
 \end{tabular}
  \caption{(Color online) Contributions from each two-body reaction process
  to the $K^-d\rightarrow \pi\Sigma n$ differential cross
 section 
 {$d\sigma/dM_{\pi\Sigma}$} 
 using the E-dep. model.
 (a) $\pi^+\Sigma^- n$; (b) $\pi^-\Sigma^+ n$; (c)
 $\pi^0\Sigma^0 n$ in the final state.
 The solid curve represents the summation of all reaction processes
  of Eq.~(\ref{eq:T_Kd_piSn});
 dashed curve: $X_{Y_K,d}$ component;
 dotted curve: $X_{Y_\pi,d}$ component;
 dotted-dotted curve: $X_{N^*,d}$ component;
 dashed-two-dotted curve: $X_{d_y,d}$ component, respectively.
 The initial kaon momentum is set to $p_{\rm lab}= 1$~GeV.
}
  \label{fig:comp_dep}
 \end{center}
\end{figure*}

We proceed now to investigate the dependence of the differential
cross section, $d\sigma/dM_{\pi\Sigma}$ of Eq.~(\ref{eq:differential2}),
as a function of the invariant mass of the final $\pi\Sigma$ state.
Results for the different $\pi\Sigma$ charge combinations,
$K^-d\rightarrow \pi^+\Sigma^-n$, $K^-d\rightarrow \pi^-\Sigma^+n$, and
$K^-d\rightarrow \pi^0\Sigma^0n$ are shown in Fig.~\ref{fig:inv_mas} (a)
and (b) for the E-dep. and E-indep. models, respectively, whereby the
isospin basis states have been decomposed into charge basis states using
Clebsch-Gordan coefficients.
In view of the planned J-PARC experiment, the initial $K^-$ momentum is
chosen as $p_K^{\rm lab}=1$~GeV/$c$, corresponding to the $K^-d$ total energy
$\sqrt{s}=W=2817$~MeV.

The differential cross section 
is an order of magnitude smaller than that
calculated by assuming a two-step
process~\cite{Jido:2009jf,Miyagawa:2012xz,Jido:2012cy,YamagataSekihara:2012yv}.
Well-defined maxima are found at 
$M_{\pi\Sigma}\sim 1420$-$1430$ MeV for the E-dep. model 
in all charge
combinations of $\pi\Sigma$ in the final state~\footnote{The difference among the spectra in the charge basis is due to the interference effect with the $I=1$ amplitude~\cite{Nacher:1998mi}.}.
The positions of the peak structures are close to the  calculated
quasi bound $\bar{K}N$ pole position
($M_{\pi\Sigma}\sim 1429$ MeV). 
In the E-dep. model, the second pole with its large width,
$\Gamma\simeq 98$~MeV, barely affects the differential cross
section.
On the other hand, no resonance structure is seen for the E-indep. model.
The magnitude of the differential cross section and the interference patterns 
with backgrounds are evidently different for the E-dep. and E-indep. models.
This suggests that the $K^-  d\rightarrow \pi  \Sigma  n$ reaction can indeed provide
significant information on the $\bar{K}N$-$\pi Y$ subsystem.

Next, we show the contributions of each reaction process to the
differential cross section (Fig.~\ref{fig:comp_dep}).
As can be seen in Eq.~(\ref{eq:T_Kd_piSn}), the reaction dynamics involves
the quasi-two-body processes characterized by the amplitudes
$X_{Y_K,d}$, $X_{Y_\pi,d}$, $X_{N^*,d}$, and $X_{d_y,d}$.
The $X_{Y_K,d}$ amplitude which contains the $\bar{K}N\rightarrow
\pi\Sigma$ final state interaction turns out to be the dominant contribution to the
cross section.
The contribution from $X_{d_y,d}$ modifies the cross
section for $\pi^-\Sigma^+ n$ and $\pi^0\Sigma^0 n$ final states,
while its influence is small for the $\pi^+\Sigma^- n$
final state.
This is because the $X_{d_y,d}^{(I=1/2)(I=0)}$ component has
Clebsch-Gordan coefficients which cancel for $\pi^+\Sigma^- n$
final state.


\subsection{Partial waves and angular dependence of the reaction}

\begin{figure*}[tbh]
\begin{tabular}{c}
   \includegraphics[width=0.33\textwidth]{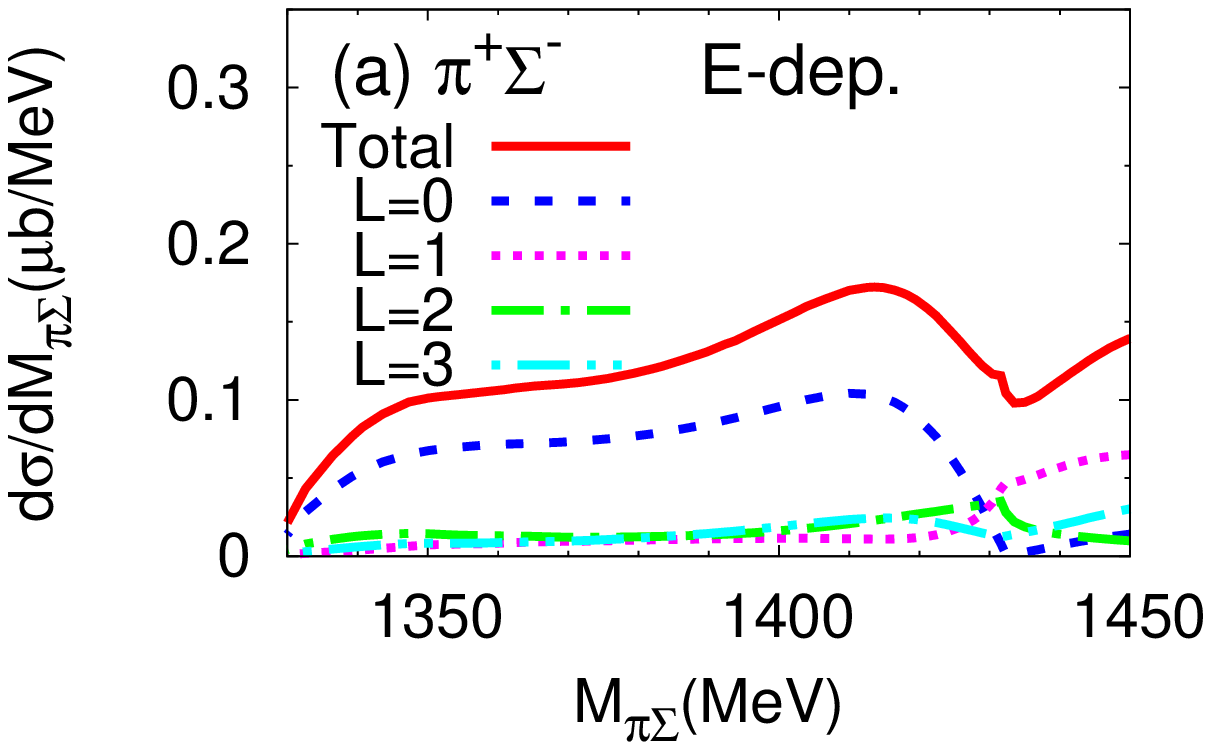}
   \includegraphics[width=0.33\textwidth]{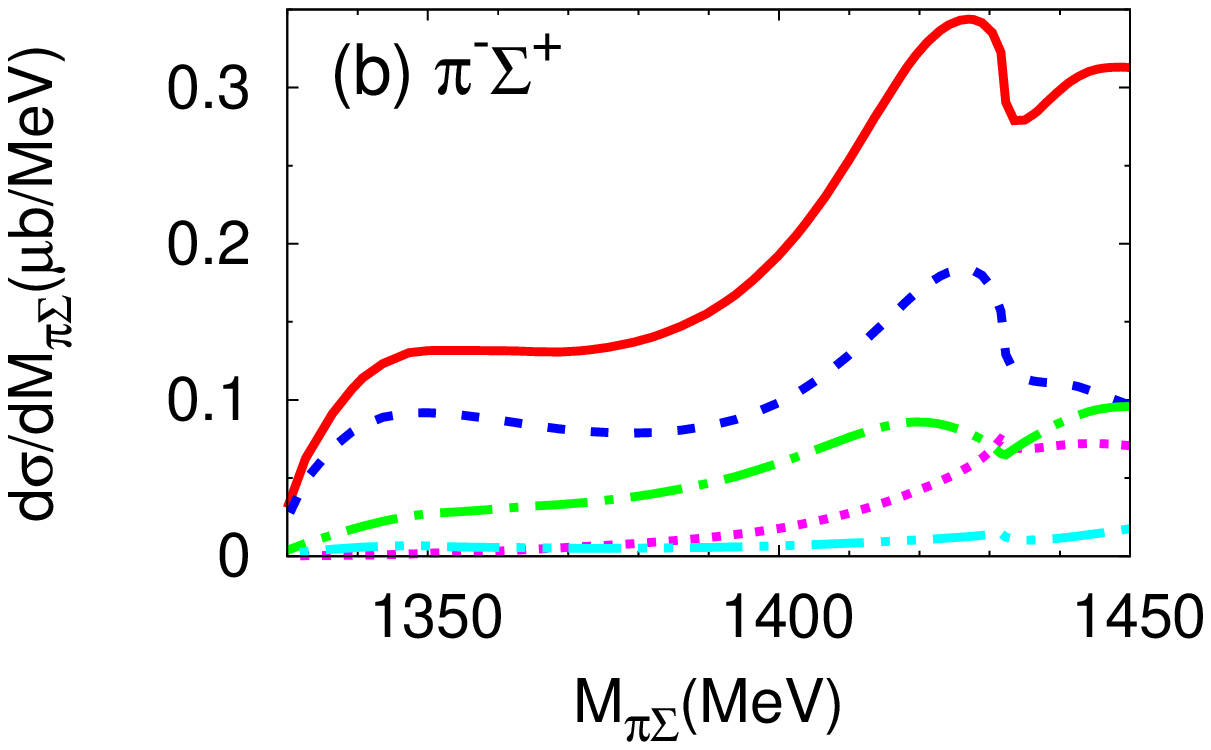}
   \includegraphics[width=0.33\textwidth]{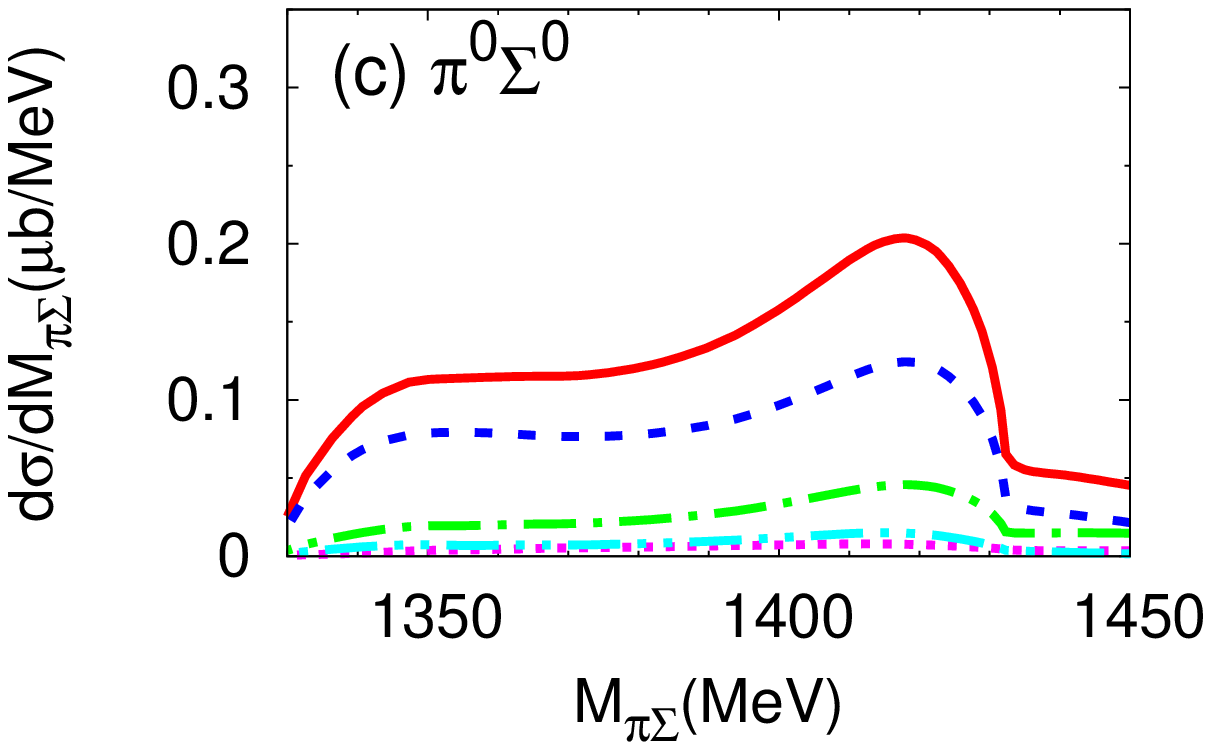}
\end{tabular}
 \caption{(Color online) 
Contributions of partial wave components to the differential cross
 section 
 {$d\sigma/dM_{\pi\Sigma}$} 
 for the E-dep. model.
 (a) $\pi^+\Sigma^- n$; (b) $\pi^-\Sigma^+ n$; (c)
 $\pi^0\Sigma^0 n$ in the final state.
 The solid curve represents the summation of total orbital angular
 momentum $L=0$ to $10$.
 The dashed curve represents the $L=0$ component.
 The dotted curve represents the $L=1$ component.
 The dashed-dotted curve represents the $L=2$ componentl
 The dashed-two-dotted curve represents the $L=3$ component.
 The contribution from $L\ge 4$ components which we omit is small.
 The initial kaon momentum is set to $p_{\rm lab}= 1$~GeV.}
 \label{fig:partial_dep}
\end{figure*}


\begin{figure*}[tbh]
 \begin{center}
 \begin{tabular}{c}
 \includegraphics[width=0.33\textwidth]{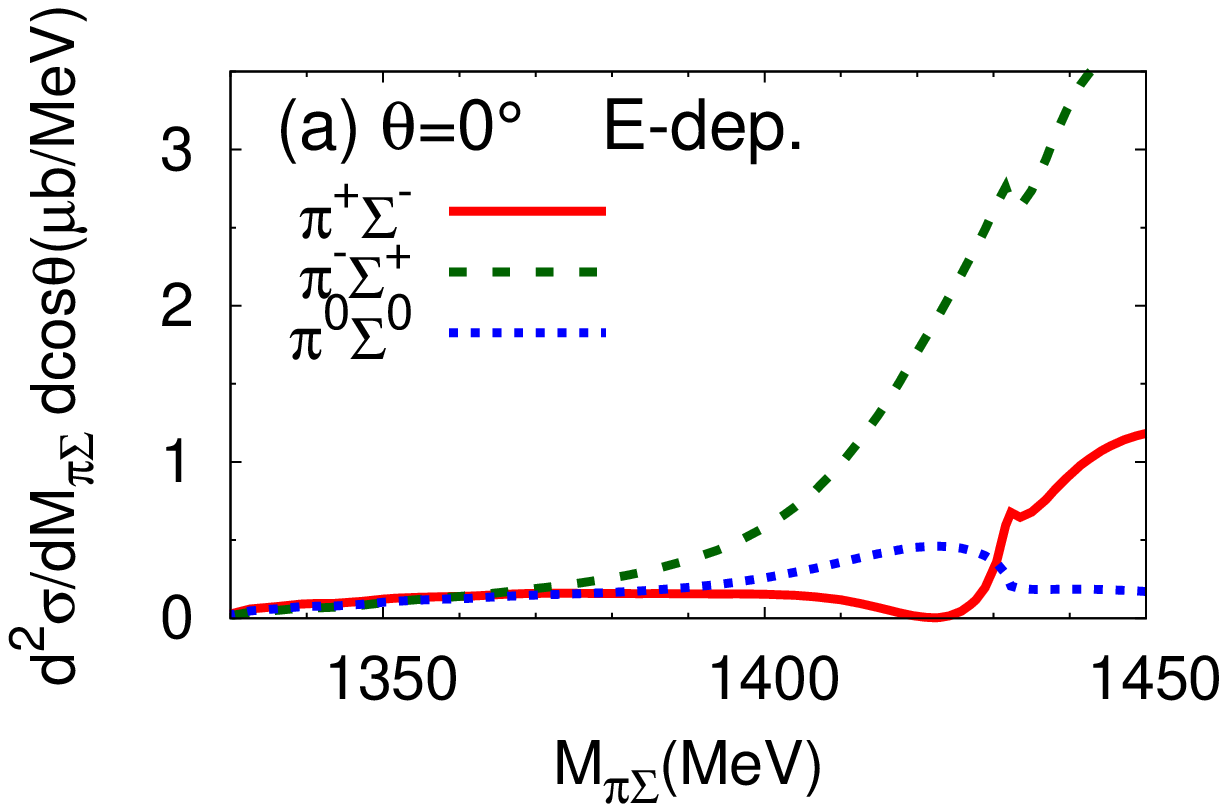}
 \includegraphics[width=0.33\textwidth]{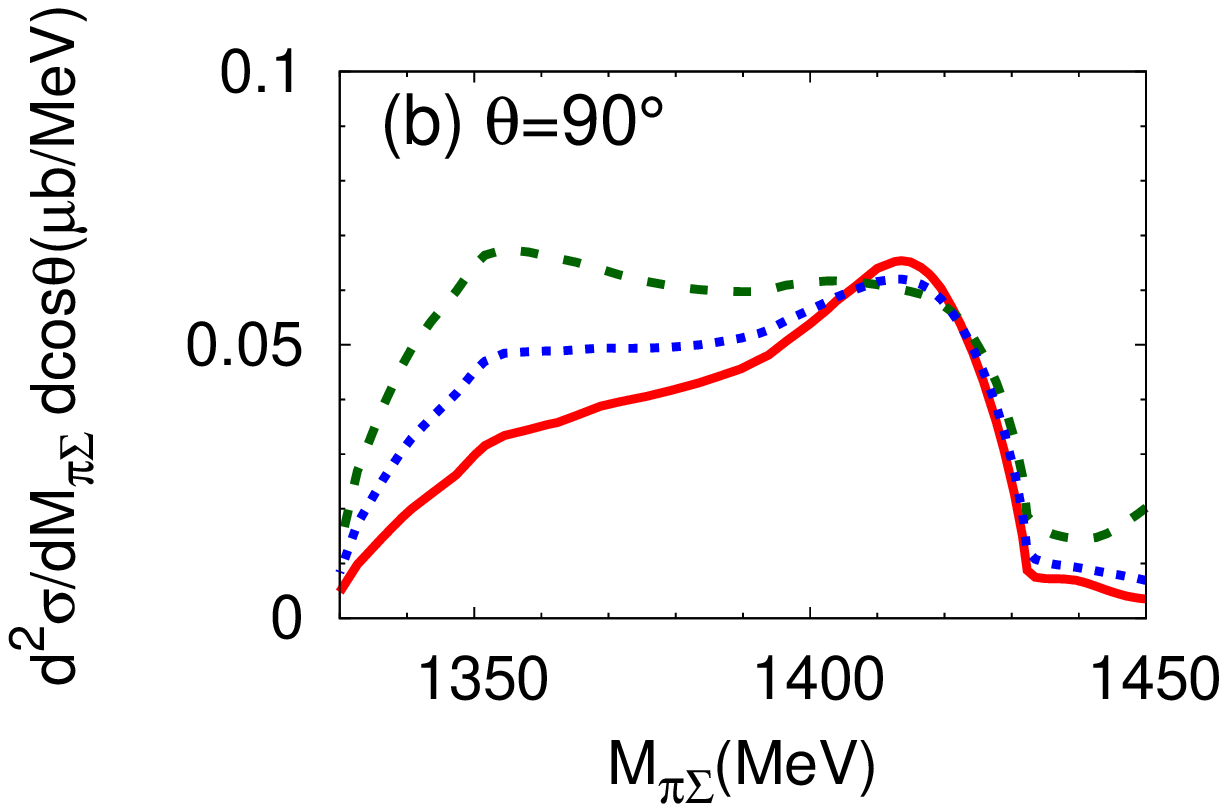}
 \includegraphics[width=0.33\textwidth]{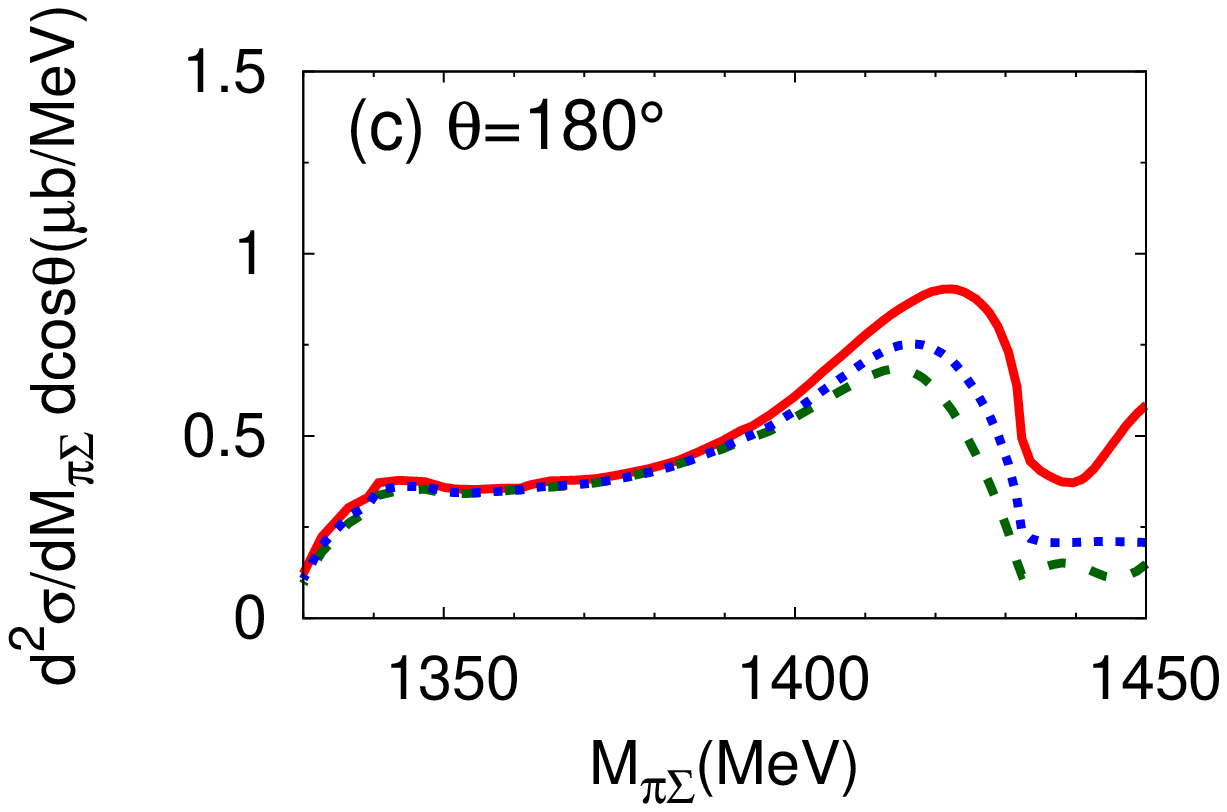}
 \end{tabular}
  \caption{(Color online) 
  Angular dependence of the double differential cross sections
  {$d^2\sigma/dM_{\pi\Sigma} d \cos \theta_{p_N}$} 
  for the E-dep. model.
  (a) $\theta_{p_N} = 0^\circ$; (b) $\theta_{p_N} = 90^\circ$; (c)
  $\theta_{p_N} = 180^\circ$, respectively.
  Here $\theta_{p_N}$ is scattering angle of the neutron in the center-of-mass frame.
  The illustration of curves and the initial kaon momentum are the same
  as those in Fig.~\ref{fig:inv_mas}.}
  \label{fig:ang_dep}
 \end{center}
\end{figure*}

\begin{figure*}[tbh]
 \begin{center}
  \begin{tabular}{c}
   \includegraphics[width=0.33\textwidth]{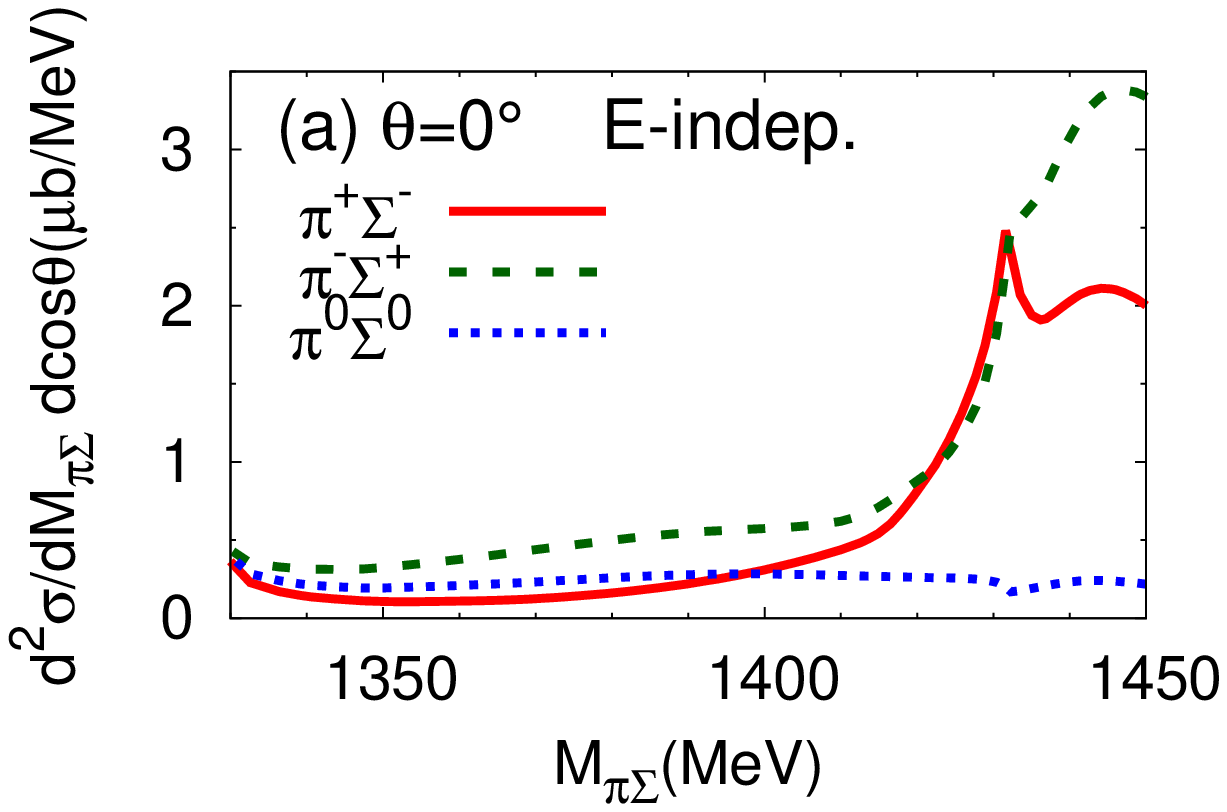}
   \includegraphics[width=0.33\textwidth]{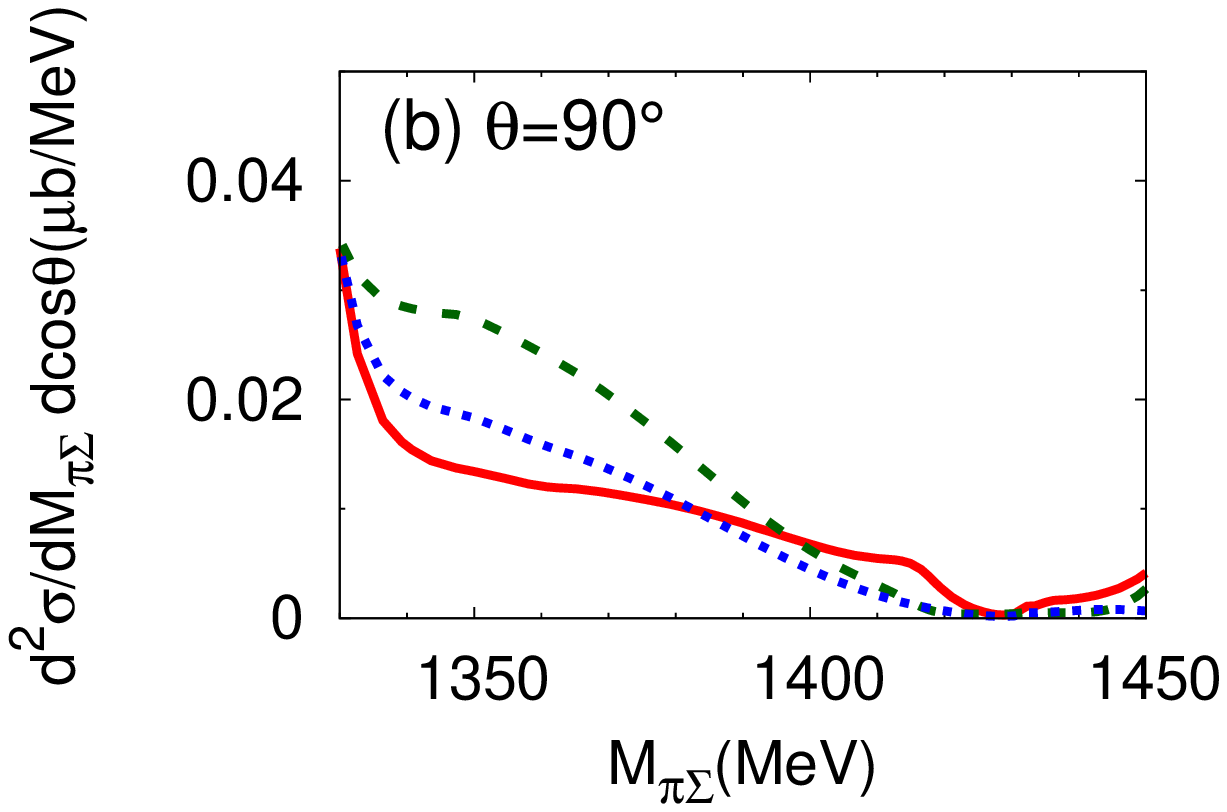}
   \includegraphics[width=0.33\textwidth]{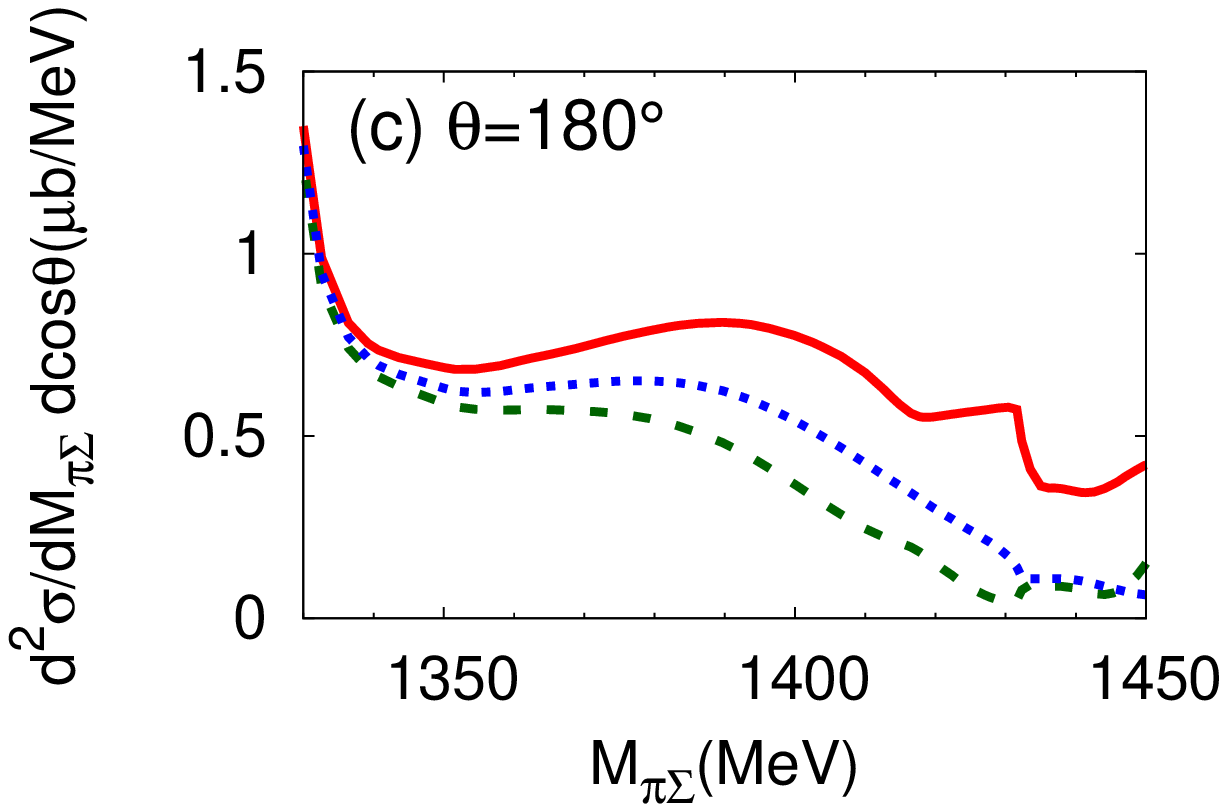}
  \end{tabular}
  \caption{(Color online) 
  Angular dependence of the double differential cross sections
  {$d^2\sigma/dM_{\pi\Sigma} d \cos \theta_{p_N}$} 
  for the E-indep. model.
  (a) $\theta_{p_N} = 0^\circ$; (b) $\theta_{p_N} = 90^\circ$; (c)
  $\theta_{p_N} = 180^\circ$.
  The illustration of curves and the initial kaon momentum are the same
  as those in Fig.~\ref{fig:inv_mas}.}
 \label{fig:ang_indep}
\end{center}
\end{figure*}


Consider now
the contributions from each partial wave component with
orbital angular momentum $L$ to the
differential cross section for E-dep. model.
The solid curve in Fig.~\ref{fig:partial_dep} show the results with
the total orbital angular momentum summed up to $L=10$, which correspond to
those shown in Fig.~\ref{fig:inv_mas}~(a), respectively.
The decomposition into angular momentum contributions with
$L=0,~1,~2$, and $3$ displayed in this figure demonstrates the convergence
of the partial wave expansion.
The large incident $K^-$ energy implies that there are sizable
contributions with $L\ne 0$.
The $s$- and $d$-wave components dominate in the 
region below $\bar{K}N$ threshold.
Around the threshold 
the $p$-wave component also becomes important for the $ \pi^+\Sigma^- n$ and
$ \pi^-\Sigma^+n$ channels.

It is instructive to investigate the angular dependence of the double differential cross
section, $d^2\sigma/dM_{\pi\Sigma} d \cos \theta_{p_N}$ defined in
Eq.~(\ref{eq:differential}).
In Fig.~\ref{fig:ang_dep} (Fig.~\ref{fig:ang_indep}),
we present the double differential cross section for neutron
scattering angles (a) $\theta_{p_N} = 0^\circ$, (b) $\theta_{p_N} = 90^\circ$, and
(c) $\theta_{p_N} = 180^\circ$ for E-dep. model (E-indep. model).
Here $\theta_{p_N}$ is the neutron scattering angle in the center-of-mass frame.
At $\theta_{p_N}=0^\circ$
one finds a strong dependence
on the final state. The $\bar{K}N$
 threshold cusp effect is enhanced in the $\pi^+\Sigma^- n$ and
 $\pi^-\Sigma^+ n$ channels.
The detailed channel dependence is closely related to the interference of the
isospin $I=0$ and $I=1$ components of the $\pi\Sigma$ in the final
state. The forward $K^- d \rightarrow \pi\Sigma n$
reaction thus provides
information on the $\bar{K}N$-$\pi Y$ interaction not only
in the $I=0$ but also in $I=1$ channel.

At $\theta_{p_N}=90^\circ$ the differential cross section is
strongly suppressed in both the E-dep. and E-indep. models.
It remains relatively flat at $\theta_{p_N}=180^\circ$.
Clearly, the interesting physics information is expected to be
observable primarily with neutrons produced in forward direction.
In the actual experiment the neutron will be detected in a forward cone
around $\theta_{p_N}=0^\circ$.
We have checked that the differential cross section integrated over an
angle interval from $\theta_{p_N}=0^\circ$ to $30^\circ$ does not change
much from the pattern seen at $\theta_{p_N}=0^\circ$.

\subsection{Cross sections above the $\bar{K}N$ threshold energy
  and cutoff dependence}
\begin{figure*}[tbh]
\begin{tabular}{c}
   \includegraphics[width=0.5\textwidth]{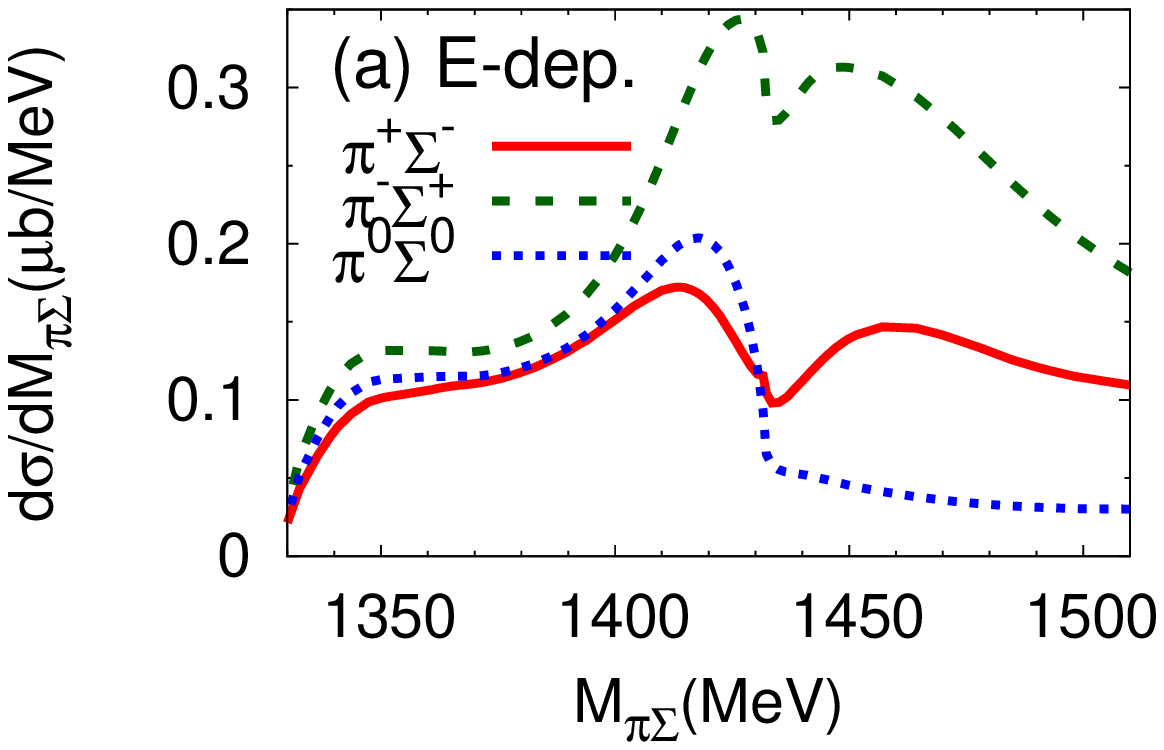}
   \includegraphics[width=0.5\textwidth]{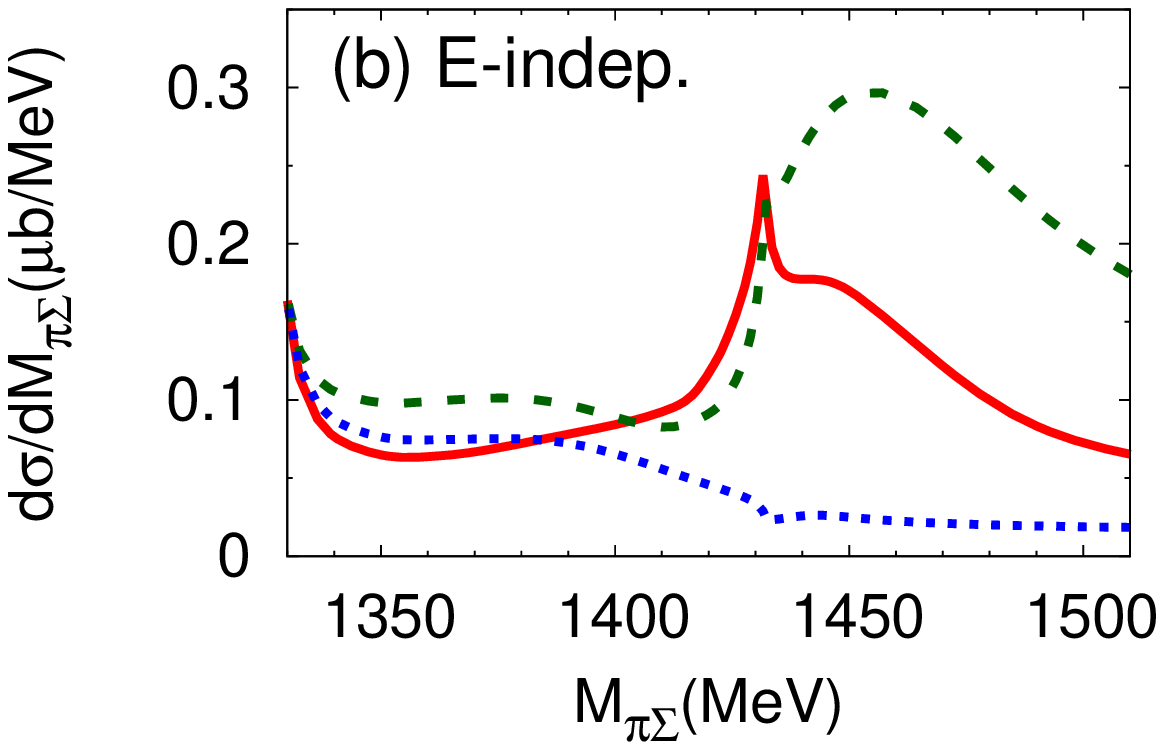}
\end{tabular}
 \caption{(Color online) Differential cross sections 
 {$d\sigma/dM_{\pi\Sigma}$} 
 for $K^-  d\rightarrow \pi\Sigma n$.
 (a) The E-dep. model; (b) the E-indep. model.
  The illustration of curves and the initial kaon momentum are the same
  as those in Fig.~\ref{fig:inv_mas}.
 }
 \label{fig:cross_above}
\end{figure*}

\begin{figure*}[tbh]
\begin{tabular}{c}
   \includegraphics[width=0.5\textwidth]{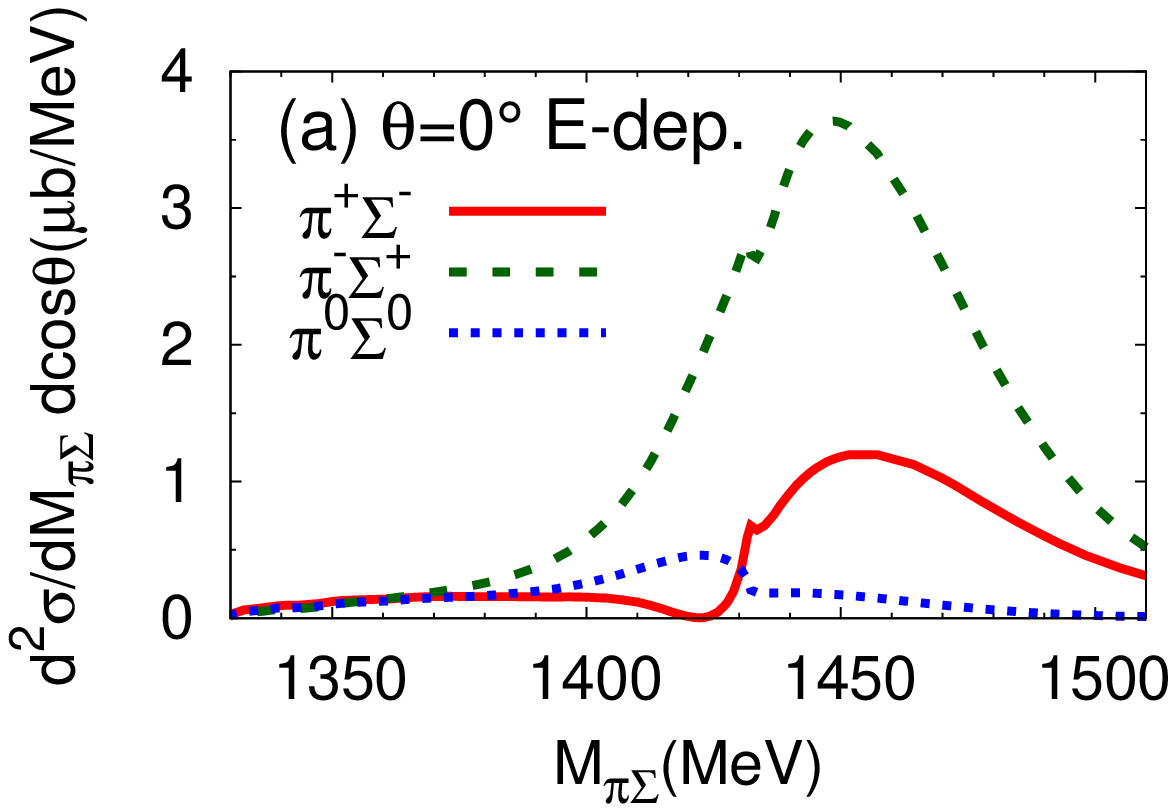}
   \includegraphics[width=0.5\textwidth]{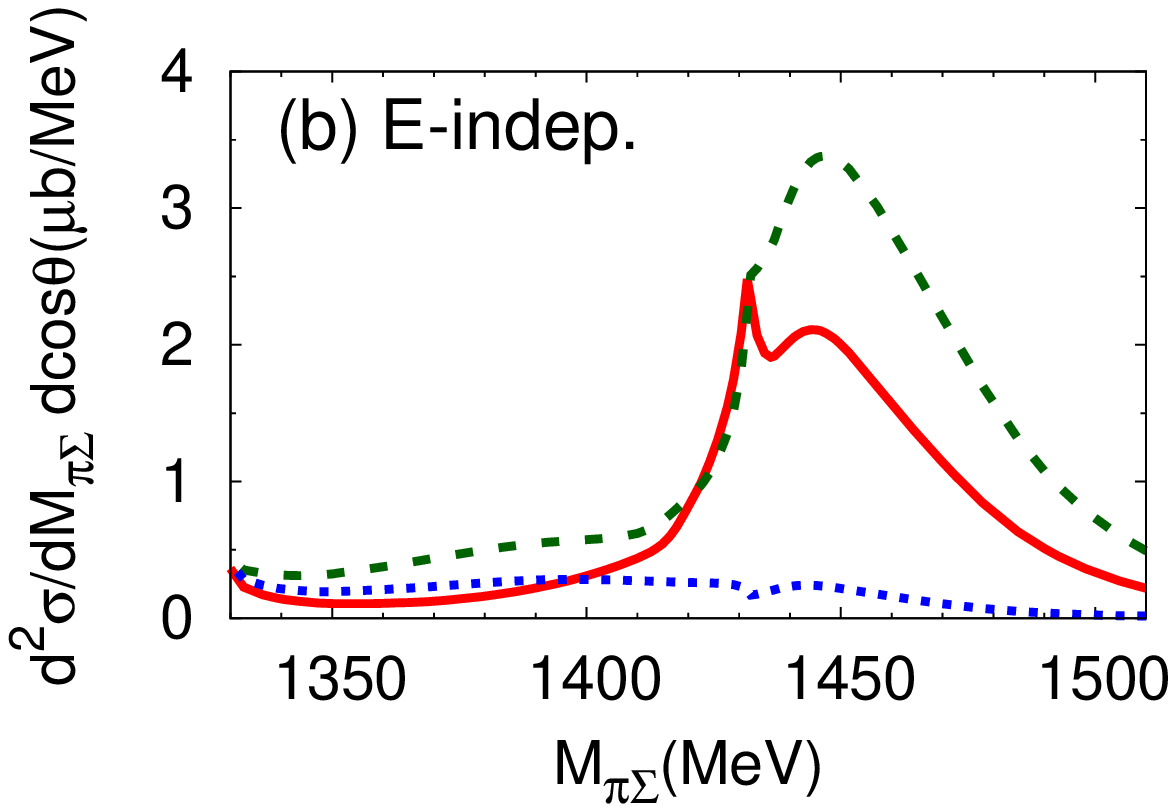}
\end{tabular}
 \caption{(Color online) Double differential cross sections 
 {$d^2\sigma/dM_{\pi\Sigma}d\cos \theta_{p_N}$} 
 for $K^- d\rightarrow \pi\Sigma n$ with the neutron emitted in
 forward direction, $\theta_{p_N}=0^\circ$.
 (a) The E-dep. model; (b) the E-indep. model.
  The illustration of curves are the same
  as in Fig.~\ref{fig:inv_mas}.
 The incident $K^-$ momentum is $p_{\rm lab}= 1$~GeV.
 }
 \label{fig:cross_above_0}
\end{figure*}

\begin{figure*}[tbh]
 \begin{center}
  \begin{tabular}{c}
   \includegraphics[width=0.33\textwidth]{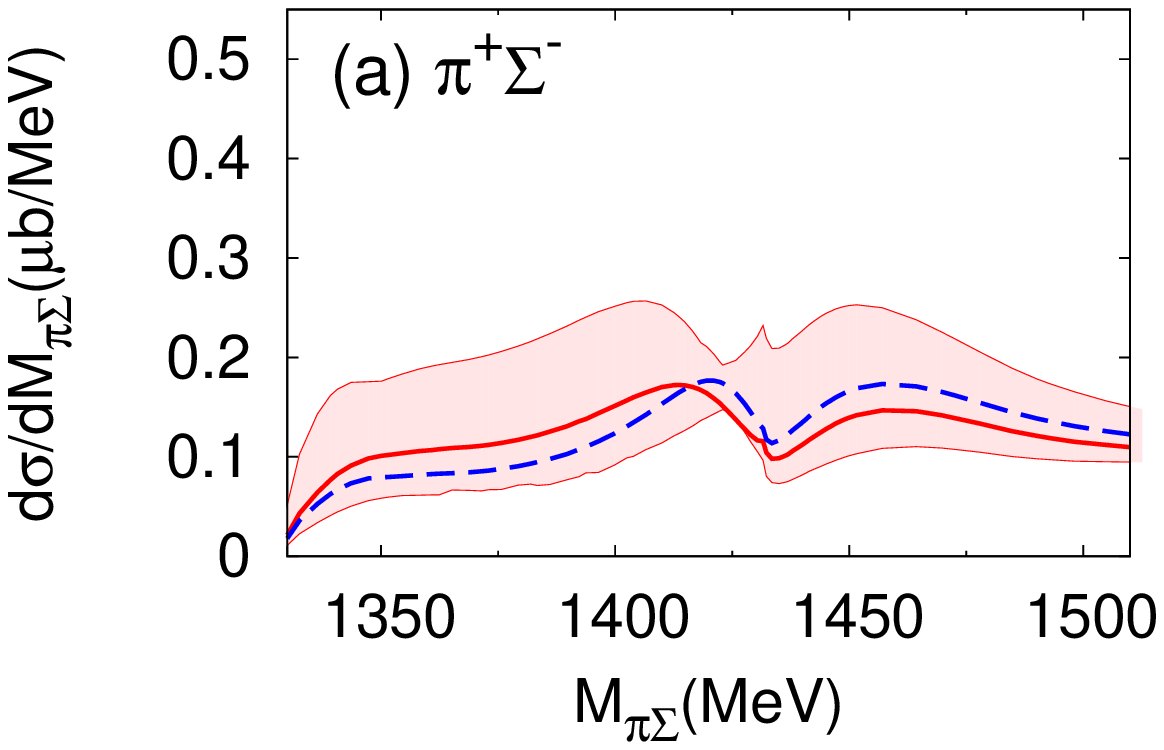}
   \includegraphics[width=0.33\textwidth]{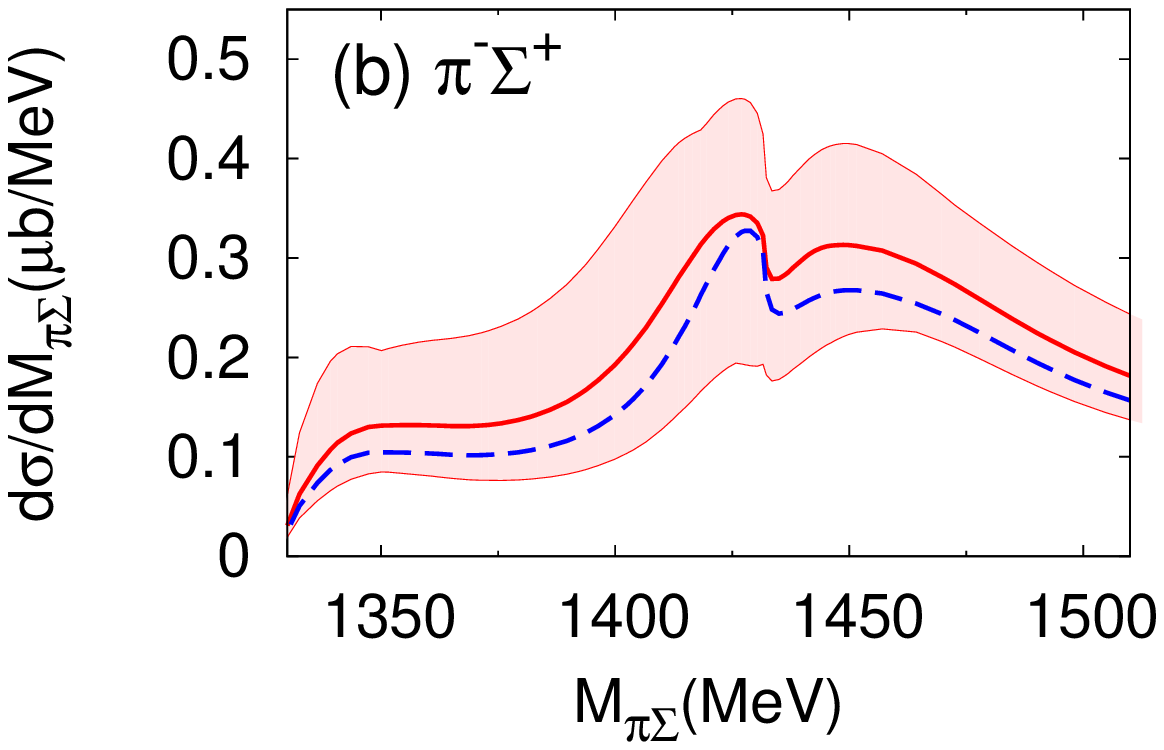}
   \includegraphics[width=0.33\textwidth]{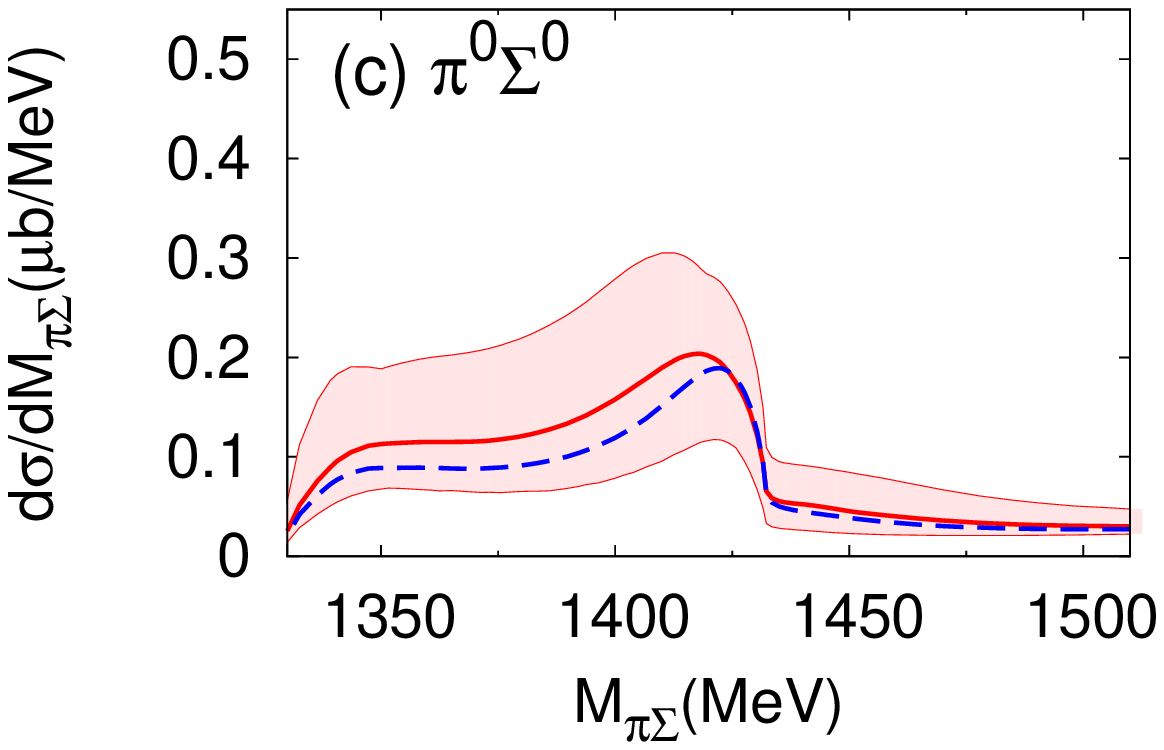}
  \end{tabular}
  \caption{(Color online) 
  Uncertainty bands reflecting cutoff
  parameter dependence of
  the differential cross section 
 {$d\sigma/dM_{\pi\Sigma}$} 
 for the E-dep. model.
 (a) $\pi^+\Sigma^- n$; (b) $\pi^-\Sigma^+ n$; (c)
 $\pi^0\Sigma^0 n$ in the final state.
  The initial kaon momentum is set to $p_{\rm lab}= 1$~GeV.
  The dashed curves refer to the choice of ``optimized'' fit to the
  $K^-p\rightarrow\pi^-\Sigma^+$ cross section in Fig.~\ref{cross_eb}.
  }
 \label{fig:cross_eb}
\end{center}
\end{figure*}

\begin{figure*}[tbh]
 \begin{center}
  \begin{tabular}{c}
   \includegraphics[width=0.33\textwidth]{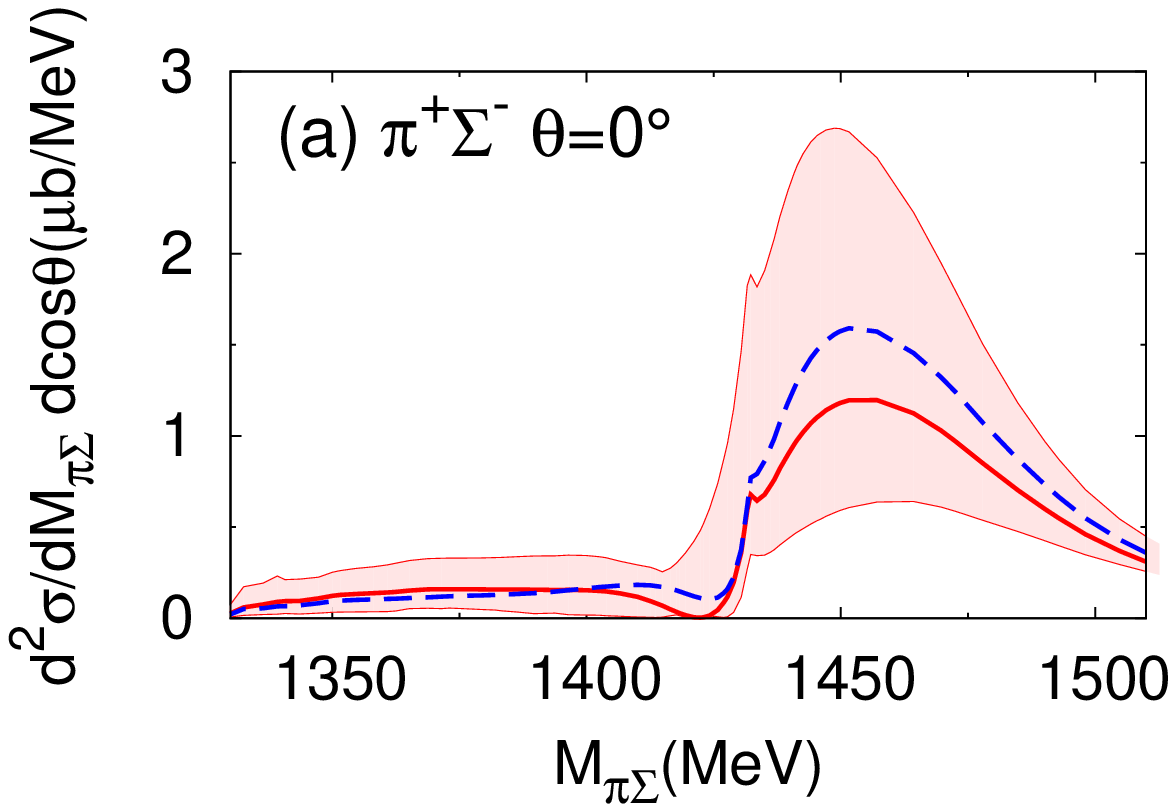}
   \includegraphics[width=0.33\textwidth]{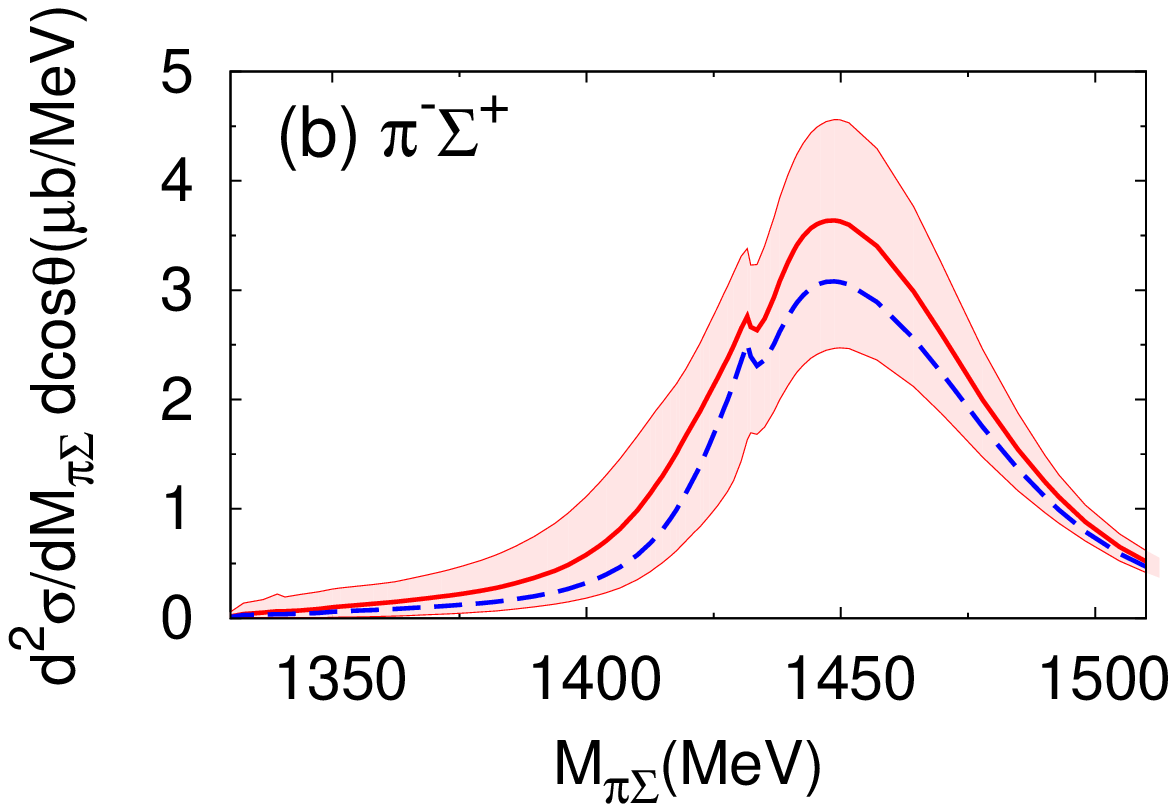}
   \includegraphics[width=0.33\textwidth]{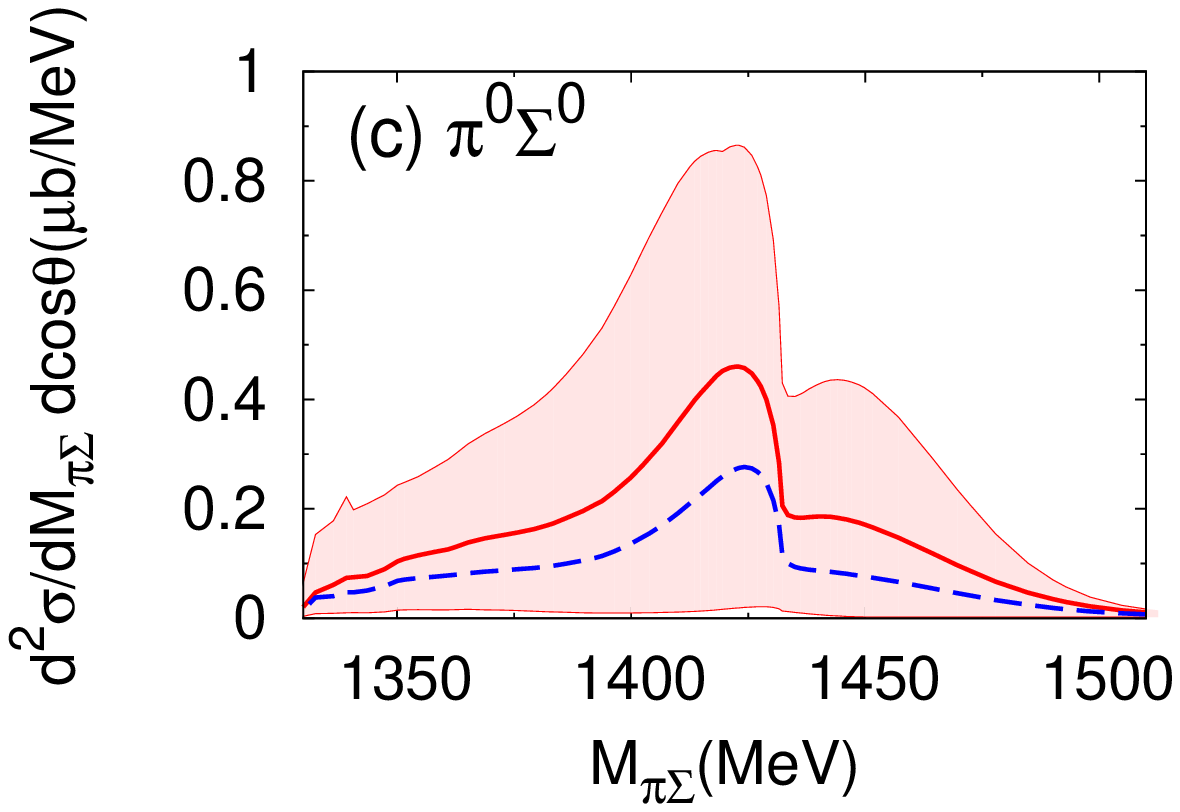}
  \end{tabular}
  \caption{(Color online) 
  Uncertainty bands reflecting cutoff
  parameter dependence of 
  the double differential cross section 
 {$d^2\sigma/dM_{\pi\Sigma}d\cos\theta_{p_N}$} 
 for the E-dep. model, with neutrons emitted in forward direction.
 (a) $\pi^+\Sigma^- n$; (b) $\pi^-\Sigma^+ n$; (c)
 $\pi^0\Sigma^0 n$ in the final state.
  The initial kaon momentum is set to $p_{\rm lab}= 1$~GeV.
    The dashed curves refer to the choice of ``optimized'' fit to the
  $K^-p\rightarrow\pi^-\Sigma^+$ cross section in Fig.~\ref{cross_eb}.
}
 \label{fig:cross_forward_eb}
\end{center}
\end{figure*}

While the primary focus in this study is on the $\bar{K}N$ subthreshold
region and the two-body $\bar{K}N$-$\pi Y$ dynamics governing the
$\Lambda(1405)$ formation in the $\bar{K}NN$ three-body system, it is
also instructive to explore $\pi\Sigma$ invariant mass spectra above
$\bar{K}N$ threshold in $K^-d\rightarrow (\pi\Sigma)n$.
In fact our calculation (see Fig.~\ref{fig:cross_above}) yields
pronounced structures, especially for the charged channels
($\pi^+\Sigma^-$ and $\pi^-\Sigma^+$), in the angle-integrated
differential cross section. These structures are qualitatively different
in the E-dep. and E-indep. approaches.

The $K^-d\rightarrow(\pi\Sigma)n$ double differential cross section with
the neutron emitted in forward direction is of special interest (see Fig.~\ref{fig:cross_above_0}).
Our three-body calculations predict a strongly developed maximum around
$M_{\pi\Sigma}=1.45$~GeV for both the E-dep. and the E-indep. models, a
feature that should be well observable.
A less pronounced effect is seen in the $K^-d\rightarrow \pi^+\Sigma^-n$
channel which requires both charge exchange and strangeness exchange mechanisms.

The appearance of the prominent maximum in
$d^2\sigma/dM_{\pi\Sigma}d\cos\theta_{p_N}$ around
$M_{\pi\Sigma}=1.45$~GeV can be traced to a combination of subtle
three-body mechanisms in the coupled $K^-pn$-$\pi\Sigma n$ system: the
nucleon exchange process between the incident $K^-$ and the deuteron,
with a propagating $\bar{K}N$ pair and a spectator nucleon, and
subsequent $\bar{K}$ exchange leading to the final $\pi\Sigma$ and
neutron.
The momentum matching between these two basic processes in the
three-body system produces the pronounced enhancement in the
$(\pi^-\Sigma^+)n$ channel about 20~MeV above $\bar{K}N$ threshold.

The appearance of such a structure in
$d^2\sigma/dM_{\pi\Sigma}d\cos\theta_{p_N}$ raises of course the
question of model dependence and sensitivity to cutoff variations in the
two-body amplitudes.
This cutoff dependence turns out indeed to be stronger in the three-body
system with its off-shell dynamics, as compared to the two-body
subsystems.
In order to examine this issue, we have performed calculations of the
$K^-d\rightarrow \pi\Sigma n$ differential cross sections using the
acceptable range of cutoff scales at the two-body vertices discussed
previously and listed in Table~\ref{cutoff_eb}.
This leads to the theoretical uncertainty bands displayed in
Fig.~\ref{fig:cross_eb} for $d\sigma/dM_{\pi\Sigma}$ and in
Fig.~\ref{fig:cross_forward_eb} for the double differential cross section with
forward-emitted neutron, both taken at an incident $K^-$ momentum
$p_{\rm lab}= 1$~GeV.
In particular, we have examined the influence of using the optimized
cross section for $K^-p\rightarrow \pi^-\Sigma^+$ while maintaining the
other cross sections well reproduced within uncertainties (compare the solid
and dashed curves in Figs.~\ref{fig:cross_eb} and \ref{fig:cross_forward_eb}).
While the absolute magnitudes of the differential cross sections are
indeed subject to uncertainties, the structural patterns of the forward
double differential cross sections for $\pi^-\Sigma^+$ and
$\pi^+\Sigma^-$ final states remains quite stable with respect to cutoff
variations, with the exception of the neutral ($\pi^0\Sigma^0$)
combination for which no prediction is possible.

At the same time as this genuine three-body dynamical structure in the 
$\pi^-\Sigma^+n$ final state appears around $M_{\pi\Sigma}\sim1450$~MeV,
it is quite remarkable that $d^2\sigma/dM_{\pi\Sigma}d\cos\theta_{p_N}$
with forward neutrons does not display a $\Lambda(1405)$ signal any more
(whereas it is still visible in the angle-integrated
$d\sigma/dM_{\pi\Sigma}$ for $K^-d\rightarrow\pi^-\Sigma^+ n$).
This is a consequence of interferences of three-body mechanisms in $I=0$
and $I=1$ amplitudes which screen the $\bar{K}N$ pole contribution.


\section{Summary}
\label{sec:summary}
Within the framework of the coupled-channels AGS equations,
we have investigated how the signature of the $\Lambda(1405)$ appears in
differential cross sections of
$K^-  d \rightarrow \pi\Sigma n$ reactions.
Two
types of meson-baryon interactions, the E-dep. and E-indep. models, have been
considered to illustrate how the difference of the subthreshold 
behaviors translates into the $ \pi\Sigma n$ spectra.
The E-dep. approach is generally favored because of its foundation in
chiral SU(3) effective field theory.

Characteristic structures reflecting the formation and dynamics of the
$\Lambda(1405)$ in the $\bar{K}NN$-$\pi\Sigma n$ three-body system are
found in differential cross section as a function of the $\pi\Sigma$
invariant mass.
By comparison of results using E-dep. and E-indep. models, it may be
possible to discriminate between these two approaches, especially in
comparison with separately measured $\pi^-\Sigma^+$ and $\pi^+\Sigma^-$
invariant mass spectra.
Of particular interest in this context are double differential cross
sections with detection of the emitted neutron in forward directions, to
be measured in a forthcoming experiment at J-PARC.
Detailed final state channel dependence originates from the interference
of $I=0$ and $I=1$ components of the final $\pi\Sigma$ states, providing
important information not only on the $I=0$ but also the $I=1$
$\bar{K}N$-$\pi\Sigma$ interactions.

Three-body dynamics in the coupled $K^-pn$-$\pi\Sigma n$ system is
predicted to generate a pronounced maximum in the $K^-d\rightarrow
\pi^-\Sigma^+n$ double-differential cross section with a forward-emitted
neutron at a $\pi\Sigma$ invariant mass $M_{\pi\Sigma}\simeq 1.45$~GeV.
Further detailed studies exploring this structure are under way.

\begin{acknowledgments}
The authors thank H. Noumi and A. Hosaka for helpful comments and discussions.
The numerical calculation has been performed on a supercomputer (NEC SX8R) at the Research Center
for Nuclear Physics, Osaka University.
This work was partly supported by the Grants-in-Aid for Scientific
 Research on Innovative Areas from MEXT (Grant No. 2404:24105008),
by RIKEN Junior Research Associate Program, 
by RIKEN iTHES Project,
by the Yukawa International Program for Quark-Hadron Sciences (YIPQS),
by JSPS KAKENHI Grants Nos. 23224006, 24740152 and 25800170,
and by DFG through CRC 110.
\end{acknowledgments}


\begin{thebibliography}{66}%
\makeatletter
\providecommand \@ifxundefined [1]{%
 \@ifx{#1\undefined}
}%
\providecommand \@ifnum [1]{%
 \ifnum #1\expandafter \@firstoftwo
 \else \expandafter \@secondoftwo
 \fi
}%
\providecommand \@ifx [1]{%
 \ifx #1\expandafter \@firstoftwo
 \else \expandafter \@secondoftwo
 \fi
}%
\providecommand \natexlab [1]{#1}%
\providecommand \enquote  [1]{``#1''}%
\providecommand \bibnamefont  [1]{#1}%
\providecommand \bibfnamefont [1]{#1}%
\providecommand \citenamefont [1]{#1}%
\providecommand \href@noop [0]{\@secondoftwo}%
\providecommand \href [0]{\begingroup \@sanitize@url \@href}%
\providecommand \@href[1]{\@@startlink{#1}\@@href}%
\providecommand \@@href[1]{\endgroup#1\@@endlink}%
\providecommand \@sanitize@url [0]{\catcode `\\12\catcode `\$12\catcode
  `\&12\catcode `\#12\catcode `\^12\catcode `\_12\catcode `\%12\relax}%
\providecommand \@@startlink[1]{}%
\providecommand \@@endlink[0]{}%
\providecommand \url  [0]{\begingroup\@sanitize@url \@url }%
\providecommand \@url [1]{\endgroup\@href {#1}{\urlprefix }}%
\providecommand \urlprefix  [0]{URL }%
\providecommand \Eprint [0]{\href }%
\providecommand \doibase [0]{http://dx.doi.org/}%
\providecommand \selectlanguage [0]{\@gobble}%
\providecommand \bibinfo  [0]{\@secondoftwo}%
\providecommand \bibfield  [0]{\@secondoftwo}%
\providecommand \translation [1]{[#1]}%
\providecommand \BibitemOpen [0]{}%
\providecommand \bibitemStop [0]{}%
\providecommand \bibitemNoStop [0]{.\EOS\space}%
\providecommand \EOS [0]{\spacefactor3000\relax}%
\providecommand \BibitemShut  [1]{\csname bibitem#1\endcsname}%
\let\auto@bib@innerbib\@empty
\bibitem [{\citenamefont {Dalitz}\ and\ \citenamefont
  {Tuan}(1959)}]{Dalitz:1959dn}%
  \BibitemOpen
  \bibfield  {author} {\bibinfo {author} {\bibfnamefont {R.}~\bibnamefont
  {Dalitz}}\ and\ \bibinfo {author} {\bibfnamefont {S.}~\bibnamefont {Tuan}},\
  }\href {\doibase 10.1103/PhysRevLett.2.425} {\bibfield  {journal} {\bibinfo
  {journal} {Phys.~Rev.~Lett.}\ }\textbf {\bibinfo {volume} {2}},\ \bibinfo
  {pages} {425} (\bibinfo {year} {1959})}\BibitemShut {NoStop}%
\bibitem [{\citenamefont {Dalitz}\ and\ \citenamefont
  {Tuan}(1960)}]{Dalitz:1960du}%
  \BibitemOpen
  \bibfield  {author} {\bibinfo {author} {\bibfnamefont {R.}~\bibnamefont
  {Dalitz}}\ and\ \bibinfo {author} {\bibfnamefont {S.}~\bibnamefont {Tuan}},\
  }\href {\doibase 10.1016/0003-4916(60)90001-4} {\bibfield  {journal}
  {\bibinfo  {journal} {Annals Phys.}\ }\textbf {\bibinfo {volume} {10}},\
  \bibinfo {pages} {307} (\bibinfo {year} {1960})}\BibitemShut {NoStop}%
\bibitem [{\citenamefont {Akaishi}\ and\ \citenamefont
  {Yamazaki}(2002)}]{Akaishi:2002bg}%
  \BibitemOpen
  \bibfield  {author} {\bibinfo {author} {\bibfnamefont {Y.}~\bibnamefont
  {Akaishi}}\ and\ \bibinfo {author} {\bibfnamefont {T.}~\bibnamefont
  {Yamazaki}},\ }\href {\doibase 10.1103/PhysRevC.65.044005} {\bibfield
  {journal} {\bibinfo  {journal} {Phys.~Rev.}\ }\textbf {\bibinfo {volume}
  {C65}},\ \bibinfo {pages} {044005} (\bibinfo {year} {2002})}\BibitemShut
  {NoStop}%
\bibitem [{\citenamefont {Shevchenko}(2012)}]{Shevchenko:2011ce}%
  \BibitemOpen
  \bibfield  {author} {\bibinfo {author} {\bibfnamefont {N.~V.}\ \bibnamefont
  {Shevchenko}},\ }\href {\doibase 10.1103/PhysRevC.85.034001} {\bibfield
  {journal} {\bibinfo  {journal} {Phys.~Rev.}\ }\textbf {\bibinfo {volume}
  {C85}},\ \bibinfo {pages} {034001} (\bibinfo {year} {2012})},\ \Eprint
  {http://arxiv.org/abs/1103.4974} {arXiv:1103.4974 [nucl-th]} \BibitemShut
  {NoStop}%
\bibitem [{\citenamefont {Kaiser}\ \emph {et~al.}(1995)\citenamefont {Kaiser},
  \citenamefont {Siegel},\ and\ \citenamefont {Weise}}]{Kaiser:1995eg}%
  \BibitemOpen
  \bibfield  {author} {\bibinfo {author} {\bibfnamefont {N.}~\bibnamefont
  {Kaiser}}, \bibinfo {author} {\bibfnamefont {P.}~\bibnamefont {Siegel}}, \
  and\ \bibinfo {author} {\bibfnamefont {W.}~\bibnamefont {Weise}},\ }\href
  {\doibase 10.1016/0375-9474(95)00362-5} {\bibfield  {journal} {\bibinfo
  {journal} {Nucl.~Phys.}\ }\textbf {\bibinfo {volume} {A594}},\ \bibinfo
  {pages} {325} (\bibinfo {year} {1995})},\ \Eprint
  {http://arxiv.org/abs/nucl-th/9505043} {arXiv:nucl-th/9505043 [nucl-th]}
  \BibitemShut {NoStop}%
\bibitem [{\citenamefont {Oset}\ and\ \citenamefont
  {Ramos}(1998)}]{Oset:1997it}%
  \BibitemOpen
  \bibfield  {author} {\bibinfo {author} {\bibfnamefont {E.}~\bibnamefont
  {Oset}}\ and\ \bibinfo {author} {\bibfnamefont {A.}~\bibnamefont {Ramos}},\
  }\href {\doibase 10.1016/S0375-9474(98)00170-5} {\bibfield  {journal}
  {\bibinfo  {journal} {Nucl.~Phys.}\ }\textbf {\bibinfo {volume} {A635}},\
  \bibinfo {pages} {99} (\bibinfo {year} {1998})},\ \Eprint
  {http://arxiv.org/abs/nucl-th/9711022} {arXiv:nucl-th/9711022 [nucl-th]}
  \BibitemShut {NoStop}%
\bibitem [{\citenamefont {Oller}\ and\ \citenamefont
  {Meissner}(2001)}]{Oller:2000fj}%
  \BibitemOpen
  \bibfield  {author} {\bibinfo {author} {\bibfnamefont {J.}~\bibnamefont
  {Oller}}\ and\ \bibinfo {author} {\bibfnamefont {U.~G.}\ \bibnamefont
  {Meissner}},\ }\href {\doibase 10.1016/S0370-2693(01)00078-8} {\bibfield
  {journal} {\bibinfo  {journal} {Phys.~Lett.}\ }\textbf {\bibinfo {volume}
  {B500}},\ \bibinfo {pages} {263} (\bibinfo {year} {2001})},\ \Eprint
  {http://arxiv.org/abs/hep-ph/0011146} {arXiv:hep-ph/0011146 [hep-ph]}
  \BibitemShut {NoStop}%
\bibitem [{\citenamefont {Hyodo}\ and\ \citenamefont
  {Jido}(2012)}]{Hyodo:2011ur}%
  \BibitemOpen
  \bibfield  {author} {\bibinfo {author} {\bibfnamefont {T.}~\bibnamefont
  {Hyodo}}\ and\ \bibinfo {author} {\bibfnamefont {D.}~\bibnamefont {Jido}},\
  }\href {\doibase 10.1016/j.ppnp.2011.07.002} {\bibfield  {journal} {\bibinfo
  {journal} {Prog.~Part.~Nucl.~Phys.}\ }\textbf {\bibinfo {volume} {67}},\
  \bibinfo {pages} {55} (\bibinfo {year} {2012})},\ \Eprint
  {http://arxiv.org/abs/1104.4474} {arXiv:1104.4474 [nucl-th]} \BibitemShut
  {NoStop}%
\bibitem [{\citenamefont {Hyodo}\ and\ \citenamefont
  {Weise}(2008)}]{Hyodo:2007jq}%
  \BibitemOpen
  \bibfield  {author} {\bibinfo {author} {\bibfnamefont {T.}~\bibnamefont
  {Hyodo}}\ and\ \bibinfo {author} {\bibfnamefont {W.}~\bibnamefont {Weise}},\
  }\href {\doibase 10.1103/PhysRevC.77.035204} {\bibfield  {journal} {\bibinfo
  {journal} {Phys.~Rev.}\ }\textbf {\bibinfo {volume} {C77}},\ \bibinfo {pages}
  {035204} (\bibinfo {year} {2008})},\ \Eprint {http://arxiv.org/abs/0712.1613}
  {arXiv:0712.1613 [nucl-th]} \BibitemShut {NoStop}%
\bibitem [{\citenamefont {Jido}\ \emph {et~al.}(2003)\citenamefont {Jido},
  \citenamefont {Oller}, \citenamefont {Oset}, \citenamefont {Ramos},\ and\
  \citenamefont {Meissner}}]{Jido:2003cb}%
  \BibitemOpen
  \bibfield  {author} {\bibinfo {author} {\bibfnamefont {D.}~\bibnamefont
  {Jido}}, \bibinfo {author} {\bibfnamefont {J.}~\bibnamefont {Oller}},
  \bibinfo {author} {\bibfnamefont {E.}~\bibnamefont {Oset}}, \bibinfo {author}
  {\bibfnamefont {A.}~\bibnamefont {Ramos}}, \ and\ \bibinfo {author}
  {\bibfnamefont {U.}~\bibnamefont {Meissner}},\ }\href {\doibase
  10.1016/S0375-9474(03)01598-7} {\bibfield  {journal} {\bibinfo  {journal}
  {Nucl.~Phys.}\ }\textbf {\bibinfo {volume} {A725}},\ \bibinfo {pages} {181}
  (\bibinfo {year} {2003})},\ \Eprint {http://arxiv.org/abs/nucl-th/0303062}
  {arXiv:nucl-th/0303062 [nucl-th]} \BibitemShut {NoStop}%
\bibitem [{\citenamefont {Yamazaki}\ and\ \citenamefont
  {Akaishi}(2002)}]{Yamazaki:2002uh}%
  \BibitemOpen
  \bibfield  {author} {\bibinfo {author} {\bibfnamefont {T.}~\bibnamefont
  {Yamazaki}}\ and\ \bibinfo {author} {\bibfnamefont {Y.}~\bibnamefont
  {Akaishi}},\ }\href {\doibase 10.1016/S0370-2693(02)01738-0} {\bibfield
  {journal} {\bibinfo  {journal} {Phys.~Lett.}\ }\textbf {\bibinfo {volume}
  {B535}},\ \bibinfo {pages} {70} (\bibinfo {year} {2002})}\BibitemShut
  {NoStop}%
\bibitem [{\citenamefont {Shevchenko}\ \emph
  {et~al.}(2007{\natexlab{a}})\citenamefont {Shevchenko}, \citenamefont {Gal},\
  and\ \citenamefont {Mares}}]{Shevchenko:2006xy}%
  \BibitemOpen
  \bibfield  {author} {\bibinfo {author} {\bibfnamefont {N.~V.}\ \bibnamefont
  {Shevchenko}}, \bibinfo {author} {\bibfnamefont {A.}~\bibnamefont {Gal}}, \
  and\ \bibinfo {author} {\bibfnamefont {J.}~\bibnamefont {Mares}},\ }\href
  {\doibase 10.1103/PhysRevLett.98.082301} {\bibfield  {journal} {\bibinfo
  {journal} {Phys.~Rev.~Lett.}\ }\textbf {\bibinfo {volume} {98}},\ \bibinfo
  {pages} {082301} (\bibinfo {year} {2007}{\natexlab{a}})},\ \Eprint
  {http://arxiv.org/abs/nucl-th/0610022} {arXiv:nucl-th/0610022 [nucl-th]}
  \BibitemShut {NoStop}%
\bibitem [{\citenamefont {Ikeda}\ and\ \citenamefont
  {Sato}(2007)}]{Ikeda:2007nz}%
  \BibitemOpen
  \bibfield  {author} {\bibinfo {author} {\bibfnamefont {Y.}~\bibnamefont
  {Ikeda}}\ and\ \bibinfo {author} {\bibfnamefont {T.}~\bibnamefont {Sato}},\
  }\href {\doibase 10.1103/PhysRevC.76.035203} {\bibfield  {journal} {\bibinfo
  {journal} {Phys.~Rev.}\ }\textbf {\bibinfo {volume} {C76}},\ \bibinfo {pages}
  {035203} (\bibinfo {year} {2007})},\ \Eprint {http://arxiv.org/abs/0704.1978}
  {arXiv:0704.1978 [nucl-th]} \BibitemShut {NoStop}%
\bibitem [{\citenamefont {Shevchenko}\ \emph
  {et~al.}(2007{\natexlab{b}})\citenamefont {Shevchenko}, \citenamefont {Gal},
  \citenamefont {Mares},\ and\ \citenamefont {Revai}}]{Shevchenko:2007ke}%
  \BibitemOpen
  \bibfield  {author} {\bibinfo {author} {\bibfnamefont {N.~V.}\ \bibnamefont
  {Shevchenko}}, \bibinfo {author} {\bibfnamefont {A.}~\bibnamefont {Gal}},
  \bibinfo {author} {\bibfnamefont {J.}~\bibnamefont {Mares}}, \ and\ \bibinfo
  {author} {\bibfnamefont {J.}~\bibnamefont {Revai}},\ }\href {\doibase
  10.1103/PhysRevC.76.044004} {\bibfield  {journal} {\bibinfo  {journal}
  {Phys.~Rev.}\ }\textbf {\bibinfo {volume} {C76}},\ \bibinfo {pages} {044004}
  (\bibinfo {year} {2007}{\natexlab{b}})},\ \Eprint
  {http://arxiv.org/abs/0706.4393} {arXiv:0706.4393 [nucl-th]} \BibitemShut
  {NoStop}%
\bibitem [{\citenamefont {Yamazaki}\ and\ \citenamefont
  {Akaishi}(2007)}]{Yamazaki:2007cs}%
  \BibitemOpen
  \bibfield  {author} {\bibinfo {author} {\bibfnamefont {T.}~\bibnamefont
  {Yamazaki}}\ and\ \bibinfo {author} {\bibfnamefont {Y.}~\bibnamefont
  {Akaishi}},\ }\href {\doibase 10.1103/PhysRevC.76.045201} {\bibfield
  {journal} {\bibinfo  {journal} {Phys.~Rev.}\ }\textbf {\bibinfo {volume}
  {C76}},\ \bibinfo {pages} {045201} (\bibinfo {year} {2007})},\ \Eprint
  {http://arxiv.org/abs/0709.0630} {arXiv:0709.0630 [nucl-th]} \BibitemShut
  {NoStop}%
\bibitem [{\citenamefont {Dote}\ \emph {et~al.}(2008)\citenamefont {Dote},
  \citenamefont {Hyodo},\ and\ \citenamefont {Weise}}]{Dote:2008in}%
  \BibitemOpen
  \bibfield  {author} {\bibinfo {author} {\bibfnamefont {A.}~\bibnamefont
  {Dote}}, \bibinfo {author} {\bibfnamefont {T.}~\bibnamefont {Hyodo}}, \ and\
  \bibinfo {author} {\bibfnamefont {W.}~\bibnamefont {Weise}},\ }\href
  {\doibase 10.1016/j.nuclphysa.2008.02.001} {\bibfield  {journal} {\bibinfo
  {journal} {Nucl.~Phys.}\ }\textbf {\bibinfo {volume} {A804}},\ \bibinfo
  {pages} {197} (\bibinfo {year} {2008})},\ \Eprint
  {http://arxiv.org/abs/0802.0238} {arXiv:0802.0238 [nucl-th]} \BibitemShut
  {NoStop}%
\bibitem [{\citenamefont {Dote}\ \emph {et~al.}(2009)\citenamefont {Dote},
  \citenamefont {Hyodo},\ and\ \citenamefont {Weise}}]{Dote:2008hw}%
  \BibitemOpen
  \bibfield  {author} {\bibinfo {author} {\bibfnamefont {A.}~\bibnamefont
  {Dote}}, \bibinfo {author} {\bibfnamefont {T.}~\bibnamefont {Hyodo}}, \ and\
  \bibinfo {author} {\bibfnamefont {W.}~\bibnamefont {Weise}},\ }\href
  {\doibase 10.1103/PhysRevC.79.014003} {\bibfield  {journal} {\bibinfo
  {journal} {Phys.~Rev.}\ }\textbf {\bibinfo {volume} {C79}},\ \bibinfo {pages}
  {014003} (\bibinfo {year} {2009})},\ \Eprint {http://arxiv.org/abs/0806.4917}
  {arXiv:0806.4917 [nucl-th]} \BibitemShut {NoStop}%
\bibitem [{\citenamefont {Wycech}\ and\ \citenamefont
  {Green}(2009)}]{Wycech:2008wf}%
  \BibitemOpen
  \bibfield  {author} {\bibinfo {author} {\bibfnamefont {S.}~\bibnamefont
  {Wycech}}\ and\ \bibinfo {author} {\bibfnamefont {A.~M.}\ \bibnamefont
  {Green}},\ }\href {\doibase 10.1103/PhysRevC.79.014001} {\bibfield  {journal}
  {\bibinfo  {journal} {Phys.~Rev.}\ }\textbf {\bibinfo {volume} {C79}},\
  \bibinfo {pages} {014001} (\bibinfo {year} {2009})},\ \Eprint
  {http://arxiv.org/abs/0808.3329} {arXiv:0808.3329 [nucl-th]} \BibitemShut
  {NoStop}%
\bibitem [{\citenamefont {Ikeda}\ and\ \citenamefont
  {Sato}(2009)}]{Ikeda:2008ub}%
  \BibitemOpen
  \bibfield  {author} {\bibinfo {author} {\bibfnamefont {Y.}~\bibnamefont
  {Ikeda}}\ and\ \bibinfo {author} {\bibfnamefont {T.}~\bibnamefont {Sato}},\
  }\href {\doibase 10.1103/PhysRevC.79.035201} {\bibfield  {journal} {\bibinfo
  {journal} {Phys.~Rev.}\ }\textbf {\bibinfo {volume} {C79}},\ \bibinfo {pages}
  {035201} (\bibinfo {year} {2009})},\ \Eprint {http://arxiv.org/abs/0809.1285}
  {arXiv:0809.1285 [nucl-th]} \BibitemShut {NoStop}%
\bibitem [{\citenamefont {Ikeda}\ \emph {et~al.}(2010)\citenamefont {Ikeda},
  \citenamefont {Kamano},\ and\ \citenamefont {Sato}}]{Ikeda:2010tk}%
  \BibitemOpen
  \bibfield  {author} {\bibinfo {author} {\bibfnamefont {Y.}~\bibnamefont
  {Ikeda}}, \bibinfo {author} {\bibfnamefont {H.}~\bibnamefont {Kamano}}, \
  and\ \bibinfo {author} {\bibfnamefont {T.}~\bibnamefont {Sato}},\ }\href
  {\doibase 10.1143/PTP.124.533} {\bibfield  {journal} {\bibinfo  {journal}
  {Prog.~Theor.~Phys.}\ }\textbf {\bibinfo {volume} {124}},\ \bibinfo {pages}
  {533} (\bibinfo {year} {2010})},\ \Eprint {http://arxiv.org/abs/1004.4877}
  {arXiv:1004.4877 [nucl-th]} \BibitemShut {NoStop}%
\bibitem [{\citenamefont {Barnea}\ \emph {et~al.}(2012)\citenamefont {Barnea},
  \citenamefont {Gal},\ and\ \citenamefont {Liverts}}]{Barnea:2012qa}%
  \BibitemOpen
  \bibfield  {author} {\bibinfo {author} {\bibfnamefont {N.}~\bibnamefont
  {Barnea}}, \bibinfo {author} {\bibfnamefont {A.}~\bibnamefont {Gal}}, \ and\
  \bibinfo {author} {\bibfnamefont {E.}~\bibnamefont {Liverts}},\ }\href
  {\doibase 10.1016/j.physletb.2012.04.055} {\bibfield  {journal} {\bibinfo
  {journal} {Phys.~Lett.}\ }\textbf {\bibinfo {volume} {B712}},\ \bibinfo
  {pages} {132} (\bibinfo {year} {2012})},\ \Eprint
  {http://arxiv.org/abs/1203.5234} {arXiv:1203.5234 [nucl-th]} \BibitemShut
  {NoStop}%
\bibitem [{\citenamefont {Ohnishi}\ \emph {et~al.}(2013)\citenamefont
  {Ohnishi}, \citenamefont {Ikeda}, \citenamefont {Kamano},\ and\ \citenamefont
  {Sato}}]{Ohnishi:2013rix}%
  \BibitemOpen
  \bibfield  {author} {\bibinfo {author} {\bibfnamefont {S.}~\bibnamefont
  {Ohnishi}}, \bibinfo {author} {\bibfnamefont {Y.}~\bibnamefont {Ikeda}},
  \bibinfo {author} {\bibfnamefont {H.}~\bibnamefont {Kamano}}, \ and\ \bibinfo
  {author} {\bibfnamefont {T.}~\bibnamefont {Sato}},\ }\href {\doibase
  10.1103/PhysRevC.88.025204} {\bibfield  {journal} {\bibinfo  {journal}
  {Phys.~Rev.}\ }\textbf {\bibinfo {volume} {C88}},\ \bibinfo {pages} {025204}
  (\bibinfo {year} {2013})},\ \Eprint {http://arxiv.org/abs/1302.2301}
  {arXiv:1302.2301 [nucl-th]} \BibitemShut {NoStop}%
\bibitem [{\citenamefont {Humphrey}\ and\ \citenamefont
  {Ross}(1962)}]{Humphrey:1962zz}%
  \BibitemOpen
  \bibfield  {author} {\bibinfo {author} {\bibfnamefont {W.~E.}\ \bibnamefont
  {Humphrey}}\ and\ \bibinfo {author} {\bibfnamefont {R.~R.}\ \bibnamefont
  {Ross}},\ }\href {\doibase 10.1103/PhysRev.127.1305} {\bibfield  {journal}
  {\bibinfo  {journal} {Phys.Rev.}\ }\textbf {\bibinfo {volume} {127}},\
  \bibinfo {pages} {1305} (\bibinfo {year} {1962})}\BibitemShut {NoStop}%
\bibitem [{\citenamefont {Sakitt}\ \emph {et~al.}(1965)\citenamefont {Sakitt},
  \citenamefont {Day}, \citenamefont {Glasser}, \citenamefont {Seeman},
  \citenamefont {Friedman} \emph {et~al.}}]{Sakitt:1965kh}%
  \BibitemOpen
  \bibfield  {author} {\bibinfo {author} {\bibfnamefont {M.}~\bibnamefont
  {Sakitt}}, \bibinfo {author} {\bibfnamefont {T.}~\bibnamefont {Day}},
  \bibinfo {author} {\bibfnamefont {R.}~\bibnamefont {Glasser}}, \bibinfo
  {author} {\bibfnamefont {N.}~\bibnamefont {Seeman}}, \bibinfo {author}
  {\bibfnamefont {J.}~\bibnamefont {Friedman}},  \emph {et~al.},\ }\href
  {\doibase 10.1103/PhysRev.139.B719} {\bibfield  {journal} {\bibinfo
  {journal} {Phys.Rev.}\ }\textbf {\bibinfo {volume} {139}},\ \bibinfo {pages}
  {B719} (\bibinfo {year} {1965})}\BibitemShut {NoStop}%
\bibitem [{\citenamefont {Kim}(1965)}]{Kim:1965zz}%
  \BibitemOpen
  \bibfield  {author} {\bibinfo {author} {\bibfnamefont {J.}~\bibnamefont
  {Kim}},\ }\href {\doibase 10.1103/PhysRevLett.14.29} {\bibfield  {journal}
  {\bibinfo  {journal} {Phys.Rev.Lett.}\ }\textbf {\bibinfo {volume} {14}},\
  \bibinfo {pages} {29} (\bibinfo {year} {1965})}\BibitemShut {NoStop}%
\bibitem [{\citenamefont {Kittel}\ \emph {et~al.}(1966)\citenamefont {Kittel},
  \citenamefont {Otter},\ and\ \citenamefont {Wacek}}]{Kittel:1966zz}%
  \BibitemOpen
  \bibfield  {author} {\bibinfo {author} {\bibfnamefont {W.}~\bibnamefont
  {Kittel}}, \bibinfo {author} {\bibfnamefont {G.}~\bibnamefont {Otter}}, \
  and\ \bibinfo {author} {\bibfnamefont {I.}~\bibnamefont {Wacek}},\ }\href
  {\doibase 10.1016/0031-9163(66)90845-6} {\bibfield  {journal} {\bibinfo
  {journal} {Phys.Lett.}\ }\textbf {\bibinfo {volume} {21}},\ \bibinfo {pages}
  {349} (\bibinfo {year} {1966})}\BibitemShut {NoStop}%
\bibitem [{\citenamefont {Evans}\ \emph {et~al.}(1983)\citenamefont {Evans},
  \citenamefont {Major}, \citenamefont {Rondio}, \citenamefont {Zakrzewski},
  \citenamefont {Conboy} \emph {et~al.}}]{Evans:1983hz}%
  \BibitemOpen
  \bibfield  {author} {\bibinfo {author} {\bibfnamefont {D.}~\bibnamefont
  {Evans}}, \bibinfo {author} {\bibfnamefont {J.}~\bibnamefont {Major}},
  \bibinfo {author} {\bibfnamefont {E.}~\bibnamefont {Rondio}}, \bibinfo
  {author} {\bibfnamefont {J.~A.}\ \bibnamefont {Zakrzewski}}, \bibinfo
  {author} {\bibfnamefont {J.}~\bibnamefont {Conboy}},  \emph {et~al.},\ }\href
  {\doibase 10.1088/0305-4616/9/8/011} {\bibfield  {journal} {\bibinfo
  {journal} {J.Phys.}\ }\textbf {\bibinfo {volume} {G9}},\ \bibinfo {pages}
  {885} (\bibinfo {year} {1983})}\BibitemShut {NoStop}%
\bibitem [{\citenamefont {Tovee}\ \emph {et~al.}(1971)\citenamefont {Tovee},
  \citenamefont {Davis}, \citenamefont {Simonovic}, \citenamefont {Bohm},
  \citenamefont {Klabuhn} \emph {et~al.}}]{Tovee:1971ga}%
  \BibitemOpen
  \bibfield  {author} {\bibinfo {author} {\bibfnamefont {D.}~\bibnamefont
  {Tovee}}, \bibinfo {author} {\bibfnamefont {D.}~\bibnamefont {Davis}},
  \bibinfo {author} {\bibfnamefont {J.}~\bibnamefont {Simonovic}}, \bibinfo
  {author} {\bibfnamefont {G.}~\bibnamefont {Bohm}}, \bibinfo {author}
  {\bibfnamefont {J.}~\bibnamefont {Klabuhn}},  \emph {et~al.},\ }\href
  {\doibase 10.1016/0550-3213(71)90302-6} {\bibfield  {journal} {\bibinfo
  {journal} {Nucl.Phys.}\ }\textbf {\bibinfo {volume} {B33}},\ \bibinfo {pages}
  {493} (\bibinfo {year} {1971})}\BibitemShut {NoStop}%
\bibitem [{\citenamefont {Nowak}\ \emph {et~al.}(1978)\citenamefont {Nowak},
  \citenamefont {Armstrong}, \citenamefont {Davis}, \citenamefont {Miller},
  \citenamefont {Tovee} \emph {et~al.}}]{Nowak:1978au}%
  \BibitemOpen
  \bibfield  {author} {\bibinfo {author} {\bibfnamefont {R.}~\bibnamefont
  {Nowak}}, \bibinfo {author} {\bibfnamefont {J.}~\bibnamefont {Armstrong}},
  \bibinfo {author} {\bibfnamefont {D.}~\bibnamefont {Davis}}, \bibinfo
  {author} {\bibfnamefont {D.}~\bibnamefont {Miller}}, \bibinfo {author}
  {\bibfnamefont {D.}~\bibnamefont {Tovee}},  \emph {et~al.},\ }\href {\doibase
  10.1016/0550-3213(78)90179-7} {\bibfield  {journal} {\bibinfo  {journal}
  {Nucl.Phys.}\ }\textbf {\bibinfo {volume} {B139}},\ \bibinfo {pages} {61}
  (\bibinfo {year} {1978})}\BibitemShut {NoStop}%
\bibitem [{\citenamefont {Iwasaki}\ \emph {et~al.}(1997)\citenamefont
  {Iwasaki}, \citenamefont {Hayano}, \citenamefont {Ito}, \citenamefont
  {Nakamura}, \citenamefont {Terada} \emph {et~al.}}]{Iwasaki:1997wf}%
  \BibitemOpen
  \bibfield  {author} {\bibinfo {author} {\bibfnamefont {M.}~\bibnamefont
  {Iwasaki}}, \bibinfo {author} {\bibfnamefont {R.}~\bibnamefont {Hayano}},
  \bibinfo {author} {\bibfnamefont {T.}~\bibnamefont {Ito}}, \bibinfo {author}
  {\bibfnamefont {S.}~\bibnamefont {Nakamura}}, \bibinfo {author}
  {\bibfnamefont {T.}~\bibnamefont {Terada}},  \emph {et~al.},\ }\href
  {\doibase 10.1103/PhysRevLett.78.3067} {\bibfield  {journal} {\bibinfo
  {journal} {Phys.Rev.Lett.}\ }\textbf {\bibinfo {volume} {78}},\ \bibinfo
  {pages} {3067} (\bibinfo {year} {1997})}\BibitemShut {NoStop}%
\bibitem [{\citenamefont {Ito}\ \emph {et~al.}(1998)\citenamefont {Ito},
  \citenamefont {Hayano}, \citenamefont {Nakamura}, \citenamefont {Terada},
  \citenamefont {Iwasaki} \emph {et~al.}}]{Ito:1998yi}%
  \BibitemOpen
  \bibfield  {author} {\bibinfo {author} {\bibfnamefont {T.}~\bibnamefont
  {Ito}}, \bibinfo {author} {\bibfnamefont {R.}~\bibnamefont {Hayano}},
  \bibinfo {author} {\bibfnamefont {S.}~\bibnamefont {Nakamura}}, \bibinfo
  {author} {\bibfnamefont {T.}~\bibnamefont {Terada}}, \bibinfo {author}
  {\bibfnamefont {M.}~\bibnamefont {Iwasaki}},  \emph {et~al.},\ }\href
  {\doibase 10.1103/PhysRevC.58.2366} {\bibfield  {journal} {\bibinfo
  {journal} {Phys.Rev.}\ }\textbf {\bibinfo {volume} {C58}},\ \bibinfo {pages}
  {2366} (\bibinfo {year} {1998})}\BibitemShut {NoStop}%
\bibitem [{\citenamefont {Beer}\ \emph {et~al.}(2005)\citenamefont {Beer} \emph
  {et~al.}}]{Beer:2005qi}%
  \BibitemOpen
  \bibfield  {author} {\bibinfo {author} {\bibfnamefont {G.}~\bibnamefont
  {Beer}} \emph {et~al.} (\bibinfo {collaboration} {DEAR}),\ }\href {\doibase
  10.1103/PhysRevLett.94.212302} {\bibfield  {journal} {\bibinfo  {journal}
  {Phys.Rev.Lett.}\ }\textbf {\bibinfo {volume} {94}},\ \bibinfo {pages}
  {212302} (\bibinfo {year} {2005})}\BibitemShut {NoStop}%
\bibitem [{\citenamefont {Bazzi}\ \emph {et~al.}(2011)\citenamefont {Bazzi},
  \citenamefont {Beer}, \citenamefont {Bombelli}, \citenamefont {Bragadireanu},
  \citenamefont {Cargnelli} \emph {et~al.}}]{Bazzi:2011zj}%
  \BibitemOpen
  \bibfield  {author} {\bibinfo {author} {\bibfnamefont {M.}~\bibnamefont
  {Bazzi}}, \bibinfo {author} {\bibfnamefont {G.}~\bibnamefont {Beer}},
  \bibinfo {author} {\bibfnamefont {L.}~\bibnamefont {Bombelli}}, \bibinfo
  {author} {\bibfnamefont {A.}~\bibnamefont {Bragadireanu}}, \bibinfo {author}
  {\bibfnamefont {M.}~\bibnamefont {Cargnelli}},  \emph {et~al.},\ }\href
  {\doibase 10.1016/j.physletb.2011.09.011} {\bibfield  {journal} {\bibinfo
  {journal} {Phys. Lett.}\ }\textbf {\bibinfo {volume} {B704}},\ \bibinfo
  {pages} {113} (\bibinfo {year} {2011})},\ \Eprint
  {http://arxiv.org/abs/1105.3090} {arXiv:1105.3090 [nucl-ex]} \BibitemShut
  {NoStop}%
\bibitem [{\citenamefont {Bazzi}\ \emph {et~al.}(2012)\citenamefont {Bazzi},
  \citenamefont {Beer}, \citenamefont {Bombelli}, \citenamefont {Bragadireanu},
  \citenamefont {Cargnelli} \emph {et~al.}}]{Bazzi:2012eq}%
  \BibitemOpen
  \bibfield  {author} {\bibinfo {author} {\bibfnamefont {M.}~\bibnamefont
  {Bazzi}}, \bibinfo {author} {\bibfnamefont {G.}~\bibnamefont {Beer}},
  \bibinfo {author} {\bibfnamefont {L.}~\bibnamefont {Bombelli}}, \bibinfo
  {author} {\bibfnamefont {A.}~\bibnamefont {Bragadireanu}}, \bibinfo {author}
  {\bibfnamefont {M.}~\bibnamefont {Cargnelli}},  \emph {et~al.},\ }\href
  {\doibase 10.1016/j.nuclphysa.2011.12.008} {\bibfield  {journal} {\bibinfo
  {journal} {Nucl.Phys.}\ }\textbf {\bibinfo {volume} {A881}},\ \bibinfo
  {pages} {88} (\bibinfo {year} {2012})},\ \Eprint
  {http://arxiv.org/abs/1201.4635} {arXiv:1201.4635 [nucl-ex]} \BibitemShut
  {NoStop}%
\bibitem [{\citenamefont {Ikeda}\ \emph {et~al.}(2011)\citenamefont {Ikeda},
  \citenamefont {Hyodo},\ and\ \citenamefont {Weise}}]{Ikeda:2011pi}%
  \BibitemOpen
  \bibfield  {author} {\bibinfo {author} {\bibfnamefont {Y.}~\bibnamefont
  {Ikeda}}, \bibinfo {author} {\bibfnamefont {T.}~\bibnamefont {Hyodo}}, \ and\
  \bibinfo {author} {\bibfnamefont {W.}~\bibnamefont {Weise}},\ }\href
  {\doibase 10.1016/j.physletb.2011.10.068} {\bibfield  {journal} {\bibinfo
  {journal} {Phys. Lett.}\ }\textbf {\bibinfo {volume} {B706}},\ \bibinfo
  {pages} {63} (\bibinfo {year} {2011})},\ \Eprint
  {http://arxiv.org/abs/1109.3005} {arXiv:1109.3005 [nucl-th]} \BibitemShut
  {NoStop}%
\bibitem [{\citenamefont {Ikeda}\ \emph {et~al.}(2012)\citenamefont {Ikeda},
  \citenamefont {Hyodo},\ and\ \citenamefont {Weise}}]{Ikeda:2012au}%
  \BibitemOpen
  \bibfield  {author} {\bibinfo {author} {\bibfnamefont {Y.}~\bibnamefont
  {Ikeda}}, \bibinfo {author} {\bibfnamefont {T.}~\bibnamefont {Hyodo}}, \ and\
  \bibinfo {author} {\bibfnamefont {W.}~\bibnamefont {Weise}},\ }\href
  {\doibase 10.1016/j.nuclphysa.2012.01.029} {\bibfield  {journal} {\bibinfo
  {journal} {Nucl. Phys.}\ }\textbf {\bibinfo {volume} {A881}},\ \bibinfo
  {pages} {98} (\bibinfo {year} {2012})},\ \Eprint
  {http://arxiv.org/abs/1201.6549} {arXiv:1201.6549 [nucl-th]} \BibitemShut
  {NoStop}%
\bibitem [{\citenamefont {Ahn}(2003)}]{Ahn:2003mv}%
  \BibitemOpen
  \bibfield  {author} {\bibinfo {author} {\bibfnamefont {J.}~\bibnamefont
  {Ahn}} (\bibinfo {collaboration} {LEPS Collaboration}),\ }\href {\doibase
  10.1016/S0375-9474(03)01164-3} {\bibfield  {journal} {\bibinfo  {journal}
  {Nucl.Phys.}\ }\textbf {\bibinfo {volume} {A721}},\ \bibinfo {pages} {715}
  (\bibinfo {year} {2003})}\BibitemShut {NoStop}%
\bibitem [{\citenamefont {Niiyama}\ \emph {et~al.}(2008)\citenamefont
  {Niiyama}, \citenamefont {Fujimura}, \citenamefont {Ahn}, \citenamefont
  {Ahn}, \citenamefont {Ajimura} \emph {et~al.}}]{Niiyama:2008rt}%
  \BibitemOpen
  \bibfield  {author} {\bibinfo {author} {\bibfnamefont {M.}~\bibnamefont
  {Niiyama}}, \bibinfo {author} {\bibfnamefont {H.}~\bibnamefont {Fujimura}},
  \bibinfo {author} {\bibfnamefont {D.}~\bibnamefont {Ahn}}, \bibinfo {author}
  {\bibfnamefont {J.}~\bibnamefont {Ahn}}, \bibinfo {author} {\bibfnamefont
  {S.}~\bibnamefont {Ajimura}},  \emph {et~al.},\ }\href {\doibase
  10.1103/PhysRevC.78.035202} {\bibfield  {journal} {\bibinfo  {journal} {Phys.
  Rev.}\ }\textbf {\bibinfo {volume} {C78}},\ \bibinfo {pages} {035202}
  (\bibinfo {year} {2008})},\ \Eprint {http://arxiv.org/abs/0805.4051}
  {arXiv:0805.4051 [hep-ex]} \BibitemShut {NoStop}%
\bibitem [{\citenamefont {Moriya}\ \emph
  {et~al.}(2013{\natexlab{a}})\citenamefont {Moriya} \emph
  {et~al.}}]{Moriya:2013eb}%
  \BibitemOpen
  \bibfield  {author} {\bibinfo {author} {\bibfnamefont {K.}~\bibnamefont
  {Moriya}} \emph {et~al.} (\bibinfo {collaboration} {CLAS}),\ }\href {\doibase
  10.1103/PhysRevC.87.035206} {\bibfield  {journal} {\bibinfo  {journal} {Phys.
  Rev.}\ }\textbf {\bibinfo {volume} {C87}},\ \bibinfo {pages} {035206}
  (\bibinfo {year} {2013}{\natexlab{a}})},\ \Eprint
  {http://arxiv.org/abs/1301.5000} {arXiv:1301.5000 [nucl-ex]} \BibitemShut
  {NoStop}%
\bibitem [{\citenamefont {Moriya}\ \emph
  {et~al.}(2013{\natexlab{b}})\citenamefont {Moriya} \emph
  {et~al.}}]{Moriya:2013hwg}%
  \BibitemOpen
  \bibfield  {author} {\bibinfo {author} {\bibfnamefont {K.}~\bibnamefont
  {Moriya}} \emph {et~al.} (\bibinfo {collaboration} {CLAS}),\ }\href {\doibase
  10.1103/PhysRevC.88.049902, 10.1103/PhysRevC.88.045201} {\bibfield  {journal}
  {\bibinfo  {journal} {Phys. Rev.}\ }\textbf {\bibinfo {volume} {C88}},\
  \bibinfo {pages} {045201} (\bibinfo {year} {2013}{\natexlab{b}})},\ \Eprint
  {http://arxiv.org/abs/1305.6776} {arXiv:1305.6776 [nucl-ex]} \BibitemShut
  {NoStop}%
\bibitem [{\citenamefont {Agakishiev}\ \emph {et~al.}(2013)\citenamefont
  {Agakishiev} \emph {et~al.}}]{Agakishiev:2012xk}%
  \BibitemOpen
  \bibfield  {author} {\bibinfo {author} {\bibfnamefont {G.}~\bibnamefont
  {Agakishiev}} \emph {et~al.} (\bibinfo {collaboration} {HADES
  Collaboration}),\ }\href {\doibase 10.1103/PhysRevC.87.025201} {\bibfield
  {journal} {\bibinfo  {journal} {Phys.Rev.}\ }\textbf {\bibinfo {volume}
  {C87}},\ \bibinfo {pages} {025201} (\bibinfo {year} {2013})},\ \Eprint
  {http://arxiv.org/abs/1208.0205} {arXiv:1208.0205 [nucl-ex]} \BibitemShut
  {NoStop}%
\bibitem [{\citenamefont {Roca}\ and\ \citenamefont
  {Oset}(2013{\natexlab{a}})}]{Roca:2013av}%
  \BibitemOpen
  \bibfield  {author} {\bibinfo {author} {\bibfnamefont {L.}~\bibnamefont
  {Roca}}\ and\ \bibinfo {author} {\bibfnamefont {E.}~\bibnamefont {Oset}},\
  }\href {\doibase 10.1103/PhysRevC.87.055201} {\bibfield  {journal} {\bibinfo
  {journal} {Phys. Rev.}\ }\textbf {\bibinfo {volume} {C87}},\ \bibinfo {pages}
  {055201} (\bibinfo {year} {2013}{\natexlab{a}})},\ \Eprint
  {http://arxiv.org/abs/1301.5741} {arXiv:1301.5741 [nucl-th]} \BibitemShut
  {NoStop}%
\bibitem [{\citenamefont {Roca}\ and\ \citenamefont
  {Oset}(2013{\natexlab{b}})}]{Roca:2013cca}%
  \BibitemOpen
  \bibfield  {author} {\bibinfo {author} {\bibfnamefont {L.}~\bibnamefont
  {Roca}}\ and\ \bibinfo {author} {\bibfnamefont {E.}~\bibnamefont {Oset}},\
  }\href {\doibase 10.1103/PhysRevC.88.055206} {\bibfield  {journal} {\bibinfo
  {journal} {Phys. Rev.}\ }\textbf {\bibinfo {volume} {C88}},\ \bibinfo {pages}
  {055206} (\bibinfo {year} {2013}{\natexlab{b}})},\ \Eprint
  {http://arxiv.org/abs/1307.5752} {arXiv:1307.5752 [nucl-th]} \BibitemShut
  {NoStop}%
\bibitem [{\citenamefont {Guo}\ and\ \citenamefont {Oller}(2013)}]{Guo:2012vv}%
  \BibitemOpen
  \bibfield  {author} {\bibinfo {author} {\bibfnamefont {Z.-H.}\ \bibnamefont
  {Guo}}\ and\ \bibinfo {author} {\bibfnamefont {J.~A.}\ \bibnamefont
  {Oller}},\ }\href {\doibase 10.1103/PhysRevC.87.035202} {\bibfield  {journal}
  {\bibinfo  {journal} {Phys. Rev.}\ }\textbf {\bibinfo {volume} {C87}},\
  \bibinfo {pages} {035202} (\bibinfo {year} {2013})},\ \Eprint
  {http://arxiv.org/abs/1210.3485} {arXiv:1210.3485 [hep-ph]} \BibitemShut
  {NoStop}%
\bibitem [{\citenamefont {Mai}\ and\ \citenamefont
  {Meissner}(2015)}]{Mai:2014xna}%
  \BibitemOpen
  \bibfield  {author} {\bibinfo {author} {\bibfnamefont {M.}~\bibnamefont
  {Mai}}\ and\ \bibinfo {author} {\bibfnamefont {U.-G.}\ \bibnamefont
  {Meissner}},\ }\href {\doibase 10.1140/epja/i2015-15030-3} {\bibfield
  {journal} {\bibinfo  {journal} {Eur. Phys. J.}\ }\textbf {\bibinfo {volume}
  {A51}},\ \bibinfo {pages} {30} (\bibinfo {year} {2015})},\ \Eprint
  {http://arxiv.org/abs/1411.7884} {arXiv:1411.7884 [hep-ph]} \BibitemShut
  {NoStop}%
\bibitem [{\citenamefont {Braun}\ \emph {et~al.}(1977)\citenamefont {Braun},
  \citenamefont {Grimm}, \citenamefont {Hepp}, \citenamefont {Strobele},
  \citenamefont {Thol} \emph {et~al.}}]{Braun:1977wd}%
  \BibitemOpen
  \bibfield  {author} {\bibinfo {author} {\bibfnamefont {O.}~\bibnamefont
  {Braun}}, \bibinfo {author} {\bibfnamefont {H.}~\bibnamefont {Grimm}},
  \bibinfo {author} {\bibfnamefont {V.}~\bibnamefont {Hepp}}, \bibinfo {author}
  {\bibfnamefont {H.}~\bibnamefont {Strobele}}, \bibinfo {author}
  {\bibfnamefont {C.}~\bibnamefont {Thol}},  \emph {et~al.},\ }\href {\doibase
  10.1016/0550-3213(77)90015-3} {\bibfield  {journal} {\bibinfo  {journal}
  {Nucl.~Phys.}\ }\textbf {\bibinfo {volume} {B129}},\ \bibinfo {pages} {1}
  (\bibinfo {year} {1977})}\BibitemShut {NoStop}%
\bibitem [{\citenamefont {Noumi}\ \emph {et~al.}(2009)\citenamefont {Noumi}
  \emph {et~al.}}]{Noumi}%
  \BibitemOpen
  \bibfield  {author} {\bibinfo {author} {\bibfnamefont {H.}~\bibnamefont
  {Noumi}} \emph {et~al.},\ }\href
  {http://j-parc.jp/researcher/Hadron/en/pac_0907/pdf/Noumi.pdf} {\enquote
  {\bibinfo {title} {{J-PARC proposal E31}},}\ } (\bibinfo {year}
  {2009})\BibitemShut {NoStop}%
\bibitem [{Note1()}]{Note1}%
  \BibitemOpen
  \bibinfo {note} {In this paper, we focus on the in-flight reactions with
  relatively energetic incident kaons. The same $K^-d\rightarrow \pi \Sigma n$
  process at lower energy has been studied in Ref.~\cite {Tan:1973at}. For
  theoretical studies with this kinematics, see Refs.~\cite
  {Jido:2010rx,Revai:2012fx}.}\BibitemShut {Stop}%
\bibitem [{\citenamefont {Jido}\ \emph {et~al.}(2009)\citenamefont {Jido},
  \citenamefont {Oset},\ and\ \citenamefont {Sekihara}}]{Jido:2009jf}%
  \BibitemOpen
  \bibfield  {author} {\bibinfo {author} {\bibfnamefont {D.}~\bibnamefont
  {Jido}}, \bibinfo {author} {\bibfnamefont {E.}~\bibnamefont {Oset}}, \ and\
  \bibinfo {author} {\bibfnamefont {T.}~\bibnamefont {Sekihara}},\ }\href
  {\doibase 10.1140/epja/i2009-10875-5} {\bibfield  {journal} {\bibinfo
  {journal} {Eur.~Phys.~J.}\ }\textbf {\bibinfo {volume} {A42}},\ \bibinfo
  {pages} {257} (\bibinfo {year} {2009})},\ \Eprint
  {http://arxiv.org/abs/0904.3410} {arXiv:0904.3410 [nucl-th]} \BibitemShut
  {NoStop}%
\bibitem [{\citenamefont {Miyagawa}\ and\ \citenamefont
  {Haidenbauer}(2012)}]{Miyagawa:2012xz}%
  \BibitemOpen
  \bibfield  {author} {\bibinfo {author} {\bibfnamefont {K.}~\bibnamefont
  {Miyagawa}}\ and\ \bibinfo {author} {\bibfnamefont {J.}~\bibnamefont
  {Haidenbauer}},\ }\href {\doibase 10.1103/PhysRevC.85.065201} {\bibfield
  {journal} {\bibinfo  {journal} {Phys.~Rev.}\ }\textbf {\bibinfo {volume}
  {C85}},\ \bibinfo {pages} {065201} (\bibinfo {year} {2012})},\ \Eprint
  {http://arxiv.org/abs/1202.4272} {arXiv:1202.4272 [nucl-th]} \BibitemShut
  {NoStop}%
\bibitem [{\citenamefont {Jido}\ \emph {et~al.}(2013)\citenamefont {Jido},
  \citenamefont {Oset},\ and\ \citenamefont {Sekihara}}]{Jido:2012cy}%
  \BibitemOpen
  \bibfield  {author} {\bibinfo {author} {\bibfnamefont {D.}~\bibnamefont
  {Jido}}, \bibinfo {author} {\bibfnamefont {E.}~\bibnamefont {Oset}}, \ and\
  \bibinfo {author} {\bibfnamefont {T.}~\bibnamefont {Sekihara}},\ }\href
  {\doibase 10.1140/epja/i2013-13095-6} {\bibfield  {journal} {\bibinfo
  {journal} {Eur.~Phys.~J.}\ }\textbf {\bibinfo {volume} {A49}},\ \bibinfo
  {pages} {95} (\bibinfo {year} {2013})},\ \Eprint
  {http://arxiv.org/abs/1207.5350} {arXiv:1207.5350 [nucl-th]} \BibitemShut
  {NoStop}%
\bibitem [{\citenamefont {Yamagata-Sekihara}\ \emph {et~al.}(2013)\citenamefont
  {Yamagata-Sekihara}, \citenamefont {Sekihara},\ and\ \citenamefont
  {Jido}}]{YamagataSekihara:2012yv}%
  \BibitemOpen
  \bibfield  {author} {\bibinfo {author} {\bibfnamefont {J.}~\bibnamefont
  {Yamagata-Sekihara}}, \bibinfo {author} {\bibfnamefont {T.}~\bibnamefont
  {Sekihara}}, \ and\ \bibinfo {author} {\bibfnamefont {D.}~\bibnamefont
  {Jido}},\ }\href {\doibase 10.1093/ptep/ptt009} {\bibfield  {journal}
  {\bibinfo  {journal} {PTEP}\ }\textbf {\bibinfo {volume} {2013}},\ \bibinfo
  {pages} {043D02} (\bibinfo {year} {2013})},\ \Eprint
  {http://arxiv.org/abs/1210.6108} {arXiv:1210.6108 [nucl-th]} \BibitemShut
  {NoStop}%
\bibitem [{\citenamefont {Amado}(1963)}]{PhysRev.132.485}%
  \BibitemOpen
  \bibfield  {author} {\bibinfo {author} {\bibfnamefont {R.~D.}\ \bibnamefont
  {Amado}},\ }\href {\doibase 10.1103/PhysRev.132.485} {\bibfield  {journal}
  {\bibinfo  {journal} {Phys. Rev.}\ }\textbf {\bibinfo {volume} {132}},\
  \bibinfo {pages} {485} (\bibinfo {year} {1963})}\BibitemShut {NoStop}%
\bibitem [{\citenamefont {Alt}\ \emph {et~al.}(1967)\citenamefont {Alt},
  \citenamefont {Grassberger},\ and\ \citenamefont {Sandhas}}]{Alt:1967fx}%
  \BibitemOpen
  \bibfield  {author} {\bibinfo {author} {\bibfnamefont {E.}~\bibnamefont
  {Alt}}, \bibinfo {author} {\bibfnamefont {P.}~\bibnamefont {Grassberger}}, \
  and\ \bibinfo {author} {\bibfnamefont {W.}~\bibnamefont {Sandhas}},\ }\href
  {\doibase 10.1016/0550-3213(67)90016-8} {\bibfield  {journal} {\bibinfo
  {journal} {Nucl.Phys.}\ }\textbf {\bibinfo {volume} {B2}},\ \bibinfo {pages}
  {167} (\bibinfo {year} {1967})}\BibitemShut {NoStop}%
\bibitem [{\citenamefont {Weinberg}(1966)}]{Weinberg:1966kf}%
  \BibitemOpen
  \bibfield  {author} {\bibinfo {author} {\bibfnamefont {S.}~\bibnamefont
  {Weinberg}},\ }\href {\doibase 10.1103/PhysRevLett.17.616} {\bibfield
  {journal} {\bibinfo  {journal} {Phys.~Rev.~Lett.}\ }\textbf {\bibinfo
  {volume} {17}},\ \bibinfo {pages} {616} (\bibinfo {year} {1966})}\BibitemShut
  {NoStop}%
\bibitem [{\citenamefont {Tomozawa}(1966)}]{Tomozawa:1966jm}%
  \BibitemOpen
  \bibfield  {author} {\bibinfo {author} {\bibfnamefont {Y.}~\bibnamefont
  {Tomozawa}},\ }\href {\doibase 10.1007/BF02857517} {\bibfield  {journal}
  {\bibinfo  {journal} {Nuovo Cim.}\ }\textbf {\bibinfo {volume} {A46}},\
  \bibinfo {pages} {707} (\bibinfo {year} {1966})}\BibitemShut {NoStop}%
\bibitem [{\citenamefont {Meissner}\ \emph {et~al.}(2004)\citenamefont
  {Meissner}, \citenamefont {Raha},\ and\ \citenamefont
  {Rusetsky}}]{Meissner:2004jr}%
  \BibitemOpen
  \bibfield  {author} {\bibinfo {author} {\bibfnamefont {U.~G.}\ \bibnamefont
  {Meissner}}, \bibinfo {author} {\bibfnamefont {U.}~\bibnamefont {Raha}}, \
  and\ \bibinfo {author} {\bibfnamefont {A.}~\bibnamefont {Rusetsky}},\ }\href
  {\doibase 10.1140/epjc/s2004-01859-4} {\bibfield  {journal} {\bibinfo
  {journal} {Eur. Phys. J.}\ }\textbf {\bibinfo {volume} {C35}},\ \bibinfo
  {pages} {349} (\bibinfo {year} {2004})},\ \Eprint
  {http://arxiv.org/abs/hep-ph/0402261} {arXiv:hep-ph/0402261 [hep-ph]}
  \BibitemShut {NoStop}%
\bibitem [{\citenamefont {Schroder}\ \emph {et~al.}(1999)\citenamefont
  {Schroder}, \citenamefont {Badertscher}, \citenamefont {Goudsmit},
  \citenamefont {Janousch}, \citenamefont {Leisi} \emph
  {et~al.}}]{Schroder:1999uq}%
  \BibitemOpen
  \bibfield  {author} {\bibinfo {author} {\bibfnamefont {H.}~\bibnamefont
  {Schroder}}, \bibinfo {author} {\bibfnamefont {A.}~\bibnamefont
  {Badertscher}}, \bibinfo {author} {\bibfnamefont {P.}~\bibnamefont
  {Goudsmit}}, \bibinfo {author} {\bibfnamefont {M.}~\bibnamefont {Janousch}},
  \bibinfo {author} {\bibfnamefont {H.}~\bibnamefont {Leisi}},  \emph
  {et~al.},\ }\href {\doibase 10.1016/S0370-2693(99)01237-X} {\bibfield
  {journal} {\bibinfo  {journal} {Phys.Lett.}\ }\textbf {\bibinfo {volume}
  {B469}},\ \bibinfo {pages} {25} (\bibinfo {year} {1999})}\BibitemShut
  {NoStop}%
\bibitem [{\citenamefont {Stoks}\ \emph {et~al.}(1994)\citenamefont {Stoks},
  \citenamefont {Klomp}, \citenamefont {Terheggen},\ and\ \citenamefont
  {de~Swart}}]{Stoks:1994wp}%
  \BibitemOpen
  \bibfield  {author} {\bibinfo {author} {\bibfnamefont {V.~G.~J.}\
  \bibnamefont {Stoks}}, \bibinfo {author} {\bibfnamefont {R.~A.~M.}\
  \bibnamefont {Klomp}}, \bibinfo {author} {\bibfnamefont {C.~P.~F.}\
  \bibnamefont {Terheggen}}, \ and\ \bibinfo {author} {\bibfnamefont {J.~J.}\
  \bibnamefont {de~Swart}},\ }\href {\doibase 10.1103/PhysRevC.49.2950}
  {\bibfield  {journal} {\bibinfo  {journal} {Phys.Rev.}\ }\textbf {\bibinfo
  {volume} {C49}},\ \bibinfo {pages} {2950} (\bibinfo {year} {1994})},\ \Eprint
  {http://arxiv.org/abs/nucl-th/9406039} {arXiv:nucl-th/9406039 [nucl-th]}
  \BibitemShut {NoStop}%
\bibitem [{\citenamefont {Torres}\ \emph {et~al.}(1986)\citenamefont {Torres},
  \citenamefont {Dalitz},\ and\ \citenamefont {Deloff}}]{Torres:1986mr}%
  \BibitemOpen
  \bibfield  {author} {\bibinfo {author} {\bibfnamefont {M.}~\bibnamefont
  {Torres}}, \bibinfo {author} {\bibfnamefont {R.}~\bibnamefont {Dalitz}}, \
  and\ \bibinfo {author} {\bibfnamefont {A.}~\bibnamefont {Deloff}},\ }\href
  {\doibase 10.1016/0370-2693(86)90744-6} {\bibfield  {journal} {\bibinfo
  {journal} {Phys.Lett.}\ }\textbf {\bibinfo {volume} {B174}},\ \bibinfo
  {pages} {213} (\bibinfo {year} {1986})}\BibitemShut {NoStop}%
\bibitem [{\citenamefont {Haidenbauer}\ and\ \citenamefont
  {Meissner}(2005)}]{Haidenbauer:2005zh}%
  \BibitemOpen
  \bibfield  {author} {\bibinfo {author} {\bibfnamefont {J.}~\bibnamefont
  {Haidenbauer}}\ and\ \bibinfo {author} {\bibfnamefont {U.-G.}\ \bibnamefont
  {Meissner}},\ }\href {\doibase 10.1103/PhysRevC.72.044005} {\bibfield
  {journal} {\bibinfo  {journal} {Phys. Rev.}\ }\textbf {\bibinfo {volume}
  {C72}},\ \bibinfo {pages} {044005} (\bibinfo {year} {2005})},\ \Eprint
  {http://arxiv.org/abs/nucl-th/0506019} {arXiv:nucl-th/0506019 [nucl-th]}
  \BibitemShut {NoStop}%
\bibitem [{Note2()}]{Note2}%
  \BibitemOpen
  \bibinfo {note} {The difference among the spectra in the charge basis is due
  to the interference effect with the $I=1$ amplitude~\cite
  {Nacher:1998mi}.}\BibitemShut {Stop}%
\bibitem [{\citenamefont {Tan}(1973)}]{Tan:1973at}%
  \BibitemOpen
  \bibfield  {author} {\bibinfo {author} {\bibfnamefont {T.~H.}\ \bibnamefont
  {Tan}},\ }\href {\doibase 10.1103/PhysRevD.7.600} {\bibfield  {journal}
  {\bibinfo  {journal} {Phys. Rev.}\ }\textbf {\bibinfo {volume} {D7}},\
  \bibinfo {pages} {600} (\bibinfo {year} {1973})}\BibitemShut {NoStop}%
\bibitem [{\citenamefont {Jido}\ \emph {et~al.}(2011)\citenamefont {Jido},
  \citenamefont {Oset},\ and\ \citenamefont {Sekihara}}]{Jido:2010rx}%
  \BibitemOpen
  \bibfield  {author} {\bibinfo {author} {\bibfnamefont {D.}~\bibnamefont
  {Jido}}, \bibinfo {author} {\bibfnamefont {E.}~\bibnamefont {Oset}}, \ and\
  \bibinfo {author} {\bibfnamefont {T.}~\bibnamefont {Sekihara}},\ }\href
  {\doibase 10.1140/epja/i2011-11042-3} {\bibfield  {journal} {\bibinfo
  {journal} {Eur. Phys. J.}\ }\textbf {\bibinfo {volume} {A47}},\ \bibinfo
  {pages} {42} (\bibinfo {year} {2011})},\ \Eprint
  {http://arxiv.org/abs/1008.4423} {arXiv:1008.4423 [nucl-th]} \BibitemShut
  {NoStop}%
\bibitem [{\citenamefont {Revai}(2013)}]{Revai:2012fx}%
  \BibitemOpen
  \bibfield  {author} {\bibinfo {author} {\bibfnamefont {J.}~\bibnamefont
  {Revai}},\ }\href {\doibase 10.1007/s00601-013-0619-z,
  10.1007/s00601-013-0694-1} {\bibfield  {journal} {\bibinfo  {journal} {Few
  Body Syst.}\ }\textbf {\bibinfo {volume} {54}},\ \bibinfo {pages} {1865}
  (\bibinfo {year} {2013})},\ \Eprint {http://arxiv.org/abs/1203.1813}
  {arXiv:1203.1813 [nucl-th]} \BibitemShut {NoStop}%
\bibitem [{\citenamefont {Nacher}\ \emph {et~al.}(1999)\citenamefont {Nacher},
  \citenamefont {Oset}, \citenamefont {Toki},\ and\ \citenamefont
  {Ramos}}]{Nacher:1998mi}%
  \BibitemOpen
  \bibfield  {author} {\bibinfo {author} {\bibfnamefont {J.}~\bibnamefont
  {Nacher}}, \bibinfo {author} {\bibfnamefont {E.}~\bibnamefont {Oset}},
  \bibinfo {author} {\bibfnamefont {H.}~\bibnamefont {Toki}}, \ and\ \bibinfo
  {author} {\bibfnamefont {A.}~\bibnamefont {Ramos}},\ }\href {\doibase
  10.1016/S0370-2693(99)00380-9} {\bibfield  {journal} {\bibinfo  {journal}
  {Phys.Lett.}\ }\textbf {\bibinfo {volume} {B455}},\ \bibinfo {pages} {55}
  (\bibinfo {year} {1999})},\ \Eprint {http://arxiv.org/abs/nucl-th/9812055}
  {arXiv:nucl-th/9812055 [nucl-th]} \BibitemShut {NoStop}%
\end{thebibliography}%

%

\end{document}